\documentclass[11pt,showpacs,amsmath,amssymb,nofootinbib]{revtex4}

\usepackage{graphicx}
\usepackage{amsfonts}
\usepackage{calrsfs}
\DeclareMathAlphabet{\pazocal}{OMS}{zplm}{m}{n}
\usepackage{bm}
\usepackage{amsmath}
\DeclareMathOperator\arctanh{arctanh}
\newcommand{\be}{\begin{equation}}
\newcommand{\ee}{\end{equation}}
\newcommand{\bea}{\begin{eqnarray}}
\newcommand{\eea}{\end{eqnarray}}
\newcommand{\beas}{\begin{eqnarray*}}
\newcommand{\eeas}{\end{eqnarray*}}
\newcommand{\bi}{\begin{itemize}}
\newcommand{\ei}{\end{itemize}}
\newcommand{\bn}{\begin{enumerate}}
\newcommand{\en}{\end{enumerate}}

\begin{document}

\title{Model Dynamics for Quantum Computing}
\author{Frank Tabakin}
 \affiliation{ Department of Physics and
Astronomy, University of Pittsburgh, PA 15260, USA}
\date{\today}                                         
\begin{abstract}
A model master equation suitable for quantum computing dynamics is presented.
In an ideal quantum computer (QC), a system of qubits evolves in time unitarily and, by virtue of their entanglement,  interfere  quantum mechanically to solve otherwise intractable problems.  In the real situation, a QC is subject to decoherence and attenuation effects due to interaction with an environment and with possible short-term random disturbances and gate deficiencies.  The stability of a QC under such attacks is a key issue for the development of realistic devices.  We assume that the influence of the environment can be incorporated by a master equation that includes unitary evolution with gates,  supplemented by a Lindblad term.
Lindblad operators of various types are explored; namely, steady, pulsed, gate friction, and measurement operators.  In the master equation, we use the Lindblad term to describe short time intrusions by random Lindblad pulses.
The phenomenological master equation is then extended to include a nonlinear Beretta term that describes the evolution of a closed system with increasing entropy.  An external Bath environment is stipulated by a fixed temperature in two different ways.  Here we explore the case of a simple one-qubit system in preparation for generalization to multi-qubit, qutrit and hybrid qubit-qutrit systems. 
This model master equation can be used to test the stability of memory and the  efficacy of quantum gates. The properties of such hybrid master equations are explored, with emphasis on the role of thermal equilibrium and entropy constraints.  Several significant  properties of time-dependent qubit evolution are revealed by this simple study.  \end{abstract}

\maketitle
%\tableofcontents
\section{Introduction}\label{sec1}
A quantum computer (QC) is a physical device that uses quantum interference to enhance the probability of getting an answer to an otherwise intractable problem~\cite{Nielsen,Preskill}.  A quantum system's ability to interfere depends on its entanglement and on maintenance of its coherent phase relations.  In a real system, there are always environmental effects and also random disturbances that can cause the quantum system to lose its ability to display quantum interference.  That process is called decoherence, as is discussed in an extensive literature~\cite{Joos} on how a quantum system becomes classical, often rapidly, due to its interaction with an external environment.   That process might also be viewed as a measuring device~\cite{Zurek,Hornberger}.  A major concern in the development of a realistic quantum computer is to understand, control,  and/or correct for detrimental environmental effects.

A general theory of how such ``open systems" evolve in time is provided by the operator sum representation (OSR),  which replaces the unitary evolution of a closed system by a more general form that accounts for the fact that the system under study (the quantum computer) is affected by an environment.  That general form involves Kraus~\cite{Kraus}  operators and is often described as a mapping. 

To gain insight, models for the environment and its interaction with the QC have been discussed~\cite{Carmichael}.  These studies  use the 
Dyson series for the time evolution and reduce the dynamics to the QC subsystem using a variety of approximations.  One approximation,  truncating the Dyson series ( a Born approximation) is used since an exact solution is  generally not available.  An oft-used  approximation is that in the
 system-environment interaction the environment restores itself rapidly to its initial condition,
 and therefore only the present situation of the environment is relevant.  That is,  one invokes a Markov approximation, which has the environment affecting the system, but the system's effect on the environment vanishes rapidly.  It is assumed that the system and environment are initially uncorrelated and are described as a product  state.

The Markov approximation is not always applicable;  that depends on the dynamics of the environment and its interaction with the system.  It is physically possible that the system affects an environment that is able to  partially preserve that influence and feed part of it back to the system.
 Indeed, there are  important papers that indicate that the Markov approximation is in doubt~\cite{Alicki}.  Nevertheless, our initial approach is to adopt the Markov approximation,  with plans to  test its applicability.

A vast literature exists on deducing a general master equation for the time evolution of  a subsystem's density matrix.  One method is to examine the above Dyson series methods, as described in~\cite{Carmichael}.   Another approach is to replace unitary evolution~\footnote{$\rho(t) = U(t) \rho(0) U^\dagger(t) .$}, by a Kraus subsystem
 form of  evolution~\footnote{ $ \rho(t) = \sum\limits_\alpha K_\alpha (t)\,  \rho(0)\,  K^\dagger_\alpha (t)$ with $\sum\limits_\alpha K_\alpha (t) K^\dagger_\alpha=1$ 
and $ \rho^\dagger(t)=\rho\,  ; \   {\rm Tr}\rho(t)=1$ .}.   Then a ``generalized infinitesimal" expansion of the Kraus operators is used to deduce a differential equation for the density matrix,  yielding a linear in density matrix master equation.  The ``generalized infinitesimal" is related to Ito calculus,  which involves  how to define variations  for stochastic, or rapidly fluctuating functions.  Several papers~\cite{Adler,Peres} discuss this procedure and  ways to deduce a master equation for the subsystem's density matrix,  which prove to be of the form deduced earlier by Gorini, Kossakowski, Sudarshan and Lindblad ~\cite{ Lindblad1,Lindblad2,Lindblad3,Lindblad4},  who used  other general considerations.  The Lindblad form can be deduced in general  from 
requiring preservation of density matrix Hermiticity, unit trace, and positive-definite properties, without the restriction to time independent  Lindblad operators or to
particular initial states.  The Lindblad form can be used to describe all environmental effects. However, we introduce special terms to isolate and to gain insight into specific effects, such as the  Bath-system dynamics.
  Lindblad's equation for the time evolution of a subsystem's density matrix has many appealing features, as discussed later.

Profound papers by Beretta et al.~\cite{Beretta1, Beretta2,  Beretta3} provide a general master equation  based on novel concepts of  non-equilibrium statistical mechanics as applied to quantum systems. The Beretta master equation  for describing an open system has not been used much for QC perhaps because it is nonlinear and it alters entropy addition rules.    An excellent rendition of the basic nature of the Beretta  (and other nonlinear master equations ) is  provided in papers by Korsch et al.~\cite{Korsch}. 

  The Beretta et al. description  includes definitions of entropy,  work, and heat for non-equilibrium systems.  Their resultant master equation has many important features, as illustrated in our study.  Indeed, we incorporate their master equation for QC and extend it using a phenomenological  viewpoint.

In section~\ref{sec2},  we  discuss the basic idea of a density matrix from  two traditional points of view.  One view is 
to form a classical ensemble average over an ensemble of quantum systems.  The second viewpoint is that  a single quantum system is prepared  in a state averaged over  its production possibilities.  After discussing  general properties of the density matrix,  we stress that the dynamics of the density matrix can be best visualized in terms of
the time development of its spin polarization and spin correlation observables.

In section~\ref{sec3}, we describe how the density matrix evolves by unitary evolution driven by a  time-dependent Hamiltonian that includes level splitting and ideal gate pulses.  During a gate pulse, a bias pulse is used to impose temporary degeneracy to halt precession and  avoid awkward phase accumulation.  Then a model master equation is introduced in section~\ref{sec4} along with an analysis of its several terms. 
These terms are: (1) Lindblad form used to include random noise and gate friction effects, (2) a Beretta term to describe a closed-system with increasing entropy,
and (3) a Bath term for contact with an environment of specific temperature.  We make a heuristic assumption that the Lindblad and Beretta forms are both useful, and  we adopt a phenomenological or hybrid viewpoint.  That viewpoint,  which is not derived but postulated,  is to use Lindblad terms to describe random, short-time intrusions on the QC,  which could be caused by for example a passing particle.  And we use the nonlinear Beretta term to describe the overall trend for a closed system to steadily increase in entropy.  A Beretta description of Bath-system (open) dynamics is also included.
The system is simultaneously driven towards equilibrium with an environment or bath of a specified temperature.   On top of these  major effects,  we describe the time-evolution of quantum gates by Hamiltonian pulses.  If the Lindblad and Bath terms are  both set to zero,  which is equivalent to removing the environment,  and the closed-system Beretta term is omitted, then the master equation reduces necessarily to ordinary unitary evolution.

In section~\ref{sec4c},  we introduce the Lindblad master equation and discuss several aspects of its general features, and limitations.
We then proceed to a similar analysis of the Beretta and Bath terms~\ref{sec4d}.
Finally, in section~\ref{sec5} properties of the full master equation are examined and conclusions and future plans are
stated in section~\ref{sec6}.

The above assumptions provide a practical dynamic framework for examining  not only the influence of an environment on the efficacy of a QC,  but also the loss of reliability in the action of gates or the general loss of coherence. The master equation we design incorporates the main features of a density matrix; namely, Hermiticity, unit trace and positive definite character,  while also including the evolution of a closed system and the effects of gates,  noise and of an external bath.

\section{Density Operator}
\label{sec2}

The density operator, also called a density matrix, is a operator in Hilbert space that represents an ensemble of quantum systems.  As introduced by von Neumann and Landau~\cite{Neumann,Landau,Fano},  the density operator, $\rho,$  can be understood as a classical ensemble average over a collection of subsystems (the ensemble) which occur in a general state $\mid \alpha \rangle,$ with a probability ${\cal P}_\alpha.$  By a general state, we mean a state of the subsystem that is a general superposition of a complete orthonormal basis (such as eigenstates of a Hamiltonian).  For the simple case of an ensemble of spin 1/2 particles, such a state called a qubit is specified by the spinor
\begin{equation} 
\mid \alpha  \rangle = \mid \hat{n}_\alpha \rangle =
\cos(\theta_\alpha/2) e^{-i \phi_\alpha/2} \mid 0 \rangle +\sin(\theta_\alpha/2) e^{+i \phi_\alpha/2} \mid 1 \rangle
= \left(
\begin{array}{l}
\cos(\theta_\alpha/2) e^{-i \phi_\alpha/2}  \\
\sin(\theta_\alpha/2) e^{+i \phi_\alpha/2}
\end{array}\right) \, , 
\end{equation} where the computational basis $\mid 0 \rangle$ and $\mid 1 \rangle$ denote spin-up and spin-down states, respectively, and  $\alpha$  labels the Euler angles $\theta_\alpha, \phi_\alpha,$ that specify a general direction  $\hat{n}_\alpha$ in which
the qubit is pointing.   Indeed, the above state is an eigenstate of the operator $\vec{\sigma}\cdot\hat{n}_\alpha\ ,$  where the components of $\vec{\sigma}$ are the Pauli operators ( see below).
This description is readily generalized to multiparticle qubit states and also to systems that are doublets without being associated with the idea of physical spin.

The above general state is normalized but not necessarily orthogonal, $  \langle\alpha'\mid \alpha \rangle \neq \delta_{\alpha', \alpha}\  .$  The quantum rule for the expectation value of a general operator $\Omega,$ is
$ \langle\alpha \mid \Omega \mid \alpha  \rangle$ and for an ensemble of  separate quantum subsystems one can form the classical  ensemble average $ \langle\Omega \rangle$  for the  Hermitian observable $ \Omega$ by taking
\be
 \langle\Omega \rangle =  \frac{\sum_{\alpha}  \langle\alpha \mid \Omega \mid \alpha  \rangle
{\cal P}_\alpha}{\sum_{\alpha} {\cal P}_\alpha}.
\ee The ensemble average is then a simple classical average
where ${\cal P}_\alpha$ is the probability that a particular
state $\alpha$ appears in the ensemble. Summing over all
possible states of course yields $\sum\limits_{\alpha} {\cal P}_\alpha=1$.
The above expression is a combination of a classical ensemble
average with the quantum mechanical expectation value. It
contains the idea that each member of the ensemble interferes
only with itself quantum mechanically and that the ensemble
involves a simple classical average over the probability
distribution of the ensemble.

We now define the {\bf density operator} by
\be
\rho \equiv \sum_{\alpha}\ \mid \alpha \rangle \langle\alpha \mid {\cal P}_\alpha \  .
\ee
Using closure~\footnote{The closure property, which is a statement that $\mid n \rangle$ is a complete 
orthonormal basis, is $ \sum_n \mid n  \rangle \,  \langle n \mid  =1.$},  the ensemble
average can now be expressed as a ratio of
traces
\be
 \langle\Omega \rangle= \frac{{\rm Tr}(\rho\,  \Omega)}{{\rm Tr}( \rho )}\equiv{\rm Tr}( \rho\, \Omega),
\ee
which entails the properties 
\bea
{\rm Tr} ( \rho ) &=&  \sum_m  \langle m\mid \rho\mid m \rangle
= \sum_\alpha\ {\cal P}_\alpha \sum_m   \langle\alpha \mid m \rangle \langle m\mid \alpha \rangle\nonumber \\
&=& \sum_\alpha\ {\cal P}_\alpha    \langle\alpha \mid  \alpha \rangle
= \sum_\alpha\ {\cal P}_\alpha=1,
\eea  where $\mid m \rangle$ denotes a complete orthonormal basis (such as the computational basis),
and
\bea
{\rm Tr} ( \rho\, \Omega)  &=&
\sum_{m m'}   \langle m\mid \rho \mid m' \rangle \langle m'\mid \Omega\mid m \rangle \nonumber \\
&=& \sum_\alpha \sum_{m m'} {\cal P}_\alpha  \langle\alpha\mid m' \rangle
  \langle m' \mid \Omega\mid m \rangle  \langle m\mid \alpha \rangle \nonumber \\
&=&\sum_\alpha {\cal P}_\alpha  \langle\alpha\mid \Omega \mid \alpha \rangle ,
\eea
which returns the original ensemble average expression.

\subsection{Properties of the Density Matrix}

 The
definition $ \rho = \sum_{\alpha}\ \mid \alpha \rangle \langle\alpha \mid {\cal
P}_\alpha $ is  a general one, if we interpret $\alpha$
as the label for the possible characteristics of a state. Several
important general properties of a density operator follow from this definition.
The density operator is:
\begin{itemize}
\item Hermitian $ \rho^\dagger \equiv \rho \, ,$ hence its eigenvalues
are real;
\item has unit
trace, $ {\rm Tr}( \rho)\equiv 1 ,$  hence the sum of its eigenvalues equals 1;

\item  is positive definite, which
means that all of its eigenvalues are greater or equal to zero.
This, together with the fact that the density matrix has unit
trace, ensures that each eigenvalue is between zero and one, and yet sum to 1.

\item for a pure state, every member of the ensemble has
the same quantum state and only one   $\alpha_0$ 
appears and the density operator becomes $\nobreak{\rho
= \mid \alpha_0 \rangle \langle\alpha _0 \mid}$. The state $\mid \alpha_0 \rangle $ is
normalized to one and hence for a pure state $\rho^2 = \rho$. 
 Thus for a
pure state one of the  density matrix eigenvalues is 1, with all others
zero.

 \item
for a general ensemble $\rho ^2 \leq \rho$  which has a mixture of possibilities as reflected
in the probability distribution ${\cal P}_\alpha$
with the equal sign holding for pure states. 
\end{itemize}
\subsubsection{ Composite Systems and Partial Trace}

For a composite system, such as an ensemble of quantum systems
each of which is prepared with a probability distribution,
 the definition of a density matrix can be generalized to a
 product Hilbert space form involving systems of type A and B
\be \rho_{A B} \equiv  \sum_{\alpha, \beta}  {\cal P}_{\alpha,
\beta} \mid \alpha \beta \rangle \langle \alpha \beta \mid, \ee  where ${\cal
P}_{\alpha, \beta}$  is the joint probability for finding the two
systems with the attributes labelled by $ \alpha$ and $\beta.$ For
example, $\alpha$ could designate the possible directions
$\hat{n}_\alpha$ of one spin-1/2 system, while $\beta$ labels the
possible spin directions of another spin 1/2 system, $\hat{n}_\beta$.  One can
always ask about the state of just system A or B by summing over or
tracing out the other system. For example,  the density matrix of
system A is picked out of the general definition above by the following partial trace steps
 \bea
 \rho_A &=&  {\rm Tr}_B ( \rho_{A B} )  \nonumber \\
  &=&  \sum_{ \alpha, \beta} \  {\cal P}_{ \alpha, \beta} \mid \alpha \rangle  \langle \alpha\mid
  {\rm Tr}_B {\Big \lbrack}  \ \mid \beta \rangle \langle \beta \mid \ {\Big \rbrack}    \nonumber \\
&=&  \sum_{ \alpha}  ( \sum_{ \beta}  {\cal P}_{\alpha, \beta} )
\mid \alpha \rangle  \langle \alpha\mid
   \nonumber \\
   &=&  \sum_{\alpha}  {\cal P}_{\alpha}  \mid \alpha \rangle  \langle \alpha\mid .
 \eea
Here we use the product space $ \mid \alpha \beta \rangle\mapsto \mid \alpha \rangle \mid \beta \rangle$ and
we define the probability for finding system A in situation $\alpha$ by
\be
{\cal P}_{\alpha} =\sum_{ \beta}  {\cal P}_{\alpha, \beta}.
\ee
This is a standard way to get an individual probability from a joint probability.

It is easy to show that all of the other properties of a density matrix still hold
true for a composite system case. It has unit trace, it is Hermitian with real
eigenvalues and is positive definite.

\subsection{Comments about the Density Matrix}

\subsubsection{Alternate Views of the Density Matrix}

In the prior discussion, the view was taken that the density
matrix implements a classical average over an ensemble of many
quantum systems, each member of which interferes quantum
mechanically only with itself. An alternate equally
valid viewpoint is that a single quantum system is prepared, but the
preparation of this single system is not pinned down. Instead all
we know is that it is prepared in any one of the states labelled
again by a generic state label $\alpha$ with a probability
${\cal P}_{\alpha}$. Despite the change in interpretation, or
rather an application to a different situation, all of the
properties and expressions presented for the ensemble average
hold true; only the meaning of the probability is altered.

An important point concerning the density matrix is that
the ensemble average (or the average expected result for a
single system prepared as described in the previous paragraph)
can be used to obtain these averages for all observables $\Omega$.
Hence in a sense the density matrix describes a system and
the system's accessible observable quantities. It represents then
an honest statement of what we can really know about a system. On
the other hand, in Quantum Mechanics it is the wave function
that tells all about a system. Clearly, since a density matrix
is constructed as a weighted average over bilinear products
of wave functions, the density matrix has less detailed information
about a system than is contained in its wave function. Explicit
examples of these general remarks will be given later.

To some authors the fact that the density matrix has less
content than the system's wave function, causes them to avoid
use of the density matrix. Others find the density matrix description
of accessible information as appealing. Indeed, S. Weinberg in recent papers~\cite{Weinberg1,Weinberg2} 
 has advocated  an
interpretation of quantum mechanics  based on using the density matrix rather than the
state vector as a description of reality.  The  attribution of such deep physical meaning to the density operator was
advocated earlier by  Hatsopoulos and Gyftopoulos~\cite{Hatsopoulos}, who inspired by a deep analysis by Park~\cite{Park} on the nature of quantum states, adopted it as the key physical ansatz of their early theory of quantum thermodynamics, which in turn prompted Beretta to design the nonlinear master equation that we adopt below as part of our model master equation.

  We now turn to discussing the basic features of the density matrix in preparation 
 for describing its dynamic evolution by means of a model master equation.

\subsubsection{ Classical Correlations and Entanglement}

The density matrix for composite systems can take many forms depending
 on how the systems are prepared.  For example, if distinct systems A \& B are
  independently produced and observed independently,  then the density
  matrix is of product form
$\rho_{AB} \mapsto \rho_A \otimes \rho_B,$ and the observables are also
 of product form
$\Omega_{AB} \mapsto\Omega_A\otimes\Omega_B.$  For such an uncorrelated
situation,  the ensemble
average factors
\be
 \langle \Omega_{AB} \rangle = \frac{{\rm Tr}( \rho_{AB}\Omega_{AB})}
{{\rm Tr}( \rho_{AB})}= \frac{{\rm Tr}( \rho_{A}\Omega_{A})}{{\rm Tr}( \rho_{A})}
 \frac{{\rm Tr}( \rho_{B}\Omega_{B})}{{\rm Tr}( \rho_{B})}=  \langle \Omega_{A} \rangle  \langle \Omega_{B} \rangle ,
\ee  as is expected for two separate uncorrelated experiments.
This can also be expressed as having the joint probability factor
${\cal P}_{\alpha, \beta}\mapsto{\cal P}_{\alpha  }  {\cal P}_{ \beta}$
the usual probability rule for uncorrelated systems.

Another possibility for the two systems is that they are prepared in a
coordinated manner,
with each possible situation assigned a probability based on the
correlated preparation technique.  For example, consider two colliding beams, A \& B,
made up of particles with the same spin. Assume the particles are produced
in matched pairs with common spin direction $\hat{n}.$ Also
assume that the preparation of that pair in that shared direction
is produced by design with a
 classical probability distribution
${\cal P}_{\hat{n}}.$  Each pair has a density matrix $
\rho_{\hat{n}} \otimes \rho_{\hat{n}}$ since they are produced
separately,  but their spin directions are correlated classically.
The density matrix for this situation is then
\be \rho_{AB} =
\sum_{\hat{n}} {\cal P}_{\hat{n }} \  \rho_{\hat{n}} \otimes
\rho_{\hat{n}}.
\ee
This is a ``mixed state" which represents
classically correlated preparation and hence any density matrix
that can take on the above form can be reproduced by a setup using
classically correlated preparations and does {\it not} represent
the essence of Quantum Mechanics, e.g.  an entangled state.

An entangled quantum state is described by a density matrix (or by
its corresponding state vectors) that is not and can not be
transformed into the two classical forms above; namely, cast into
a product or a mixed form. For example, the two-qubit Bell state
$\frac{1}{\sqrt{2}} (\mid 01  \rangle + \mid 10 \rangle)$ has a density matrix
\be
\rho= \frac{1}{2}( \mid 01 \rangle  \langle 01 \mid + \mid 01 \rangle  \langle 10 \mid
                + \mid 10 \rangle  \langle 01 \mid + \mid 10 \rangle  \langle 10 \mid \, )
\ee
that is not of simple product or mixed form. It is the prime
example of an entangled state.

The basic idea of decoherence can be described by considering the
above Bell state case with time dependent coefficients
\be
\rho= \frac{1}{2}(a_1(t) \mid 01 \rangle  \langle 01 \mid +a_2(t) \mid 01 \rangle  \langle 10 \mid +a_2^*(t)
\mid 10 \rangle  \langle 01 \mid + (1-a_1(t)) \mid 10 \rangle  \langle 10 \mid \, ) .
\ee If the
off-diagonal terms $a_2(t)$ vanish, by attenuation and/or via
time averaging, then the above density matrix does reduce to the
mixed or classical form, 
\be
\rho= \frac{1}{2}(a_1(t) \mid 01 \rangle  \langle 01 \mid + (1-a_1(t)) \mid 10 \rangle  \langle 10 \mid \, ) ,
\ee
which is an illustration of how
decoherence leads to a classical state.

\subsection{ Observables and the Density Matrix}

Visualization of the density matrix and understanding its significance is greatly enhanced by defining associated real spin observables.
In the simplest one-qubit case, the density matrix is a $2 \times 2$ Hermitian positive definite matrix of unit trace.  Thus it is fully stipulated by three real parameters,  which are identified as the polarization vector $\vec{P}(t),$  also called the Bloch vector.  One can deduce that only three parameters are needed  for the one-qubit case from the following steps: (1) a general $2 \times 2$ matrix with complex entries involves $2(2 \times 2)=8$ real numbers; (2) the Hermitian condition
reduces the diagonal terms to 2 and the off-diagonal terms to 2, a net of 4 remaining real numbers; (3)  the unit trace reduces the count by 1, so we have 4-3=3 parameters.  These steps generalize to multi-qubit and to qutrit cases.

Operators or gates acting on a single qubit state are represented by 2 $\times$ 2   matrices.  The dimension of the  single qubit state vectors
(  $ | 0  \rangle$ and $| 1  \rangle$ ) is $N=2^{n_q}=2,$
with $n_q=1.$  The Pauli matrices provide an operator basis of all such matrices.  The Pauli-spin matrices are:
\begin{eqnarray}\label{Pauli} 
 \sigma_0=
 \left(
\begin{array}{lc}
1 & 0\\
0 & 1
\end{array}\right)   ,
 \sigma_1=
 \left(
\begin{array}{lc}
0 & 1\\
1 & 0
\end{array}\right)   ,
\sigma_2 =
 \left(
\begin{array}{lc}
0 & -i \\
i & 0
\end{array}\right)  \, ,
\sigma_3  = \left(
\begin{array}{lc}
1 & 0\\
0 & -1
\end{array}\right) .
 \end{eqnarray} These are all Hermitian traceless matrices $\sigma_i =
\sigma^\dagger_i$.  We use the labels $(1,2,3)$ to denote the directions $(x,y,z).$
The fourth Pauli matrix $\sigma_0$ is simply the unit matrix.
Any $2 \times 2$ matrix can be constructed from  these four Pauli matrices,  which therefore are an operator basis, also called the computational basis operators.
That construction applies to the density matrix $\rho(t) $ at any time t and to the Hamiltonian $H(t) ,$  and  Lindblad operators $L(t).$

\subsubsection{Polarization}

The general form of a one-qubit density matrix,  using the 4  Hermitian Pauli matrices as an operator basis is:
	\begin{eqnarray}\label{rho1} 
\rho(t)&=&\frac{1}{2} {\Big \lbrack}   \sigma_0 + P_1 (t)\   \sigma_1+ P_2(t)\   \sigma_2+P_3(t)\  \sigma_3  {\Big \rbrack} \\  \nonumber
&=& \frac{1}{2}  {\Big \lbrack}\   \sigma_0 +\vec{ P}(t)\cdot \vec{\sigma} \   {\Big \rbrack} \\  \nonumber
	 &=& 
\frac{1}{2} \left(
 \begin{array}{lc}
1+P_3(t) & P_1(t) - i P_2(t)  \\
 P_1(t) + i P_2(t)  & 1-P_3(t) 
\end{array}\right),
\end{eqnarray}
where the spin operators are
$\vec{\sigma}=\left \{  \vec{\sigma}_1,\vec{\sigma}_2,\vec{\sigma}_3 \right \},$  and  the real polarization vector is
$\vec{P}(t)=\left \{ P_1(t),P_2(t),P_3(t) \right \}.$ 
 The polarization, $  \vec{ P}(t) $ is  a real vector, which  follows from the Hermiticity of the
density matrix $\rho^\dagger(t) \equiv \rho(t)$ and from the ensemble average relation
$$
\vec{P}(t)= {\rm Tr}( \vec{\sigma} \cdot \rho(t) ) \equiv  \langle \vec{\sigma} \rangle.
$$  Thus specifying the polarization vector ( also called the Bloch vector) determines the density matrix and it  is
 convenient to view the polarization as a function of time to gain insight into qubit dynamics.

The above expression clearly satisfies the density matrix conditions that ${\rm Tr}(\rho(t))=1 ,\rho^\dagger(t)=\rho(t).$  The positive definite condition follows from
determining  that the two eigenvalues are $\lambda_1(t)= \frac{ 1+ \mathbf{P}(t)}{2}, \lambda_2= \frac{ 1- \mathbf{P}(t)}{2},$  where
 $\mathbf{P}(t)= \sqrt{  P_1(t)^2 +P_2(t)^2+P_3(t)^2 }\leq 1.  $  The unit trace condition becomes simply that the eigenvalues of $\rho$ sum to one
 \bea
 {\rm Tr}(\rho(t))= 1 &=& {\rm Tr}(U_{\rho}(t)\cdot \rho_D(t) \cdot U_{\rho}^\dagger(t))   \\ \nonumber
  &=& {\rm Tr}((U_{\rho}^\dagger(t) \cdot U_{\rho}(t)) \cdot \rho_D(t)) \\ \nonumber
   &=&  {\rm Tr}( \rho_D(t))  \\ \nonumber  
 &=&   \lambda_1(t) + \lambda_2(t) , \eea where $U_{\rho}(t)$ is the unitary matrix that diagonalizes the density matrix at time t.  The diagonal density matrix $\rho_D(t)$ has  real eigenvalues along the diagonal. The positive definite condition now asserts that each of 
 these eigenvalues is greater or equal to zero and less than or equal to one:  $0\leq \lambda_i(t)\leq 1,$   while summing to 1.
 For the one qubit case the above conditions mean that   $\lambda_1(t) + \lambda_2(t)=  1,$ and since $0\leq \lambda_i(t)\leq 1,$ the polarization vector
 must have a length between zero and one.
 
 Note that the density matrix, polarization vector and its eigenvalues in general depend on time.  Indeed, the dynamics of a one-qubit system 
 is best visualized by how the polarization or eigenvalues change in time.
 
 The polarization operator is simply $\Omega=\vec{\sigma}$ and we have the following relations for the value and time derivative of the polarization vector:
 \bea
\vec{P}(t) &=& {\rm Tr}( \, \vec{\sigma}\cdot\rho(t)\, )  =  \langle \vec{\sigma}  \rangle\\ \nonumber
\frac{d \vec{P}(t)}{dt} &=& {\rm Tr}(\,  \vec{\sigma} \cdot  \frac{ d \rho(t)}{dt} \,).
 \eea 
 
Much of what is presented here applies to multi-qubit and qutrit cases. The main difference for more qubits/qutrits is an increase in the number of polarization and
spin correlation observables.

Several other quantities are used to monitor the changing state of a quantum system.  Later energy, power, heat transfer and temperature concepts will be discussed.  Next   purity, fidelity, and  entropy  attributes will be examined. 

\subsubsection{Purity} 

%${\pazocal P}$
%${\mathcal P}$

 The purity ${\pazocal P}(t)$ is defined as
${\pazocal P}(t) =  \langle \rho(t)  \rangle = {\rm Tr}(\, \rho(t)\, \rho(t)\, ).$  It is called purity since for a pure state density matrix $\rho = |\psi \rangle  \langle\psi | ,$
$\rho^2=\rho$ and ${\rm Tr}(\, \rho^2\, )= {\rm Tr}(\, \rho)\, =1,$ but in general  ${\rm Tr}(\, \rho^2\, )\leq 1.$  For a pure state, we see that
$\rho^2=\rho,$ implies that  each eigenvalue satisfies $\lambda_i ( \lambda_i - 1) =0,$ so $ \lambda_i=0 \ \text{or}\  1.$  Since the eigenvalues sum to 1,
a pure state has one eigenvalue equal to one, all others are zero. A mixed or impure state has $ \sum_{i=1,2^{n_q}}\ 
   \lambda_i^2  < 1\, ,$ which indicates that the nonzero eigenvalues are less than 1.

For a one-qubit system, the purity is simply related to the polarization vector
\bea
{\pazocal P}(t)&=&  {\rm Tr}(\rho^2(t))= \frac{ 1 + {\mathbf P}^2(t)}{2}, \\ \nonumber
\frac{d {\pazocal P}(t)}{d t}&=& 2 \,  {\rm Tr}(\, \rho(t) \  \frac{ d \rho(t)}{dt} \,)= {\vec P}(t) \cdot  \frac{d}{dt}{\vec P}(t)\, ,
\eea where  ${\mathbf P}(t)$ is the length of the polarization vector  ${\vec P}(t) $.  Thus a pure state has a polarization vector that is on the unit Bloch sphere, whereas an
impure state's polarization vector is inside the Bloch sphere. 
The purity ranges from a minimum of 0.5 to a maximum of 1. Later we will see how dissipation and entropy changes can bring the polarization inside the Bloch sphere and hence generate impurity.

\subsubsection{ Fidelity}

Fidelity  measures the closeness of two states.   In its simplest form, this quantity can be defined as ${\cal F}^2={\rm Tr}[\,\rho_A \   \rho_B\,].$  For the special case that
$\rho_A = \mid \psi_A   \rangle   \langle \psi_A \mid \  \&\   \rho_B = \mid \psi_B   \rangle  \langle  \psi_B \mid ,$  this yields ${\cal F} \equiv   |   \langle \psi_A \mid \psi_B   \rangle |, $ which is clearly the magnitude of the overlap probability amplitude.

 To align the quantum definition of fidelity with classical probability theory,  a more general definition is invoked; namely,
\be
{\cal F}(\rho_A , \rho_B ) \equiv {\rm Tr}{\Big  \lbrack}  \sqrt{  \sqrt{\rho_A }\, \      \rho_B \, \     \sqrt{\rho_A } } \   {\Big  \rbrack}\, .
\ee  When $\rho_A$ and  $\rho_B$ commute, they can both be diagonalized by the same unitary matrix,  but with different eigenvalues.
In that limit, we have $\rho_A = \sum\limits_{i} \lambda^A_i \mid i  \rangle   \langle i \mid \ \&\ \rho_B = \sum\limits_{j} \lambda^B_j \mid j  \rangle   \langle j \mid $ and 
\be
{\cal F}(\rho_A , \rho_B ) \equiv {\rm Tr}{\Big  \lbrack}  \sqrt{ \rho_A \, \     \rho_B \, }     \   {\Big  \rbrack} \equiv \sum\limits_{i,j }\sqrt{ \lambda^A_i  \lambda^B_j}\, ,
\ee  which is the classical limit result.

We will use fidelity to monitor the efficacy or stability of any QC process, where $\rho_A(t)$ is taken as the exact result and $\rho_B(t)$ is the result including decoherence, gate friction, and
dissipation effects.  

%We also define a time-averaged fidelity by
%\be
%\tilde{ {\cal F}}(\rho_A , \rho_B ) \equiv  \frac{1}{T_L} \int^{T_L}_0 {\cal F}(\rho_A , \rho_B ) dt,
%\ee where $T_L$ is the Larmor precession period.

\subsubsection{ Entropy}

The Von Neumann~\cite{Neumann} entropy at time t is defined by
\be
S(t)=- {\rm Tr}(  \rho(t)\, \log_2 \, \rho(t) ).
\label{entropy1} 
\ee  The Hermitian density matrix can be diagonalized by a unitary matrix  $U_\rho(t)$ at  time t,  $$\rho(t)= U_\rho(t) \   \rho^D(t) \   U_\rho^\dagger(t), $$
where $ \rho_D(t)$ is diagonal matrix  $\rho^D(t)_{i,j}= \delta_{i,j} \lambda_i $ of the eigenvalues.  Then
\be
S(t)=-  ( \lambda_1 \,  \log_2\, \lambda_1+\lambda_2 \,  \log_2\, \lambda_2). 
\label{entropy1s}
\ee With a base 2 logarithm,  the maximum entropy for one qubit  is $S_{max} \equiv 1$   which occurs when the  two eigenvalues are all equal to $1/2.$  That is the
most chaotic, or least information situation.  The minimum entropy of zero obtains when one eigenvalue is one, all others being zero; 
 that is the most organized, maximum information situation.  For one qubit,  zero entropy places the polarization vector on the Bloch sphere, where the length of the polarization vector is one.
If the polarization vector moves inside the Bloch sphere,  entropy increases.  For $n_q$ qubits entropy ranges between zero and $n_q$.

For later use,  consider the time derivative of the entropy 
  \begin{eqnarray}\label{entropy2}
\frac{d S}{dt} &=&  - \sum_{i=1,2} (\frac{d \lambda_i}{dt}   \log_2(  \lambda_i) +\frac{ d\lambda_i}{dt}) \\  \nonumber
&=&  -{\rm Tr}( \frac{d  \rho}{dt}  \log_2(\rho(t) ) ).
 \end{eqnarray}  Since ${\rm Tr}(\rho) =\sum_{i=1,2} \lambda_i  \equiv 1,$ the second RHS term above vanishes.  Note that the above result is derived assuming that
 for all eigenvalues $\lambda_i/\lambda_i\rightarrow 1,$ which is ambiguous for zero eigenvalues. This is no doubt related to  divergences that could arise when say  
 $\lambda_1=0,$ and $\frac{d \lambda_1}{dt}$  is nonzero.  We will confront this issue later.
 
 Note that the eigenvalues,  purity, fidelity and entropy all depend on the length of the polarization vector ${\mathbf P}(t)$.

\section{First Steps towards a  Master Equation Model-- Unitary Evolution, Gates and Pulses }
\label{sec3}

The master equation for the time evolution of the system's density matrix is now presented.  We are interested in developing a simple model that incorporates the main features of the qubit dynamics for a quantum computer.  These main features include seeing how the dynamics evolve under the action of gates and the role of both closed system dynamics and of  open system decoherence, dissipation and the system's approach to equilibrium.  From the density matrix we can determine a variety of observables,  such as the polarization vector,  the power and heat rates,  the purity,  fidelity, and  entropy all as a function of time.

\subsection{Unitary  evolution} 
We start with the observation that the density matrix for a closed system  is driven by a Hamiltonian $H(t),$  that can be explicitly time dependent,
as $ \rho(t) = U(t) \rho(0) U^\dagger(t),$ where the unitary operator is $ U(t)= e^{ -\frac{i}{\hbar} H(t)\, t}.$  For infinitesimal time increments this yields
the unitary evolution or commutator term:
  \be  \label{Meqn1} 
\frac{d \rho}{dt} =  - \frac{i}{\hbar} {\Big  \lbrack}  H(t), \rho(t){\Big  \rbrack}.  
 \ee  This term specifies the reversible motion of a closed system.  To include dissipation, an additional operator ${\pazocal L}$ 
 will be added  $\dot{ \rho} =  - \frac{i}{\hbar} {\Big  \lbrack}  H(t), \rho(t){\Big  \rbrack}+{\pazocal L}$ which describes an irreversible open system.

 \subsection{Hamiltonian}
\begin{figure}[!bp]
 \includegraphics[width=\textwidth]{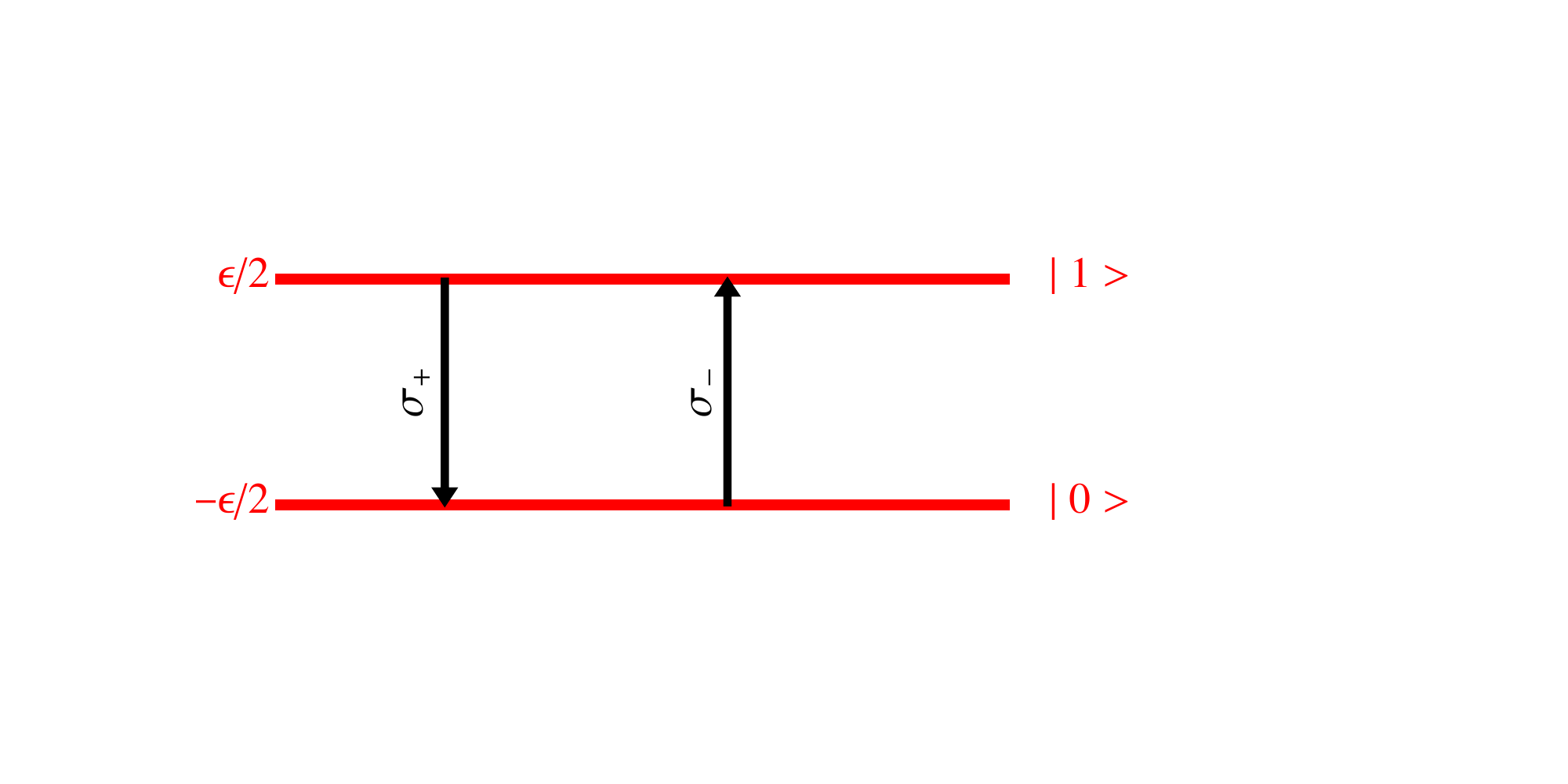}
\caption{The qubit levels with splitting  $\epsilon = \ \hbar\, \omega_L .$  With our conventions the operator $\sigma_{-}\equiv \frac{ \sigma_x - i   \sigma_y}{\sqrt{2}}$  raises the qubit to the polarization-down state $ \mid 1  \rangle,$  while $\sigma_{+}\equiv \frac{ \sigma_x + i   \sigma_y}{\sqrt{2}}$ lowers the qubit to the polarization-up ground state $ \mid 0  \rangle$.}
\protect\label{figlevels}
\end{figure}

   Our  Hamiltonian $H(t)= H_0 + V(t)$ is  an Hermitian operator in spin space;  for one qubit it is a $2 \times 2$ matrix.  
    It consists of a time independent  $ H_0,$ plus a time dependent part $V(t).$  
For $ n_q=1$, a typical Hamiltonian is  
\be
\label{Ueqn1} 
 H_0 \equiv - \frac{1}{2} \ \hbar\, \omega_L \   \vec{\sigma}\cdot \hat{z} = - \frac{1}{2} \ \hbar\, \omega_L \   \sigma_z ,
\ee
 which describes a 2 level system with eigenvalues
 $- \frac{1}{2} \ \hbar \omega_L $ for state $ \mid 0  \rangle,$ and $  + \frac{1}{2} \ \hbar \omega_L$ for state $\mid 1  \rangle,$  see Fig.~\ref{figlevels}. 
 
 The polarization vector for this case precesses about the $ \hat{z}$ direction with the Larmor  angular frequency $ \omega_L.$  This follows from the unitary evolution term
 \bea
 \frac{d \vec{P}(t)}{dt} &=& {\rm Tr}(\,  \vec{\sigma}\    \frac{ d \rho(t)}{dt} \,) \\ \nonumber
&=&  -i  \frac{  \omega_L}{2}\,   {\rm Tr}{\Big \lbrack}  \vec{\sigma} \cdot   [ \vec{\sigma}\cdot \hat{z}, \rho(t)] {\Big  \rbrack}   \\ \nonumber
&=& - \vec{\omega}_L\times \vec{P}(t),
 \eea where
$\vec{\omega}_L=\omega_L \   \hat{z},$
    which is a Larmor precession of the polarization vector about the direction $\hat{z}.$  
  The  polarization vector then has a fixed value of $P_z$ and the x and y components
vary  as   
\bea
  P_x(t)&=&  P_x(0) \cos(  \omega_L  t ) + P_y(0)  \sin(  \omega_L  t ) \\ \nonumber
 P_y(t)&=& P_y(0) \cos(  \omega_L  t ) - P_x(0)  \sin(  \omega_L  t ).
 \eea The above  is equivalent to $ \dot{P}_i =
 \sum\limits_{j=1,3}  M_{i,j} P_j,$ with
 \begin{eqnarray}\label{Mmatrix0} 
 M=
 \left(
\begin{array}{lcl}
\ \ 0 & \omega_L &\  0\\
-\omega_L & 0&\  0\\
\ \  0 &\ 0 &\  0
\end{array}\right)   .
\end{eqnarray}   This form will be extended to dissipative cases later.

\begin{figure}[!tbp]
 \includegraphics[width=\textwidth]{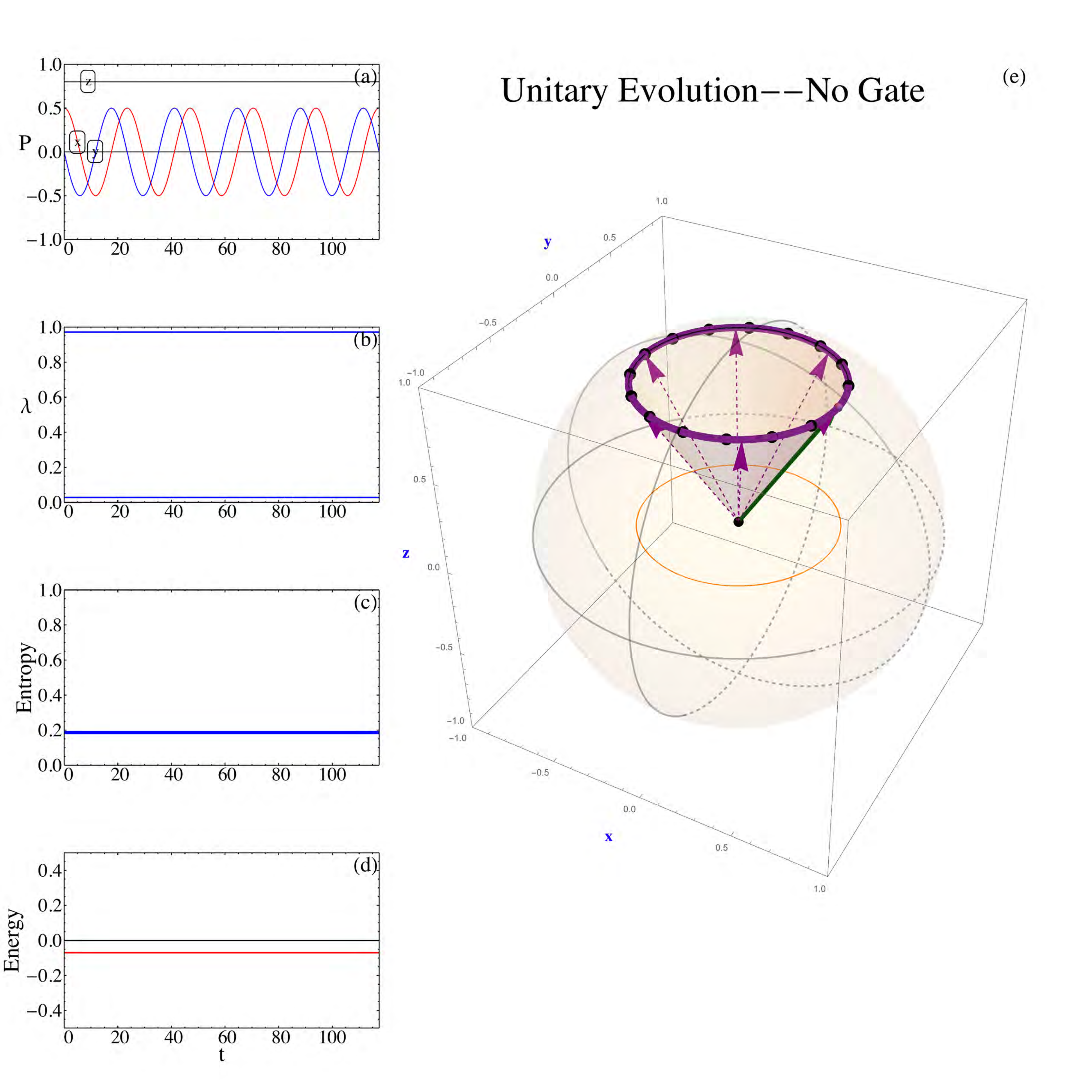}
\caption{Polarization vector precession (no gates and no dissipation): (a) fixed $P_z$ and oscillating
$P_x, P_y$ components versus time, (b) the two (fixed) eigenvalues $\lambda$ of $\rho(t),$
(c) the fixed entropy, and  (d) the fixed energy (power and heat rate are zero).
 The Bloch sphere (e) with solid vector indicating the initial location of the polarization
 (which originates from the center of the  Bloch sphere)  while the subsequent motion follows the thick path as also shown by the  dashed polarization vectors at subsequent times.
 The dots indicate equal time interval locations of the polarization vector. 
 The precession is  also  projected to the x-y plane.  The initial density matrix and level parameters are in Table~\ref{table1}.  }
\protect\label{precess1}
\end{figure}

 Thus the level splitting $ \hbar\, \omega_L$ produces a  precessing polarization  with a fixed z-axis value and circular motion in the  x-y plane ( see Fig.~\ref{precess1}). 
  The basic Hamiltonian $H_0$ is  selected to be time independent. 
The initial density matrix and level splitting parameters used in our examples~\footnote{All of the numerical examples in this paper were generated by Mathematica codes based on the QDENSITY/QCWAVE packages~\cite{qdensity}.} are listed in Table~\ref{table1}.
 Energy is in $\mu$eV,  frequency in GHz and  time is in nanoseconds (nsec).

\begin{table}[h]
\caption{Initial Density Matrix \& Level Parameters}
\centering
\begin{tabular}{c c  }
\hline\hline
Name & Value   \\ [0.5ex] 
\hline
$P_1$&  0.5  \\
$P_2$&  0.0  \\
$P_3$&  0.8  \\
${\mathbf P}$&0.943  \\
Initial Purity& 0.945  \\
Initial Entropy&0.186  \\
Initial Temperature& 0.93 mK \\ 
Larmor frequency $\omega_L $& 0.2675 GHz \\
Larmor Period $T_L$ &  23.5 nsec \\
Level split $\hbar  \omega_L $&0.1761 $\mu$eV  \\
\hline
\end{tabular}
\label{table1}
\end{table}
 Next we add a time dependence in the form of  Hamiltonian 
pulses that produce quantum gates.  

%\newpage
 \subsubsection{One-qubit ideal gates}
 
 For our QC application, the Hamiltonian $H(t)=H_0 + V(t)$ is used to incorporate two effects.  The first is the level splitting, Eq.~\ref{Ueqn1}.   Here $ \omega_L$ denotes the Larmor angular frequency  associated with the level splitting $\hbar  \omega_L ,$  which sets the  Larmor time scale $T_L = 2 \pi/\omega_L$ for the system. 
 The $V(t)$ term is used to include quantum gates $\Omega,$ which are $2 \times 2$ Hermitian matrices.  For example,  a single qubit  NOT gate is
$\Omega =\sigma_1.$  The NOT acts as: $ \sigma_x  |0  \rangle= |1 \rangle\ \&\  \sigma_x |1  \rangle= |0 \rangle.$  This basic gate is simply a spinor rotation about the $\hat{x}$ axis by $\pi$ radians.   Clearly, two NOTs return to the original state. $NOT.NOT = \sigma_1^2={\pazocal I}_2.$  

A gate operator  $\Omega$ is introduced as a  Hamiltonian generator $V_G(t)$
 \be
V_G(t) \equiv  \hbar\,  \theta_G(t)\, \Omega , 
\ee  where  $\theta_G(t)$ is a gate pulse that is centered at time $t_0$  with a width $\tau.$  The pulse $\theta_G(t)$ has inverse time units.
 Thus the pulse essentially starts at $t_1=t_0 -\tau/2 $ and ends at $t_2=t_0 +\tau/2;$  we typically take this pulse to be of Gaussian form,
 $\theta_G \rightarrow \theta_g   $
 \be
 \theta_g(t)=\frac{\sqrt{\pi}}{2 \tau}\  e^{-(\frac{t - t_0}{\tau})^2} .
 \ee   
We call $V_G(t)$ the gate generator~\footnote{ The unitary operator associated with this gate generator is:
$U_G(t,t_1) = e^{ -\frac{i}{\hbar}\int_{t_1}^{t} V_G(t') dt' } $.} since it generates the effect of a specific gate. 
The pulse function $\theta_G(t)$ is designed to generate a suitable rotation over an interval $t_1$ to $t_2.$   Since we want to have a smooth pulse, we take these pulses to be of either  Gaussian $\theta_g(t)$ or soft square 
 $\theta_s(t)$ shape.   The soft square shape is defined by
 $$\theta_s(t)= N_f \  \frac{1}{2}\,    {\Big  \lbrack} {\rm Erf}(\frac{t-t_1}{\tau})-  {\rm Erf}(\frac{t-t_2}{\tau}) {\Big  \rbrack}  ,$$ where $N_f$ is fixed by the $\int_{-\infty}^{\infty}   \theta_G(t)\  dt= \frac{\pi}{2}$ condition.

The NOT gate pulse represents a series of infinitesimal rotations about the x-axis and in order to give the correct NOT gate effect, we need to normalize the pulse by
$\int_{-\infty}^{\infty}   \theta_G(t)\  dt= \frac{\pi}{2}.$ The same form can be applied to a one qubit Hadamard  
\be
\Omega ={\mathcal H} = 
 \frac{1}{\sqrt{2}} \left(
\begin{array}{lc}
1 & 1\\
1 & -1
\end{array}\right)= \frac{\sigma_1+\sigma_3}{\sqrt{2}}, 
\ee which is a spinor rotation about the $(\hat{x} + \hat{z})/\sqrt{2}$   axis by $\pi$ radians.

 \subsubsection{Bias  gates}

 Application of such a gate pulse does not carry out our objective of achieving a NOT gate, unless we do something to remove the level splitting at least during the action of the pulse. This corresponds to a temporary stoppage of precession.  We therefore, introduce a bias pulse which is designed to make the levels degenerate during the gate pulse. The strength of the bias is adjusted by some type of non-intrusive monitoring, or by fore-knowledge of the fixed level splitting, to temporarily establish level degeneracy.   During the action of the gate,  the levels have to be completely degenerate, otherwise disruptive phases accumulate.  Therefore, we use a soft square bias pulse that straddles the time interval of the gate pulse.  The soft square bias pulse shape  is defined by:
 $ \theta_B(t)\equiv \theta_s(t)/\theta_s(t_0) ,$ 
 which is preferred over a square pulse since it has finite derivatives and thus yields smooth variations of power as shown later. The width of the above bias pulse is $\tilde{t}_2-\tilde{t}_1,$ and $\tau$ is the thickness of the edges. To be sure that no precession occurs during a gate the time values used in the bias pulse  $\tilde{t}_2$ and $\tilde{t}_1$ are taken to be slightly larger and slightly smaller than the gate pulse values $t_2 $ and $t_1.$ 

 The bias pulse $ \theta_B(t)$ is added to the Hamiltonian  to create a temporary degeneracy as 
\be
V_B(t) =  \frac{\hbar\, \omega_L}{2} \   \theta_B(t) \  \sigma_3  ,\ee where the bias normalization is $\int_{-\infty}^{\infty}  \theta_B(t)\  dt=  1.$ Note $\theta_B$ is unitless, whereas  $\theta_G$  has 1/time units.
   
Combining these terms we have for a single pulse, with gate and bias
$$H_1(t) =  - \frac{\hbar\, \omega_L}{2} \,  ( \, 1-   \theta_B(t)\, )\  \sigma_3 + \hbar \, \theta_G(t)\ \Omega\, .$$ 
Here we see that the bias turns off precession  and the gate term generates the action of a gate $\Omega\, .$
 Without a bias pulse to produce level degeneracy, awkward phases accumulate 
   that are detrimental to clean-acting gates. Aside from intervals when the gate and the bias pulse act,  the polarization
  vector precesses at the Larmor frequency, which is zero for degenerate levels.  The bias pulse is simply an action to stop the precession,
  then the gate pulse rotates the qubit,  and subsequently precession is restored  once the bias is removed.  That process is equivalent to stopping a spinning top,
  rotate it,  and then get it spinning again,  which requires some work.   As discussed later the power supplied to the system during a gate pulse is determined by
  $$\frac{d }{dt}W(t)={\rm Tr} ( \rho(t) \ \dot{H}(t))=  +\frac{ \hbar\, \omega_L}{2} \,  \dot{\theta}_B(t)\,  P_z(t)+\hbar\, \dot{\theta}_G (t)  \langle\Omega \rangle .$$  The derivative of the Hamiltonian divides into a  gate plus a bias term.
   
\subsubsection{Gate and Bias Cases}

\begin{figure}[!tbp]       
    \includegraphics[width=\textwidth]{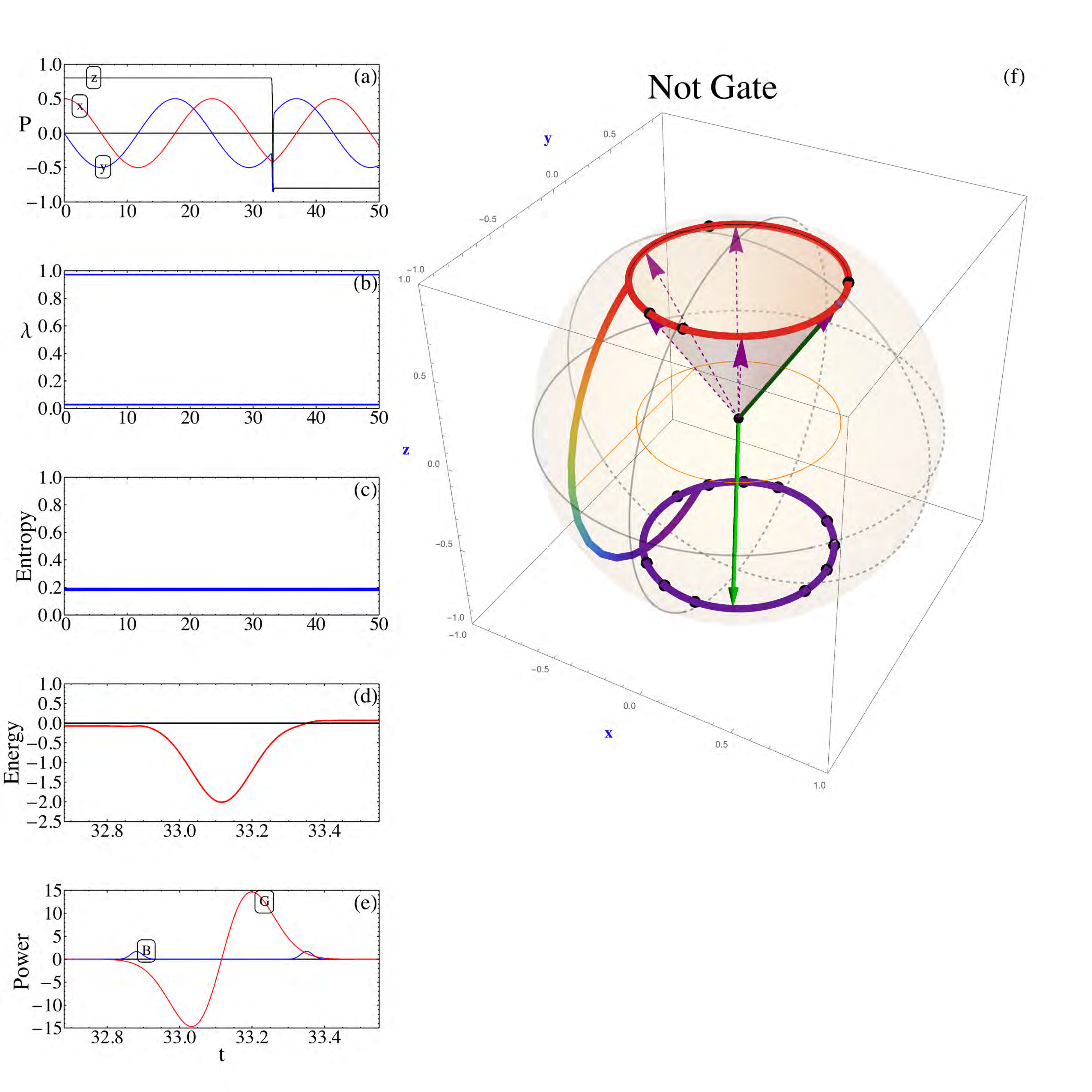}  
    \caption{Polarization vector trajectory for unitary Not Gate with level splitting and bias. Dissipation is off ${\pazocal L\rightarrow 0}.$
(a) changes in polarization with time, $ P_x \rightarrow P_x \,  , P_y \rightarrow -P_y, \,  ,P_z \rightarrow  -P_z  $ during  Not Gate pulse ; (b) the two (fixed) eigenvalues $\lambda$ of $\rho(t) ;$
(c) the fixed entropy ; (d) the energy versus time ; (e)Power by gate (G) and bias (B) (heat rate is zero). here $P_y$ is
negative during the gate ; and (f) precession about positive $\hat{z}$ is moved to  $-\hat{z}$ axis by NOT gate.  The dots indicate equal time interval locations of the polarization vector. }
\label{Not1}
\end{figure}

\begin{figure}[!tbp]       
    \includegraphics[width=\textwidth]{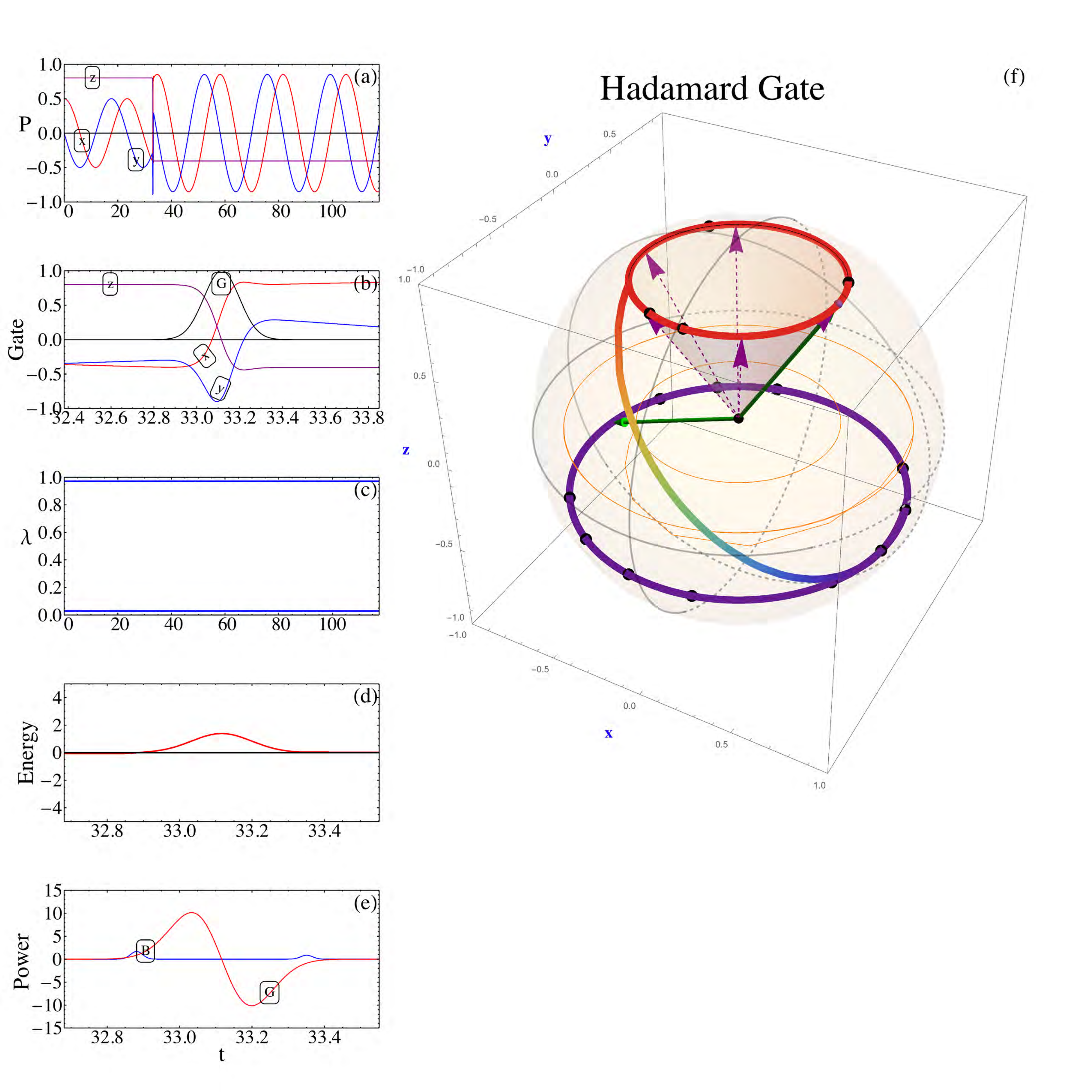}  
    \caption{Polarization vector trajectory for a unitary Hadamard gate with level splitting and bias. Dissipation is off ${\pazocal L\rightarrow 0}.$
    (a) Polarization versus time; (b) Polarization evolution $ P_x \rightarrow P_z \, , P_y \rightarrow -P_y, \, , P_z \rightarrow  P_x  $ during Hadamard gate pulse(G) when gate starts $t_1=32.9$\, nsec, (Px,Py,Pz)=(-0.402,-0.301,0.798) ; (c) the two (fixed) eigenvalues $\lambda_1,\lambda_2$ of $\rho(t) ;$
 (d) the energy versus time ; (e) Power by gate (G) and bias (B) (heat rate is zero) ; and (f) Polarization vector trajectory, precession about positive $\hat{z}$ continues after polarization is moved as shown by Hadamard gate.
}
\label{Had1}
\end{figure}

 In Figures~\ref{Not1}-\ref{Had1},   the one-qubit polarization vector motion for a NOT and a Hadamard gate are shown
 with no dissipation (${\pazocal L}\rightarrow 0$) and with a bias pulse acting during the gate pulse.  The detailed case shows that during the NOT pulse
 one gets the expected  change of $ P_x \rightarrow P_x ; \  P_y \rightarrow -P_y ; \ P_z \rightarrow  -P_z\   .$
  The power supplied to the system during the NOT gate is also displayed separately for the  gate power  and the bias power.
 These are explained by the bias power $=  (\hbar\, \omega_L/2 )\  \dot{\theta}_B(t) P_z(t)\, $ and the gate power
  $= \hbar\,  \dot{\theta}_G\ P_x(t)  ,$
 where the x-polarization is fixed during the NOT gate,  but the z-polarization flips.  The values of the polarization from the time $t_1$ when the gate pulse starts
 to its end at $t_2$ explain the shapes seen in Fig~\ref{Not1}.

During the Hadamard pulse
one gets the expected change of $ P_x\rightarrow P_z ; \   P_y\rightarrow -P_y ; \  P_z\rightarrow P_x .$
The power supplied to the system during the Hadamard gate is also displayed separately for the  gate power  and the bias power.
 These are explained by the bias power $= P_z(t)\  (\hbar\,  \omega_L/2 )\  \dot{\theta}_B(t)$ and
 the gate power $=\hbar\,  (P_x(t) + P_z(t)\, )\   \dot{\theta}_G\, / \sqrt{2},$
 where the y-polarization flips during the Hadamard gate,  and  the z and x polarization interchange.  The values of the polarization during the pulse
  explain the shapes seen in  Fig~\ref{Had1}.
   
  The gate pulses can produce net work done on the system. No heat transfer occurs by way of the gate or bias,  that exchange arises later from dissipation.
  After the gate pulses are complete, the precession continues about the $\hat{z}$ axis.

\begin{figure}[!tbp]       
    \includegraphics[width=\textwidth]{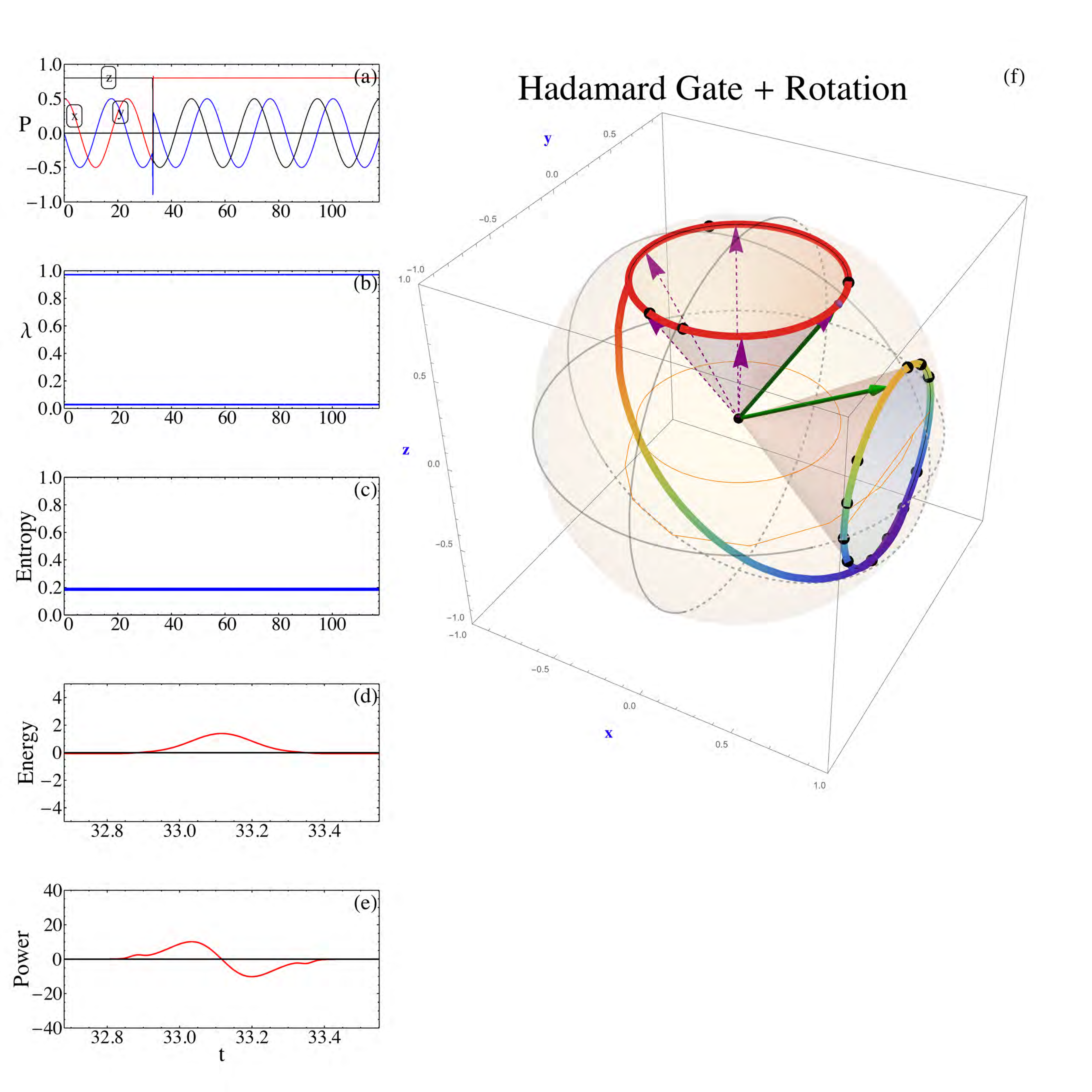}  
    \caption{Polarization vector trajectory for unitary Hadamard Gate plus bias and with ${\pazocal L}$ off.
    The precession axis is changed from the $\hat{z}$ to the $\hat{x}$ axis during the gate pulse.
(a) changes in polarization  $ P_x \rightarrow P_x \, , P_y \rightarrow -P_y, \, P_z \rightarrow  -P_z  $ during Hadamard gate pulse; 
 (b) the two (fixed) eigenvalues $\lambda_1,\lambda_2$ of $\rho(t) ;$
(c) the fixed entropy ; (d) the energy versus time changes due to gate, bias and Hamiltonian axis rotation; (e) Power by gate , bias  and Hamiltonian rotation (heat rate is zero) ;  (f) precession starts about $\hat{z}$ axis and continues about $\hat{x}$ axis after polarization is moved as shown by Hadamard gate.  The dots indicate equal time interval locations of the polarization vector. }
\protect\label{Had2}
\end{figure}

Another case of a Hadamard gate is shown in Fig.~\ref{Had2}.  In this case,  the Hamiltonian is smoothly rotated from $\hat{z} $ to $ \hat{x}$ during the Hadamard gate pulse.  This Hamiltonian rotation, which is equivalent to rotating a level splitting  magnetic field from the z to x direction,
is accomplished by setting:
\bea
H_1(t) &\rightarrow&  - \frac{\hbar\, \omega_L}{2} \,  ( \, 1-   \theta_B(t)\, ) (\eta (t_0-t)\  \sigma_3 + \eta(t-t_0)\   \sigma_1 ) + \hbar\, \theta_G(t)\  {\pazocal H}, \\ \nonumber
\eta(t)&\equiv& \frac{ 1 +{\rm Erf}(t/a)}{2}.
\eea   Here $\eta$ is a smooth step function of width $a.$  As a result the precession which  started around the $\hat{z}$ continues about the $\hat{x}$ axis after the Hadamard  gate pulse as shown in Fig.~\ref{Had2}.  

\newpage
\subsubsection{Gate pulses and instantaneous  gates}

To fully replicate the results obtained when a set of instantaneous  gates act, as in a QC algorithm, it is necessary to invoke additional steps.  One possible step is to apply a bias pulse over the full set of gate pulses,  thereby making the qubits degenerate during a QC action,  including final measurements.  Another way, which we prefer,  is to let the precession continue between gate pulses,  which means that each gate acts with an associated bias pulse, as illustrated earlier.  Then  one needs to design the gate pulses and associated measurements to act at appropriate times to replicate  the standard
description of instantaneous gates.  For example, we define a delay time $T_D$ as an integer $n_D$  multiple of the Larmor period $T_L.$  The first gate starts at a time $t^{(1)}_1\equiv T_D = n_D T_L.$    The first pulse ends at a time $t^{(1)}_2\equiv t^{(1)}_1+\tau.$
The next gate starts  at a time $t^{(2)}_1\equiv t^{(1)}_1 +T_D$ and ends at a time $t^{(2)}_2\equiv t^{(2)}_1 +\tau.$  This setup repeats for $N_G$ gates and yields the final time that we use to define the completion of the QC process as $T_f\equiv N_G\, T_D + \tau  =N_G\, T_D + t_2 - t_1.$  At the time $T_f$ the action of the gates is complete and the corresponding density matrix $\rho(T_f)$ is the same as the instantaneous, static gate result $U_G \rho(0)  U^\dagger_G,$  where $U_G$ is a product of the $N_G$ gate operators.   The general result for the final time is
$$
T_f = N_G\  T_D +  \tau  +  n_T\  T_L   ,
$$ 

\begin{figure}[!htbp]       
\includegraphics[width=\textwidth]{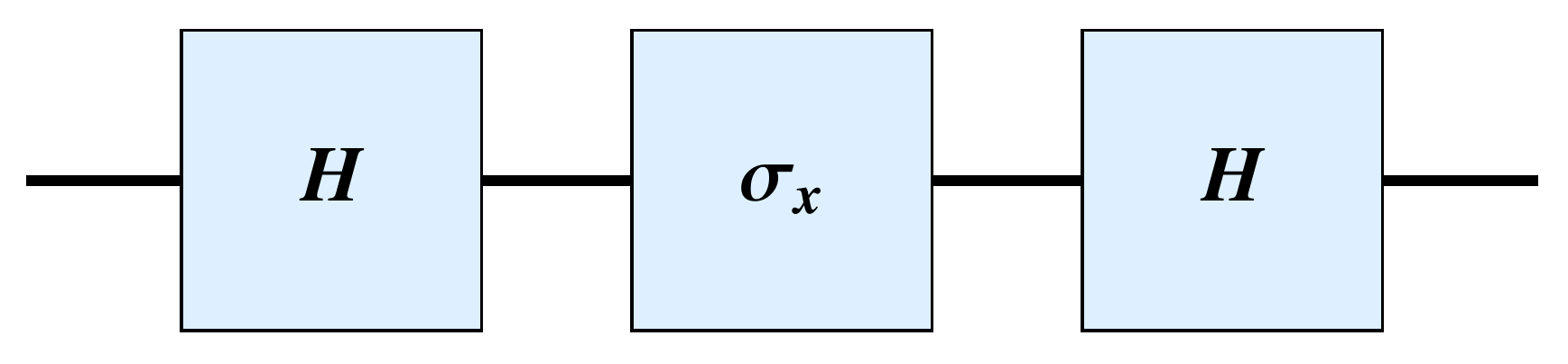}  
\includegraphics[width=\textwidth]{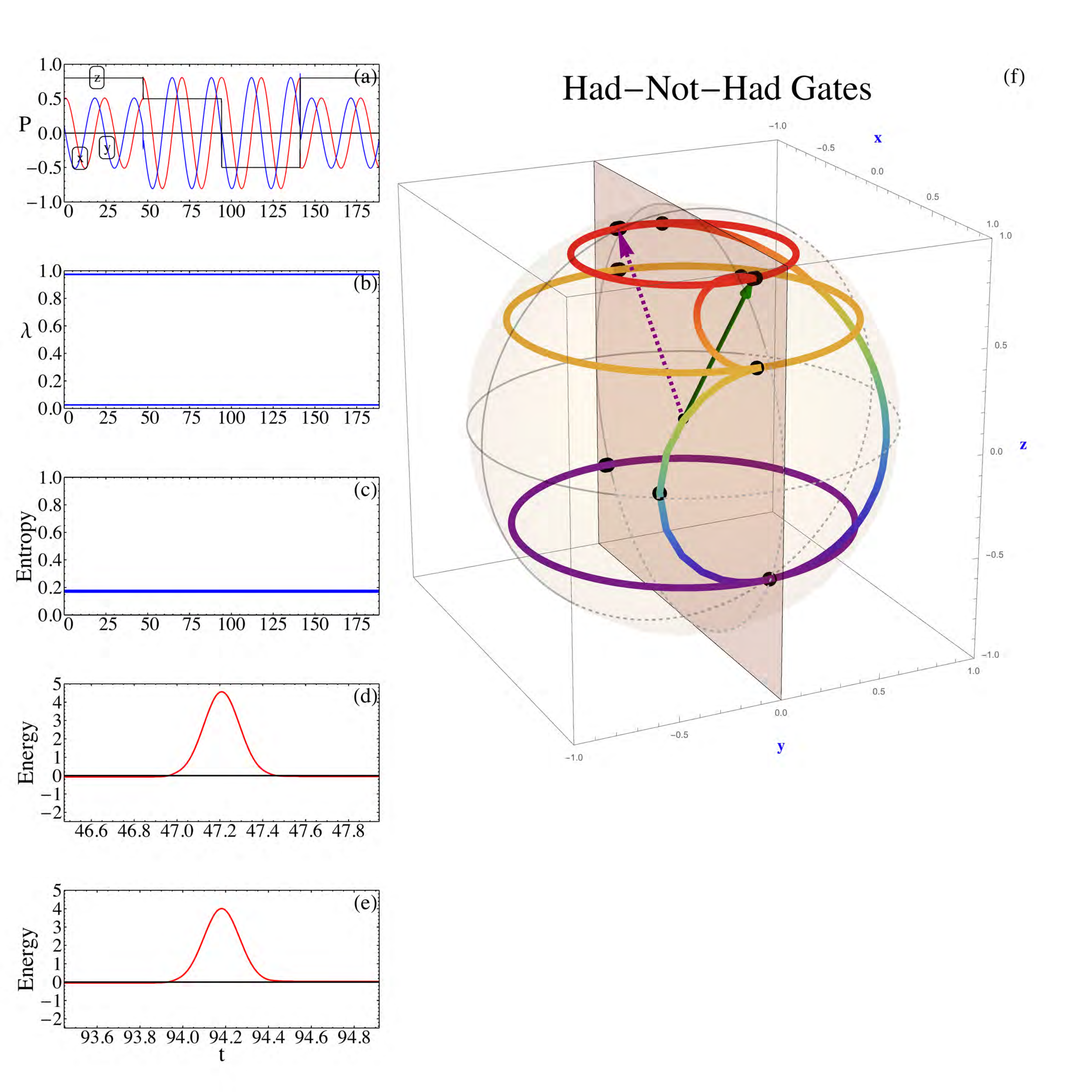} 
 \caption{Polarization for Hadamard-Not-Hadamard pulse gates. (a) Polarization vectors, (b) Eigenvalues, (c) Entropy,  (d) Energy of first pulse, (e) Energy of second pulse, all versus time.  Then in (f) Trajectories and initial and final polarization vectors, along with the three rapid gate pulses. The initial  polarization vector (solid green arrow ) $(P_x,P_y,P_z)=( 0.5 ,0.1 , 0.8)$  and final polarization vector (dashed purple arrow )  $( -0.5 ,-0.1 , 0.8)$ are seen to give the expected $\sigma_z$ gate result. }
\protect\label{FIGHNH}
\end{figure}

\begin{figure}[ht]       
\includegraphics[width=\textwidth]{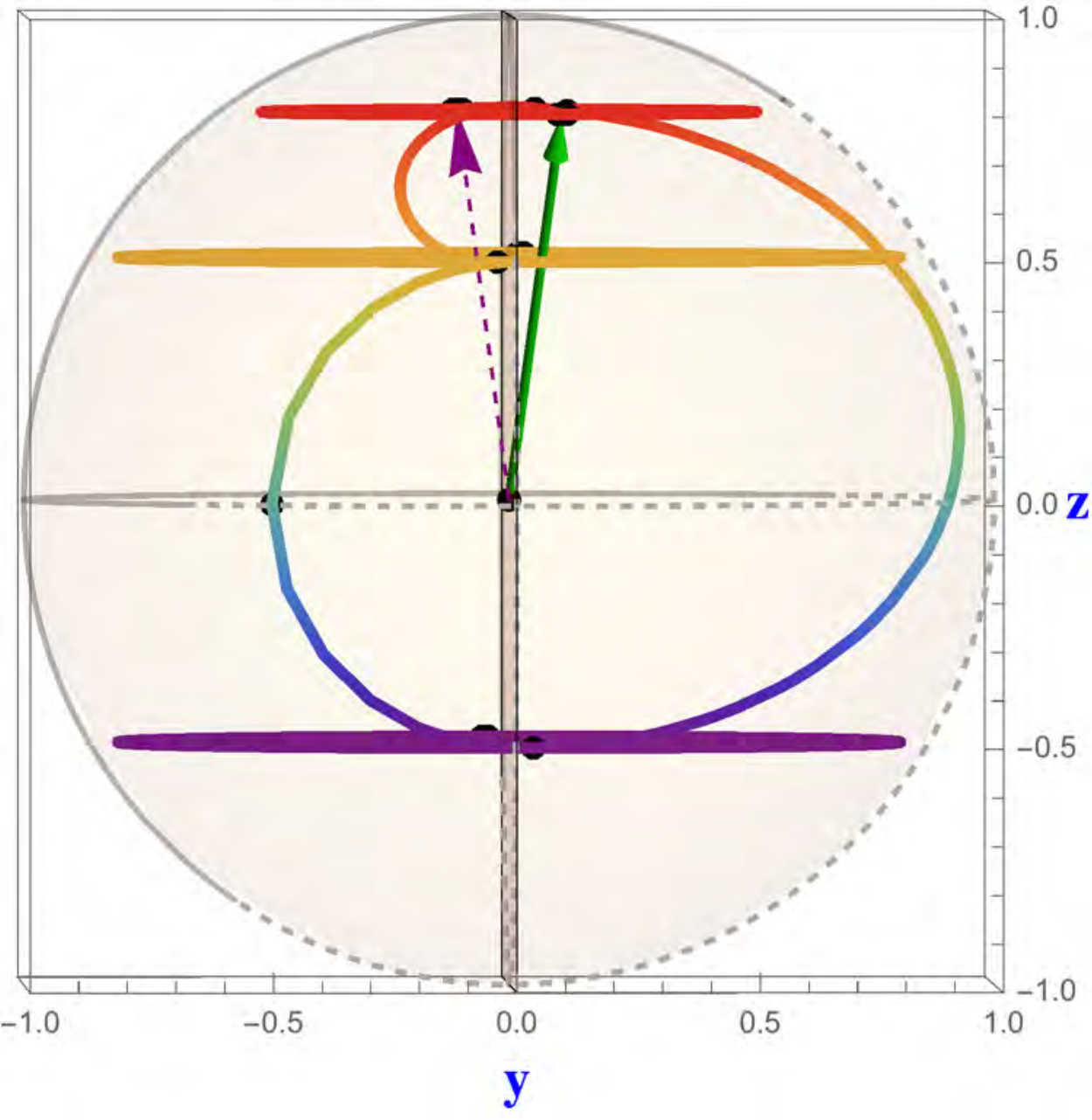} 
    \caption{ Polarization trajectories for Hadamard-Not-Hadamard pulse gates projected onto the y-z plane.
     The first Hadamard downward pulse is seen with the next Not downward pulse, and then the final upward Hadamard  pulse.    In this example,  the initial  polarization vector (solid green arrow ) $(P_x,P_y,P_z)=( 0.5 ,0.1 , 0.8)$  and final polarization vector (dashed purple arrow )  $( -0.5 ,-0.1 , 0.8)$ are seen  projected onto the y-z plane.} \protect\label{FIGHNH2}
\end{figure}

For example, consider a three gate case for one qubit  $U_G\equiv   {\pazocal H}   \cdot  \sigma_x\cdot  {\pazocal H}\equiv \sigma_z,$  which is a three gate $N_G=3$ Hadamard-Not-Hadamard sequence.  This case is illustrated in Fig.~\ref{FIGHNH}.  At the first stage before the gate acts,  the polarization vector precesses about the z-axis,  then the Hadamard acts  at time $t^{(1)}_1$ and the polarization path moves rapidly  to the second lower precession circle at time 
$t^{(1)}_2.$  After a few precessions, the Not gate brings the path to the negative z region at time $t^{(2)}_2.$  Finally, the final Hadamard lifts the path back up to the original precession cone, but with a phase change.
The final result  at time $T_f= 3 \, T_D + t_2-t_1$ is obtained by the transformation $P_x \rightarrow - P_x , P_y \rightarrow - P_y , P_z \rightarrow  P_z, $
which is, as it should be,  equivalent to the action of a single $\sigma_z$  gate.  The projected version also shown in 
 Fig.~\ref{FIGHNH2} displays this process and the finite time dynamic gate actions.

This process can be implemented for any set of gate pulses and can be generalized to multi-qubit/qutrit cases.  Thus the pulsed gate approach  can replicate the standard instantaneous static gate description by carefully designing the timing of the gates on the Larmor precession time grid.  This requires examining or measuring the gates at the selected time $T_f.$  If the Larmor period $T_L$ varies in time, this procedure can be generalized.

For a sequence of $N_G$ gates 
 \bea
  H(t)&=&\sum\limits_{i=1}^{N_G}   H_1(t-t^0_i) \\ \nonumber
  &=& \sum\limits_{i=1}^{N_G}  -\hbar\,  \frac{\omega_L}{2}\,  ( \, 1-   \theta_B(t-t^0_i)\, )  \sigma_3 + \theta(t-t^0_i)\ \Omega_i
     \eea where
   $\Omega_i$ is the ith gate acting at the time centered at $t_0+ t^0_i.$  This can generate a chain of gates.
   
   We conclude that one can replicate the  action of instantaneous static gates, which is central to the usual description of QC algorithms, by including a bias pulse during the gate action, and by applying the gates on the Larmor period time-grid.
\clearpage

    \subsection{  Schr\"odinger, Heisenberg, Dirac ( Interaction) and Rotating Frame Pictures}
  
    In our treatment, we use the Schr\"odinger picture for the density matrix,  so that all aspects of the dynamics are described by the density matrix through its polarization and spin correlation observables.  Other choices are to either use the Heisenberg picture, where the time development is incorporated into the Hermitian operators, or use the Dirac or Interaction picture,  wherein the operators evolve in time with the ``free" Hamiltonian $H_0(t).$ Then the Dirac picture density matrix
 $\tilde{\rho}(t)=e^{+ i H_0(t)\,  t\ / \hbar}\,  \rho(t)\, e^{- i H_0(t)\,  t\ / \hbar}$ evolves as:
 \be  \label{Meqn3} 
\frac{d \tilde{\rho}(t)}{dt}=  - \frac{i}{\hbar} {\Big  \lbrack} {\tilde{V}(t), \tilde{\rho}(t)}{\Big  \rbrack}  +  \mathcal{\tilde{L}}, 
\ee where the tilde denotes interaction picture operators.  For the choice of  $H_0(t)= -   \hbar \omega_L\, \sigma_3/2\, ,$ going to the Dirac picture
is simply going to a frame rotating about the z-axis in which frame the Larmor precession vanishes.  That is called the rotating frame.  Since we include gates into
$V(t)$ and hence $\tilde{ V}(t)$,  the gate bias pulse that we introduce in the Schr\"{o}dinger picture, corresponds to a rotating frame that stops rotating during the action of a gate.

There are advantages offered by each of these choices.  We stick with the Schr\"{o}dinger description because it most clearly reveals the full dynamics by
viewing the time evolution of the spin observables.

\section{The  Master Equation Model}
\label{sec4}

The master equation for the time evolution of the system's density matrix is now presented.  We seek a simple model that incorporates the main features of  qubit dynamics for a quantum computer.  These main features include seeing how the dynamics evolve under the action of gates and the role of both closed system dynamics and of  open system decoherence, dissipation and the system's approach to equilibrium.  From the density matrix we can determine a variety of observables,  such as the polarization vector,  the power and heat rates,  the purity,  fidelity, and  entropy all as a function of time.
\subsection{Definition of the Model  Master Equation}

To the unitary evolution, we now  we add a term $\mathcal{ L}(t) $ which is required to be Hermitian and traceless so that the  density matrix  $\rho(t)$ maintains its hermiticity and trace one properties. In addition, $\mathcal{ L}(t) $ has to keep 
$\rho(t)$ positive definite. To identify explicit physical effects, we separate  $\mathcal{ L}(t) $ into three terms:
 \bea  \label{Meqn2} 
\frac{d \rho}{dt}&=&  - \frac{i}{\hbar} {\Big  \lbrack}  H(t), \rho(t){\Big  \rbrack}  +  \mathcal{L} \\ \nonumber
\mathcal{L}&=&\mathcal{L}_1+ \mathcal{L}_2 + \mathcal{L}_3 \\ \nonumber
\mathcal{L}_1&=& \Gamma {\Big \lbrace} L(t)\   \rho(t)\   L^\dagger(t)   -   \frac{   L^\dagger(t) L(t) \ \rho(t)  + \rho(t)\  L^\dagger(t) L(t)  }{2}  {\Big \rbrace},\ \ \ \ {\rm Lindblad}\\ \nonumber
\mathcal{L}_2&=&     \gamma_2 \,  \rho(t)\, ( \tilde{S}  -    \langle\tilde{S} \rangle )
 -\gamma_2 \, \beta_2(t) \,  {\Big \lbrace} \frac{\rho(t) H(t)+ H(t)  \rho(t)}{2}  -  \rho(t)  \langle H(t) \rangle {\Big \rbrace} \ \ \ \ {\rm Beretta}  \\ \nonumber
\mathcal{L}_3&=&  \gamma_3 \,  \rho(t)\, ( \tilde{S}  -    \langle\tilde{S} \rangle )
 -\gamma_3 \, \boldsymbol{\beta_3(t)} \, {\Big \lbrace} \frac{\rho(t) H(t)+ H(t)  \rho(t)}{2}  -  \rho(t)  \langle H(t) \rangle {\Big \rbrace} ,\ \ \ \ \ \ {\rm Bath}
 \eea the operator  $\tilde{S} \equiv- \log_e \rho(t),$ involves a base e logarithm to assure that
  a Gibbs density matrix is obtained in equilibrium (see later).   The QC entropy is defined with a base 2 operator
  $\hat{S}= -\log_2 \rho(t) $ with entropy equal to $ \langle S \rangle =  \langle\hat{S} \rangle= - {\rm Tr}[ \rho(t)\, \log_2 \rho(t) ].$  The conversion factor
  is $\tilde{S} \equiv \log_e (2) \  \hat{S} $   and $  \langle\tilde{S} \rangle\equiv \log_e (2) \    \langle\hat{S} \rangle,$ with $\log_e (2) \  =0.693147.$ 
  The level splitting, gates and bias pulses are included in $H(t).$   The state dependent, and hence time dependent, functions $\beta_2(t), \boldsymbol{\beta_3(t)} $ will be defined later.
   
 When we discuss equilibrium, a form that combines the Beretta and Bath terms ${\mathcal L}_{2\,3}(t) $   is used:  
 \be  \label{Meqn2b} 
\mathcal{L}_{2\,3}(t)=    \gamma_{23} \,  \rho(t)\, ( \tilde{S}  -    \langle\tilde{S} \rangle )
 -\gamma_{23} \, \beta_{23}(t) \,  {\Big \lbrace} \frac{\rho(t) H(t)+ H(t)  \rho(t)}{2}  -  \rho(t)  \langle H(t) \rangle {\Big \rbrace},   
\ee with $ \gamma_{23}\equiv \gamma_{2}+ \gamma_{3},$ and  $ \beta_{23}(t) \equiv \frac{\gamma_{2} \beta_{2}(t) +  \gamma_{3} \boldsymbol{\beta_3(t)}  }{ \gamma_{23}  }.$

The $\mathcal{L}_1$ is of Lindblad~\cite{Lindblad2}  form, where the $ L(t)$ are 
  time-dependent Lindblad spin-space operators, which we will represent later as pulses.  
 The most important properties of the $\mathcal{L}_1 \cdots \mathcal{L}_3$ operators are that they are Hermitian and traceless, 
  which means that as the density matrix evolves in time, it remains Hermitian and of unit trace.  They also have the property of maintaining the positive definite property of the density matrix.
  Note $\Gamma$  sets the rate of the Lindblad contribution $\mathcal{L}_1,$
 in  inverse time units. In our heuristic master equation,  we use the Lindblad form to describe the impact of external noise on the system, where we represent the noise as random pulses.  In addition, we  also use the Lindblad form to describe  dissipative/friction effects on the quantum gates, by having the Lindblad pulses coincide with the action time of the gate pulses. We also show later that a strong Lindblad pulse can represent a quantum measurement.
 
The $\mathcal{L}_2,$  term is the  Beretta~~\cite{Beretta1} contribution,  which describes a closed system.  The closed system involves no heat transfer,  with motion along a path of increasing entropy, as occurs for example with a non-ideal gas in an insulated container.
 This is accomplished by a state dependent  $\beta_2(t)$ that is presented later.  Note $\gamma_2,$  sets the strength of the Beretta contribution $\mathcal{L}_2,$
 in  inverse time units  as a fraction of the Larmor angular frequency.
 
In one simple version of the  Bath contribution~\cite{Korsch}  $\mathcal{L}_3,$   the Bath temperature $T$  in Kelvin  stipulates  a fixed 
value of $\boldsymbol{\beta_3(t)} \rightarrow \beta_3 \equiv 1/ (k_B\, T)\, ,$  where $k_B$ is the Boltzmann constant (86.17 $\mu eV$/Kelvin).  A more general $\mathcal{L}_3$ Bath contribution, based on a general theory~\cite{Beretta1, Beretta2,  Beretta3}  of thermodynamics~\footnote{The author gratefully thanks one Reviewer for pointing out this very significant improvement.},  
defines a state dependent  $\boldsymbol{\beta_3(t)} $ by using a  fixed temperature  $T_Q$ to specify a fixed $\dot{Q}(t)/\dot{S}(t)$ ratio (see later). 
Note $\gamma_3$  sets the strength of the Bath contribution $\mathcal{L}_3,$
 in  inverse time units as a fraction of the Larmor angular frequency. 
 
In Table~\ref{parameters1}, typical  values of  $\omega_L, \Gamma, \gamma_2, \gamma_3, \beta_3$ used in our test  cases are shown;
these parameters are selected to focus on the role of each term. Realistic values  can be invoked for various experimental conditions.
\begin{table}[h]
\caption{Test Master Equation Parameters}
\centering
\begin{tabular}{c c c }
\hline\hline
Name & Value  & Units \\ [0.5ex] 
\hline
$\omega_L$&0.2675 &GHz \\
$\Gamma$&0.00213& GHz \\
$\gamma_2$&0.0426& GHz \\
$\gamma_3$&0.0852&GHz \\
$\beta_3$&0.000425&1/$\mu$eV  \\ [1ex]
\hline
\end{tabular}
\label{parameters1}
\end{table}

   \subsection{Some general properties of the  Master Equation}
  \subsubsection{Power and Heat rate evolution} 
  
The various terms in the master equation play different roles in the dynamics.  To examine those differing roles consider the energy of the system
 and its rate of change.  With our Hamiltonian  $ H(t)$  and a density matrix $\rho(t),$ we form the ensemble average
  $$E(t) =  \langle H(t) \rangle \equiv {\rm Tr} [  H(t)\,\rho(t)] .$$  Taking the time derivative, we obtain
  \bea
   \frac{d E(t)}{dt} &=&  {\rm Tr} [ \rho(t)\, \frac{d }{dt}H(t)] + {\rm Tr} [  H(t)\, \frac{d }{dt}\rho(t)]\\ \nonumber
   &=&\hspace{8 mm}   \frac{d }{dt}W(t) \hspace{15 mm}  + \frac{d }{dt}Q(t)\\ \nonumber
   \frac{d }{dt}W(t)&=&{\rm Tr} [ \rho(t) \, {\dot{H}}(t)] \\ \nonumber
    \frac{d }{dt}Q(t)&=& {\rm Tr} [  H(t) \, {\dot{\rho}}(t)] \, .
     \eea
   We identify the
    term  $\frac{d Q(t)}{dt}$ as the energy transfer rate and  
    $\frac{d W(t)}{dt}$ as the work per time or power,  with the convention that $Q(t) > 0$ indicates heat transferred into the system,
    and $W(t) > $ indicates work done on  the system
  The time dependence of the density matrix is given by the unitary evolution plus the ${\mathcal L}$ terms
   of  Eq.~\ref{Meqn2}.
  
   Now consider just the power term.  Since  $ \frac{dH(t)}{dt}$ is nonzero when gate pulses are active, power is invoked in the system only via the time derivatives of the gate and bias pulses (and by any temporal changes in the level splitting).

The  energy transfer rate $\frac{d Q(t)}{dt}$  can now be examined using the dynamic evolution
  \begin{eqnarray}
  \label{Heqn1}
{\rm Tr} [  H(t) \,  \frac{d \rho(t)}{dt}]&=& {\rm Tr} {\Big \lbrack}  H(t) \,  \lbrace -\frac{i}{\hbar} [ \rho(t), H(t)]+ {\pazocal L}  \rbrace  {\Big \rbrack} \\ \nonumber
 &=& {\rm Tr}Ä [  H(t) \,  {\pazocal L}] .   \end{eqnarray}
 Using the permutation invariance of the trace, the unitary evolution part does not contribute to energy transfer.   Heat  arises  from the Lindblad and Bath terms. 
 We will now see: (1) how  the closed system (Beretta term) does not generate energy transfer; it does increase entropy; and (2)  the Bath term involves heat and associated entropy transfer.
 
The Lindblad term generates energy transfer (heat) according to:~\footnote{Note that if $L(t)$ commutes with $H(t),$ the heat transfer vanishes.}
 \be
 \label{Heqn2}  \frac{d Q}{dt}=  \Gamma\  {\rm Tr}{\Big [ } H(t) \,{ \lbrace} L(t)\   \rho(t)\   L^\dagger(t)  -   \frac{   L^\dagger(t)  L(t) \ \rho(t)  + \rho(t)\  L^\dagger(t) L(t)   }{2}  { \rbrace} {\Big  ] }  \\  \nonumber
 \ee
 
 The Beretta term heat transfer equation is:
 \bea
 \label{Heqn3}  \frac{d Q}{dt} &=&   \gamma_2 \,  {\rm Tr}{\Big [ } H(t) \,    [  \rho(t)\, ( \tilde{S}  -    \langle\tilde{S} \rangle )
 - \beta_2(t) \,  {\Big \lbrace} \frac{\rho(t) H(t)+ H(t)  \rho(t)}{2}  -  \rho(t)  \langle H(t) \rangle {\Big \rbrace} ] {\Big  ] }  \\  \nonumber
 &=&   \gamma_2 \,   {\Big \lbrace}  ( \langle H(t) \tilde{S} \rangle  -     \langle H(t) \rangle  \langle\tilde{S} \rangle ) - \beta_2(t) (  \langle H^2(t) \rangle -   \langle H(t) \rangle^2   )   {\Big \rbrace}  \\  \nonumber
  &=&   \gamma_2 \,   {\Big \lbrace}  \langle \Delta E\ \Delta \tilde{S}\rangle - \beta_2(t)  \langle\Delta E\, \Delta E\rangle   {\Big \rbrace} ,
 \eea where we have defined
  \bea
 \label{Heqn4}
 \langle\Delta E\, \Delta E\rangle &\equiv&  \langle H^2(t) \rangle -   \langle H(t) \rangle^2 = \  \langle (H- \langle H  \rangle)^2 \rangle  \\  \nonumber
 \langle H^2(t)  \rangle &\equiv& {\rm Tr} [\rho(t) \,  H(t) \,  H(t)] \ \hspace{4 mm} \langle H(t)  \rangle  \equiv {\rm Tr} [\rho(t) \,  H(t)]\\ \nonumber 
  \langle\Delta E\, \Delta \tilde{S}\rangle &\equiv&  \langle H(t) \tilde{S} \rangle  -     \langle H(t) \rangle  \langle\tilde{S} \rangle  \\ \nonumber
   &=& \langle (H- \langle H \rangle)(\tilde{S}- \langle\tilde{S} \rangle) \rangle  \\ \nonumber
  \langle H(t) \tilde{S} \rangle &\equiv& {\rm Tr} [\rho(t) \, H(t)  \tilde{S}]  \ \hspace{13 mm}   \langle  \tilde{S} \rangle  \equiv  {\rm Tr} [\rho(t) \,  \tilde{S}]\\ \nonumber
\tilde{S}&\equiv& -  \log_e \rho(t).
\eea
 The major feature of Beretta's ${\pazocal L}_2$ contribution is  a state and hence time-dependent  $\beta_2(t)$ 
 \bea
 \label{Heqn5}
\beta_2(t)&\equiv&  \frac{ \langle\Delta E\, \Delta \tilde{S}\rangle} {  \langle\Delta E\, \Delta E\rangle } \\  \nonumber
&=& \frac{  \langle H(t) \tilde{S} \rangle  -     \langle H(t) \rangle  \langle\tilde{S} \rangle  }{  \langle H^2(t) \rangle -   \langle H(t) \rangle \langle H(t) \rangle }   \\  \nonumber
&=& \frac{  \langle (H- \langle H \rangle)(\tilde{S}- \langle\tilde{S} \rangle) \rangle    }{ \langle (H- \langle H \rangle)^2 \rangle  },  
  \eea
  defined so that the system is closed and 
 energy is  not transferred  to or from the system,
  $\frac{d Q}{dt}=  0.$  This choice also makes the closed system follow a path of increasing entropy (see later). 
  That increase of entropy for a closed system
 signifies that the closed system is dynamically constrained to reorder itself to maximize its entropy.
 Here $\beta_2(t)$  has a highly nonlinear dependence on the density matrix.
 for a one qubit system, $\beta_2(t)$ is of rather simple form 
 \be
 \beta_2(t) =2\,  \frac{1-{\bf P}(t)^2}{1-P_z(t)^2}\   \frac{P_z(t)}{{\bf P}(t)}\  \frac{\arctanh( {\bf P}(t))}{\hbar\, \omega_L},
 \ee for an equilibrium Gibbs density matrix with fixed z and  zero x and y polarization, this reduces to $\beta_2(t) =2\, 
  \arctanh (P_z )/(\hbar\, \omega_L)=\frac{\log_e(1+P_z)}{ \log_e(1-P_z) }/(\hbar\, \omega_L) .$

The Bath term  ${\pazocal L}_3$ does generates heat:
 \bea
 \label{Heqn6}  \frac{d Q}{dt} &=&   \gamma_3 \,  {\rm Tr}{\Big [ } H(t)) \, \rho(t)\,  {\Big \lbrace}  ( \tilde{S}  -    \langle\tilde{S} \rangle ) -\boldsymbol{\beta_3(t)} ( H(t)  -    \langle H(t) \rangle )   {\Big \rbrace}{\Big  ] }  \\  \nonumber
 &=&   \gamma_3 \,   {\Big \lbrace}  ( \langle H(t) \tilde{S} \rangle  -     \langle H(t) \rangle  \langle\tilde{S} \rangle ) - \boldsymbol{\beta_3(t)} (   \langle H^2(t) \rangle -    \langle H(t) \rangle  \langle H(t) \rangle   )   {\Big \rbrace}  \\  \nonumber
  &=&   \gamma_3 \,   {\Big \lbrace}  \langle\Delta E\, \Delta \tilde{S}\rangle -\boldsymbol{\beta_3(t)}  \langle\Delta E\, \Delta E\rangle   )   {\Big \rbrace} .
 \eea where the Korsch~\cite{Korsch} option sets $\boldsymbol{\beta_3(t)}\rightarrow \beta_3=1/(k_B\, T)$  to a constant, where $T$ is a specified bath temperature.  Entropy is changed by the bath term.  A better choice for  $\boldsymbol{\beta_3(t)}$ involves the entropy evolution, which is discussed next.
 
\subsubsection{Entropy evolution} 

   Let us  now consider the time evolution of the  base 2 entropy Eq.~\ref{entropy1}.
  Equation ~\ref{entropy2} gives the time derivative of the entropy  as  $\frac{d S}{dt} =-{\rm Tr}[  \log_2(\, \rho(t)\,  )\,   \frac{d  \rho(t)}{dt}\, ].$  
  Inserting the time derivative from Eq.~\ref{Meqn2}, 
 we again get no change from the unitary term,  just from the dissipative ${\pazocal L}$ terms:
  \begin{eqnarray}\label{Seqn2} 
\frac{d S}{dt} &=& -{\rm Tr}[  \log_2(\rho(t)) \,   \frac{d  \rho(t)}{dt}  ]  \\  \nonumber
&=&  - {\rm Tr}[  \log_2(\rho(t)) \, \mathcal{L}\   ] . 
 \end{eqnarray}  This is a general result for $n_q$  qubits.  
 
 Note that the Lindblad term generates a change in entropy according to:
 \begin{eqnarray}\label{Seqn3}  \frac{d S}{dt}&=&  - \Gamma\  {\rm Tr}{\Big [ } \log_2(\rho(t)) \, { \lbrace} L(t)\   \rho(t)\   L^\dagger(t)  -   \frac{   L^\dagger(t)  L(t) \ \rho(t)  + \rho(t)\  L^\dagger(t) L(t)   }{2}  { \rbrace} {\Big  ] }  \\  \nonumber
&=&  - \Gamma\   {\rm Tr}{\Big [ } \log_2(\rho(t)) \, { \lbrace} L(t)\   \rho(t)\   L^\dagger(t)   -  \rho(t)
 L^\dagger(t) L(t) \   { \rbrace} {\Big  ] }  \\  \nonumber
 &=&  - \Gamma\   {\rm Tr}{\Big [ } \log_2(\rho(t)) \,  [ \  L(t) \  , \rho(t) L^\dagger(t) \ ]  {\Big  ] }\, .
  \end{eqnarray}

 For the Beretta (closed system) ${\pazocal L}_2$ and the Bath  ${\pazocal L}_3$ terms,  entropy changes according to:
  \begin{eqnarray}
  \label{Seqn4}
    \frac{d S}{dt}&=&- \gamma_2 \,  {\rm Tr}{\Big  [  } \log_2(\rho(t)) \, \rho(t)\,  {\Big \lbrace}  ( \tilde{S}  -    \langle\tilde{S} \rangle ) - \beta_2(t) ( H(t)  -     \langle H(t) \rangle )   {\Big \rbrace} 
  {\Big ] } \\  \nonumber
   \frac{d S}{dt}&=& -\gamma_3 \,   {\rm Tr}{\Big  [ } \log_2(\rho(t)) \,  \rho(t)\,  {\Big \lbrace}  ( \tilde{S}  -    \langle\tilde{S} \rangle ) - \boldsymbol{\beta_3(t)}( H(t)  -     \langle H(t) \rangle )   {\Big \rbrace}
  {\Big ] } \, .  
  \end{eqnarray}
  
  For the Beretta  closed system case, with $   \langle\Delta \tilde{S}\, \Delta \tilde{S}\rangle\equiv  \langle \tilde{S}^2 \rangle  -    \langle\tilde{S} \rangle^2  = \langle ( \tilde{S}-  \langle\tilde{S} \rangle )^2 \rangle$ :
   \begin{eqnarray}
  \label{Seqn5}
    \frac{d S}{dt}&=&+ \frac{\gamma_2}{  \log_e(2)} {\Big  [  }  ( \langle \tilde{S}^2 \rangle  -    \langle\tilde{S} \rangle^2 ) - \beta_2(t) (   \langle H(t)\tilde{S} \rangle  -     \langle H(t) \rangle  \langle\tilde{S} \rangle)   
  {\Big ] } \\  \nonumber
  &=&+\frac{\gamma_2}{  \log_e(2)} {\Big  [  }\langle\Delta \tilde{S}\, \Delta \tilde{S}\rangle - \beta_2(t)  \langle\Delta E\, \Delta \tilde{S}\rangle   
  {\Big ] } \\  \nonumber
   &=&+\frac{\gamma_2}{  \log_e(2)} {\Big  [  }\langle\Delta \tilde{S}\, \Delta \tilde{S}\rangle   \langle\Delta E\, \Delta E\rangle  -  \langle\Delta E\, \Delta \tilde{S}\rangle^2    
  {\Big ] } / \langle\Delta E\, \Delta E\rangle.  
  \end{eqnarray}  Note that entropy increases for the Beretta term $\frac{d S}{dt}\geq 0,$ since $\langle\Delta \tilde{S}\, \Delta \tilde{S}\rangle\, \langle\Delta E\, \Delta E\rangle \geq \langle\Delta E\, \Delta \tilde{S}\rangle^2 \, .$

   For the Bath term, when we switch to a base ``e" entropy $\tilde{S}$,  
    \be
 \label{Seqn6}
   \frac{d \tilde{S}}{dt}= +\gamma_3 \,   {\Big  [  }    \langle\Delta \tilde{S}\, \Delta \tilde{S}\rangle   - \boldsymbol{\beta_3(t)} \langle\Delta E\, \Delta \tilde{S}\rangle      
  {\Big ] } .  
  \ee   Forming the energy rate to entropy rate ratio and setting it  equal to a fixed quantity  $k_B T_Q $,  we obtain the condition $k_B T_Q= dQ/d\tilde{S}= \dot{Q}/\dot{\tilde{S} } $

    \be
 \label{Seqn7}
   \frac{\dot{Q}} 
   {\dot{\tilde{S}} }= \frac{  \langle\Delta E\, \Delta \tilde{S}\rangle -\boldsymbol{\beta_3(t)}  \langle\Delta E\, \Delta E\rangle        } {    \langle\Delta \tilde{S}\, \Delta \tilde{S}\rangle   - \boldsymbol{\beta_3(t)} \langle\Delta E\, \Delta \tilde{S}\rangle      
    }  = k_B\, T_Q \, ,
  \ee  
   
  which yields an expression for the state dependence of   $\boldsymbol{\beta_3(t)} : $
 \be
 \label{Seqn8}
   \boldsymbol{\beta_3(t)}= \frac {   \langle\Delta E\,  \Delta \tilde{S}\rangle    -  k_B T_Q\,  \langle\Delta \tilde{S}\, \Delta \tilde{S}\rangle }
   {  \langle\Delta E\, \Delta {E}\rangle - k_B T_Q\, \langle\Delta E\,  \Delta \tilde{S}\rangle        }   \, .  
  \ee

   For the equilibrium (Gibbs density matrix) limit,  which occurs at final time $t_f, $ \, $\boldsymbol{\beta_3(t_f)}\rightarrow \beta_3= \frac{1}{k_B\, T} , $
   where $T$ is the specified Bath (Gibbs) temperature. 
  In equilibrium, the polarization vector is in the z-direction since $ H[t_f] =  H_0 \equiv -   \frac{1}{2} \ \hbar\, \omega_L \   \sigma_z \, . $  The difference between the two Bath versions for one qubit evolution. will be explored later.
 
\subsubsection{Purity evolution}

For a one qubit system, both the entropy and the purity are functions of the length of the polarization vector, with the entropy being zero and the purity being one on the Bloch sphere.  For polarization vectors inside the Bloch sphere, the entropy is increased and the purity decreased.  Note that $1-{\pazocal P}(t)$ plays a role  parallel to entropy. 
Consider the change in Purity (denoted by the symbol ${\pazocal P}   $), where 
 $${\pazocal P} (t)  \equiv {\rm Tr}[ \rho(t) \,  \rho(t) ],$$  and the rate of change of this purity is given by
 $$ \frac{d {\pazocal P}(t)}{dt} = 2\  {\rm Tr}[ \rho(t) \, \frac{ d \rho(t)}{dt}].$$   Using  Eq.~\ref{Meqn1}, we find that the purity is unchanged by the unitary term, but is altered,
 mainly diminished,  by the Lindblad term:
 \begin{eqnarray}\label{Purity} 
    \frac{d  {\pazocal P}(t )}{dt}&=& 2\  {\rm Tr}{\Big  \lbrack} \rho(t) \,  \mathcal{L}_1\ {\Big  \rbrack} \\   \nonumber
 &=&  2\ \Gamma\    {\rm Tr} {\Big  \lbrack} \rho(t) \, {\Big \lbrace} L(t)\   \rho(t)\   L^\dagger(t)   -   \frac{   L^\dagger(t) L(t) \ \rho(t)  
 + \rho(t)\  L^\dagger(t) L(t)  }{2}  {\Big \rbrace}\  {\Big  \rbrack}  \\   \nonumber
 &=&  2\ \Gamma\    {\rm Tr}{\Big  \lbrack} \rho(t) \,  {\Big \lbrace} L(t)\   \rho(t)\   L^\dagger(t)   -     L^\dagger(t) L(t) \ \rho(t)  
   {\Big \rbrace}\  {\Big  \rbrack}  \\   \nonumber
   &=&  2\ \Gamma\    {\rm Tr}{\Big  \lbrack} \rho(t) \, [ L(t)\  , \rho(t)\    L^\dagger(t) ]  {\Big  \rbrack}. 
 \end{eqnarray} 
  The trace property ${\rm Tr}[A B] = {\rm Tr}[B A]$  was again invoked to deduce that  $ {\rm Tr}{\Big  \lbrack}   \rho(t) \,  {  \lbrack}  H(t), \rho(t){ \rbrack} \  {\Big  \rbrack} \equiv 0.$

\subsubsection{Temperature}

Temperature is a basic quantity in equilibrium thermodynamics.  It is not  clear if any such concept can be applied to non-equilibrium systems.  Nevertheless, it is tempting to
invoke a ``temperature" quantity  for dynamic systems, albeit with caution.  In that spirit, we note that the inverse temperature $\beta$ does apply to the equilibrium Gibbs state.
Two possible definitions of temperature for non-equilibrium cases are based on (1) the level occupation probability ratios or (2) the ratio of the energy change and the entropy change    $k_B T= dQ/d\tilde{S}= \dot{Q}/\dot{\tilde{S} } $

The simplest case (1) occurs for a diagonal Hamiltonian (for us that occurs away from gate pulses), the ensemble average energy is:
\be
  \langle H(t) \rangle ={\rm Tr}[ H(t) \, \rho(t)] = - \frac{ \hbar \omega_L}{4} {\rm Tr}[ \sigma_3 \, P_z \sigma_3]=  -\frac{ \hbar \omega_L}{2} P_z
= \epsilon_1\, \rho_{11} +  \epsilon_2\  \rho_{22},
\ee where  $ \epsilon_1=-\hbar \omega_L/2,$ $ \ \epsilon_2=\hbar \omega_L/2 ,$ \   $ \rho_{11} =(1+P_z)/2$  and $ \  \rho_{22} =(1-P_z)/2.$ 
Here we see the connection between temperature the z-polarization,
 and that the level occupation probabilities are $n_1= \rho_{11}$
and  $n_2= \rho_{22} .$

 When the gates are acting, we need to include off-diagonal terms in $H(t).$
 So we introduce the unitary operator $\tilde{U}(t)$ that diagonalizes $H(t)=\tilde{U}^\dagger(t)\, H_d(t) \,   \tilde{U}(t)$

 \be
    \langle H(t) \rangle ={\rm Tr}[ H(t) \,  \rho(t) ] =  {\rm Tr}[H_d(t)   \,\tilde{\rho}(t)]= \tilde{\epsilon_1}\,\tilde{ \rho}_{11} +  \tilde{\epsilon}_2\  \tilde{\rho}_{22},
 \ee  where $\tilde{\epsilon_1},\tilde{\epsilon_2}$ are the eigenvalues of $H(t)$ and $ \tilde{\rho} \equiv  \tilde{U}(t) \rho(t) \tilde{U}^\dagger(t).$ In this way we can define
 level occupation probabilities including gate pulses:  $\tilde{n}_1=\tilde{ \rho}_{11}$
and   $\tilde{n}_2=\tilde{ \rho}_{22}.$ 

  Note that for a system described by a  Gibbs density matrix, that density matrix commutes with the Hamiltonian.  As a consequence, the unitary evolution vanishes.
Thus for a Gibbs equilibrium state the Hamiltonian and the density matrix are diagonalized by the same unitary matrix. 
   In addition, for a Hamiltonian
of the form $H \propto \sigma_z, $ the polarization in the x and y direction  vanish for a Gibbs  density matrix,
and $P_z = \tanh (\frac{ \hbar \omega_L}{2 k_B T}),$ which relates the z-polarization to the absolute temperature T in a rather complicated way.  A simpler form follows.

We  use the Gibbs density matrix
$$\rho_G \equiv \frac{e^{- \frac{H(t)} { k_B T}}}{{\pazocal Z}} \ \ ;  \  \  \   {\pazocal Z} \equiv {\rm Tr}( e^{- \frac{H(t)} { k_B T} }) $$ to define  absolute temperature
\be
T= \frac{1}{k_B}  \frac{\tilde{ \epsilon}_2  -\tilde{ \epsilon}_1}{ \log_e(\frac{\tilde{n}_1 }{\tilde{n}_2}) }.
\ee In the region away from the gate pulses where the Hamiltonian is diagonal, this reduces to 
\be
T=  \frac{ \hbar \omega_L}{k_B} \frac{1} { \log_e( \frac{n_1 }{n_2})}\ .
\ee  the above definition is equivalent to $P_z= \tanh (\frac{ \hbar \omega_L}{2 k_B T}).$

  Note the above temperature definition yields a positive absolute temperature for $n_1  \geq n_2,$  for larger ground state occupation,
but  negative absolute temperature for $n_2 > n_1.$  This is the well known, negative absolute temperature that appears for a small number of levels that are excited to produce a  ``level occupation/temperature  inversion"~\cite{Tinversion1,Tinversion2}.   For us that appears when the z-polarization flips sign.

Case (2) $k_B T= dQ/d\tilde{S}= \dot{Q}/\dot{\tilde{S} } $  can be evaluated from the  ratio of the time derivatives of the heat and of the entropy. The result is
$$
T=  \frac{ \hbar \omega_L}{k_B} \frac{ \dot{P}_z}{ \dot{\bf{P}}}\, \frac{1}{  \log_e( \frac{1+ \bf{P}}{ 1- \bf{P} })} \ \ 
{\underset{P_x=P_y\rightarrow0}\Longrightarrow} \ \  \frac{ \hbar \omega_L}{k_B} \frac{1}{  \log_e( \frac{1+ P_z}{ 1- P_z })}\,  ,
 $$
which reduces to the above case (1) definition when $ P_x$ and $P_y$ vanish.  Here ${\bf P}$ is the length of the polarization vector.

  These two thermodynamic definitions can simply be extended to the gate and to non-equilibrium regions by fiat.
  Non-equilibrium situations occur when  $ P_x$ and $P_y$ are non-zero. The above two quantities have the units of temperature in all regions,  but a clear meaning of temperature applies only for equilibrium. Nevertheless, it is of interest to use $T$ as defined above in non-equilibrium regions where they serve as useful indicia of dynamical changes.  
 \subsection{Lindblad }
 \label{sec4c}
 
  \subsubsection{Comments on positive definite property of Lindblad}

 An explicit demonstration for one-qubit that the Lindblad term keeps $\rho(t)$ positive definite is obtained by setting $\rho(t) = U_\rho(t) \rho_D(t)  U_\rho^\dagger(t),$
 where $\rho_D(t)$ is the diagonal matrix 
$\left(
\begin{array}{c c}
 \lambda_1& 0 \\
0 & \lambda_2
 \end{array}\right)  .  $
 It then follows that the commutator term does not alter the eigenvalues and that  the Lindblad term keeps the 
 eigenvalues positive and between zero and 1.  One finds:~\footnote{Note  $\frac{d}{dt}( U_\rho\, \rho_D\, U_\rho^\dagger) = U_\rho\, \dot{ \rho}_D\, U_\rho^\dagger +   \dot{U}_\rho\, \rho_D\, U_\rho^\dagger
  +  U_\rho\, \rho_D\, \dot{U}_\rho^\dagger$ and  $U_\rho^\dagger\, \frac{d \rho}{dt}\,  U_\rho = \dot{ \rho}_D  
   +  U_\rho^\dagger\, \dot{U}_\rho\, \rho_D 
  +  \rho_D\, \dot{U}_\rho^\dagger\, U_\rho$  For the $i\, th$ diagonal component the last two terms vanish since $\frac{d}{dt} ( U_\rho^\dagger\, U_\rho)=0.$
  Thus we arrive at  
 $  \dot{ \lambda}_i(t) = [\, U_\rho^\dagger(t)\, {\pazocal L}\,U(t)_\rho\, ]_{i\,i}\   .$
 }
   $  \dot{ \lambda}_i(t) = [\, U_\rho^\dagger(t)\, {\pazocal L}_1\, U_\rho(t)\, ]_{i\,i}\   $ 
 $$ \frac{ d \lambda_1(t)}{dt} = \Gamma | V_{1,2}|^2 \  (1 - \alpha\   \lambda_1(t) \  ),$$
 where  $ \alpha=( | V_{2,1}|^2+ | V_{1,2}|^2 ) /  | V_{1,2}|^2   \rangle 1$ and $V \equiv U_\rho^\dagger(t)\, L \, U_\rho(t). $  This form shows that  the eigenvalues stay within the zero to 1 region.
 
 For a system of $n_q  > 1$  qubits,  the above generalizes to:
  \bea
   \frac{ d \lambda_i(t)}{dt} &=& \Gamma\ {\Big  \lbrack} {\big  \lbrack}  \sum_{s }  \mid V_{i s}  \mid^2  \lambda_s {\big  \rbrack} - (V^\dagger V)_{i i}   \lambda_i  {\Big  \rbrack} \\ \nonumber
  &=& \Gamma \sum_{s \neq i } (  \mid V_{i s}  \mid^2  \lambda_s  - \mid V_{s i}  \mid^2   \lambda_i  )  ,
    \eea  where  $ i= 1 \cdots 2^{n_q}$ and the eigenvalues  are ordered  from the largest to smallest.  From this equation, it follows that the largest and smallest eigenvalues stay within the zero to one range if they start in that range and therefore all eigenvalues are within that range.  For example, as the largest eigenvalue
    $\lambda_1 \rightarrow 1,$ the other eigenvalues approach zero and hence $\frac{ d \lambda_1(t)}{dt} = - \Gamma \sum_{s \neq 1 }  \mid V_{s 1}  \mid^2     \leq 0.$
    This negative derivative bounces the eigenvalue back into the zero to one range, if it approaches one.  When the smallest eigenvalue
approaches zero the other eigenvalues are all positive and less than 1, and hence
 $\frac{ d \lambda_{n_q}(t)}{dt} = + \Gamma \sum_{s \neq n_q }  \mid V_{ n_q \, s}  \mid^2 \lambda_s    \geq 0.$
  This positive derivative bounces the eigenvalue back into the zero to one range, if it approaches zero.  This is a very simple proof that the Lindblad form preserves the positive definite nature of a density matrix.

The same steps using the diagonal density matrix basis  apply to the derivative of the entropy:
\be
\frac{d S}{dt} =- \Gamma\  \sum_{s  i}  \log_2 \lambda_i \  (  \mid V_{i s}  \mid^2  \lambda_s  - \mid V_{s i}  \mid^2   \lambda_i  ),
\ee and to the purity
\be
\frac{d {\pazocal P}(t)}{dt} =+ 2\  \Gamma\      \sum_{s  i}   \lambda_i \  (  \mid V_{i s}  \mid^2  \lambda_s  - \mid V_{s i}  \mid^2   \lambda_i  ).
\ee  The entropy expression can be recast as
\bea
\frac{d S}{dt} &=& + \Gamma\  \sum_{s i }  \mid V_{s i}  \mid^2 \lambda_i ( \log_2 \lambda_i - \log_2 \lambda_s) \\ \nonumber
 &\geq& + \Gamma\  \sum_{s  i}  \mid V_{s i}  \mid^2  (  \lambda_i -  \lambda_s) \\ \nonumber
  &\geq& + \Gamma\  \sum_{s  i}  \lambda_i   (  \mid V_{s i}  \mid^2    - \mid V_{ i s}  \mid^2   ).  
\eea  We have used $  x \log(x) \geq x-1  $ to obtain the inequality above, with the result that
$$\frac{d S}{dt} \geq  \Gamma\  {\rm Tr} {\big \lbrack }\, \rho(t) \,   [ L^\dagger(t), L(t)]\,  {\big \rbrack }.$$  Thus imposing the condition 
$[ L(t), L^\dagger(t)]=0$ yields   $\frac{d S}{dt} \geq 0$ , which is the known condition for
obtaining increasing entropy from the Lindblad form.

Applying essentially the same steps to the purity 
\bea
\frac{d {\pazocal P}(t)}{dt} &=& - 2\  \Gamma\      \sum_{s i }  \mid V_{s i}  \mid^2 \lambda_i ( \lambda_i -  \lambda_s) \\ \nonumber
&\leq&  - 2\  \Gamma\      \sum_{s i }  \mid V_{s i}  \mid^2  ( \lambda_i -  \lambda_s)  \\ \nonumber
  &\leq&- 2\   \Gamma\  \sum_{s  i}  \lambda_i   (  \mid V_{s i}  \mid^2    - \mid V_{ i s}  \mid^2   )  \\ \nonumber
  \frac{d {\cal P}(t)}{dt} & \leq  &- 2\   \Gamma\  {\rm Tr} {\big \lbrack }\, \rho(t) \, [ L^\dagger(t), L(t)]\,  {\big \rbrack }
\eea   For $[ L(t), L^\dagger(t)]=0,$  we obtain  decreasing purity  $ \frac{d {\cal P}(t)}{d t} \leq 0$ from the Lindblad form.

Above  simple proofs show that for $n_q$ qubits the Lindblad yields increased entropy and decreased purity  with Lindblad operators that
satisfy $[ L(t), L^\dagger(t)]=0\, .$  The same procedure applies to Renyi,  Tsallis, and  other  definitions of entropy.

Beretta~\cite{Beretta3} objects to the Lindblad form on the grounds that the entropy derivative diverges when the combination
$ \dot{ \lambda}_i \, \log_2 \lambda_j \  $ with $\dot{\lambda}_i  \neq 0$ and $ \lambda_j\rightarrow 0 $ occurs.   Based on this, even though it occurs for just a short time, he rejects the Lindblad form. Note this divergence is already contained in the 
equation~\ref{entropy2}, as mentioned earlier.   We nevertheless adopt a Lindblad form and simply 
avoid the occurrence of $\dot{\lambda}_i  \neq 0$ and $ \lambda_j\rightarrow 0 $  by using 
the Lindblad to incorporate noise pulses that occur after the Beretta and Bath terms  have already acted to increase the system's entropy, away from a pure one qubit state. That circumvents  the problem for one qubit; the multi-qubit case 
remains an issue.

  \subsubsection{Steady Lindblad Operator }
  
\begin{figure}[h]       
    \includegraphics[width=\textwidth]{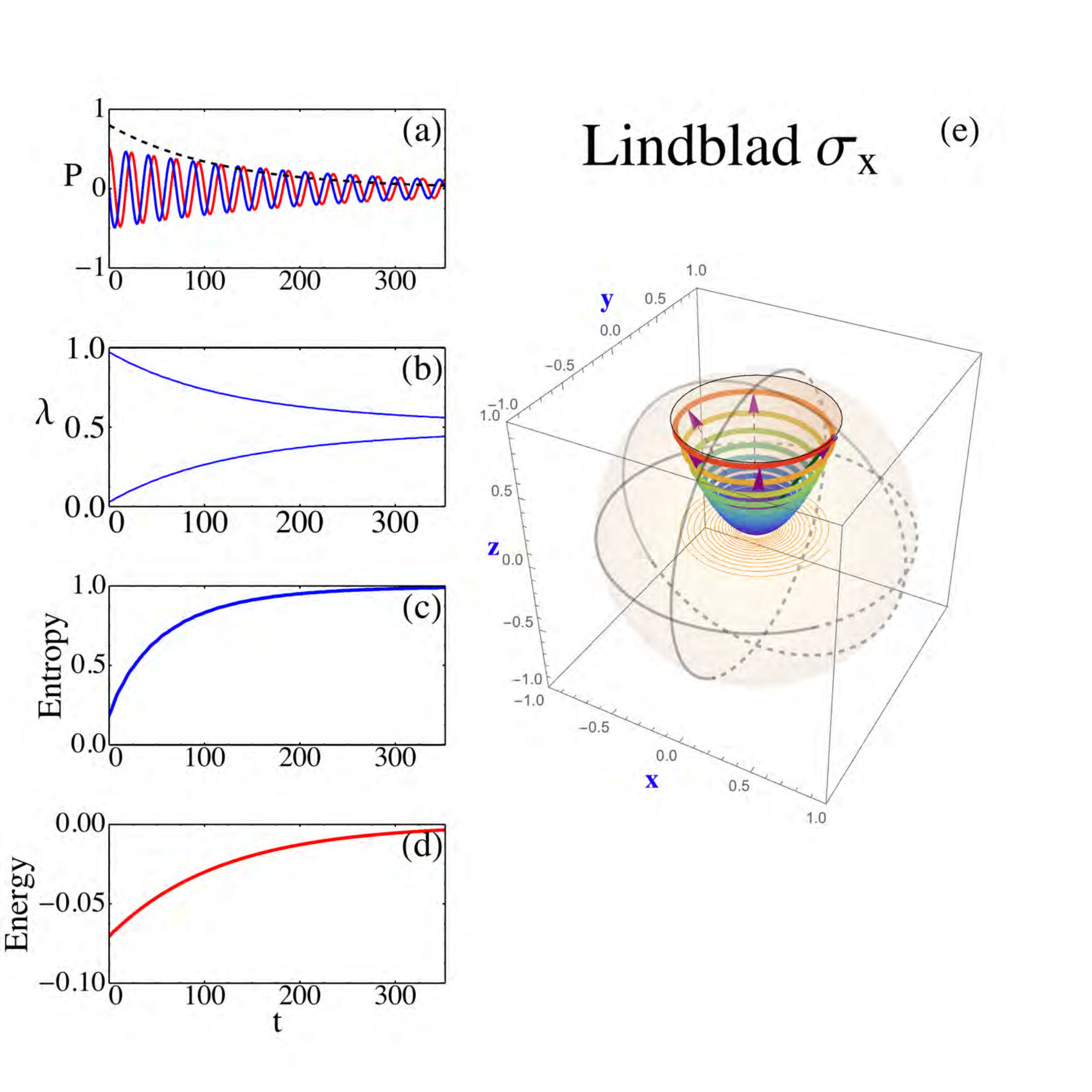}  
    \caption{Steady Lindblad. $\sigma_x$}
\protect\label{LindX}
\end{figure}

\begin{figure}[h]       
    \includegraphics[width=\textwidth]{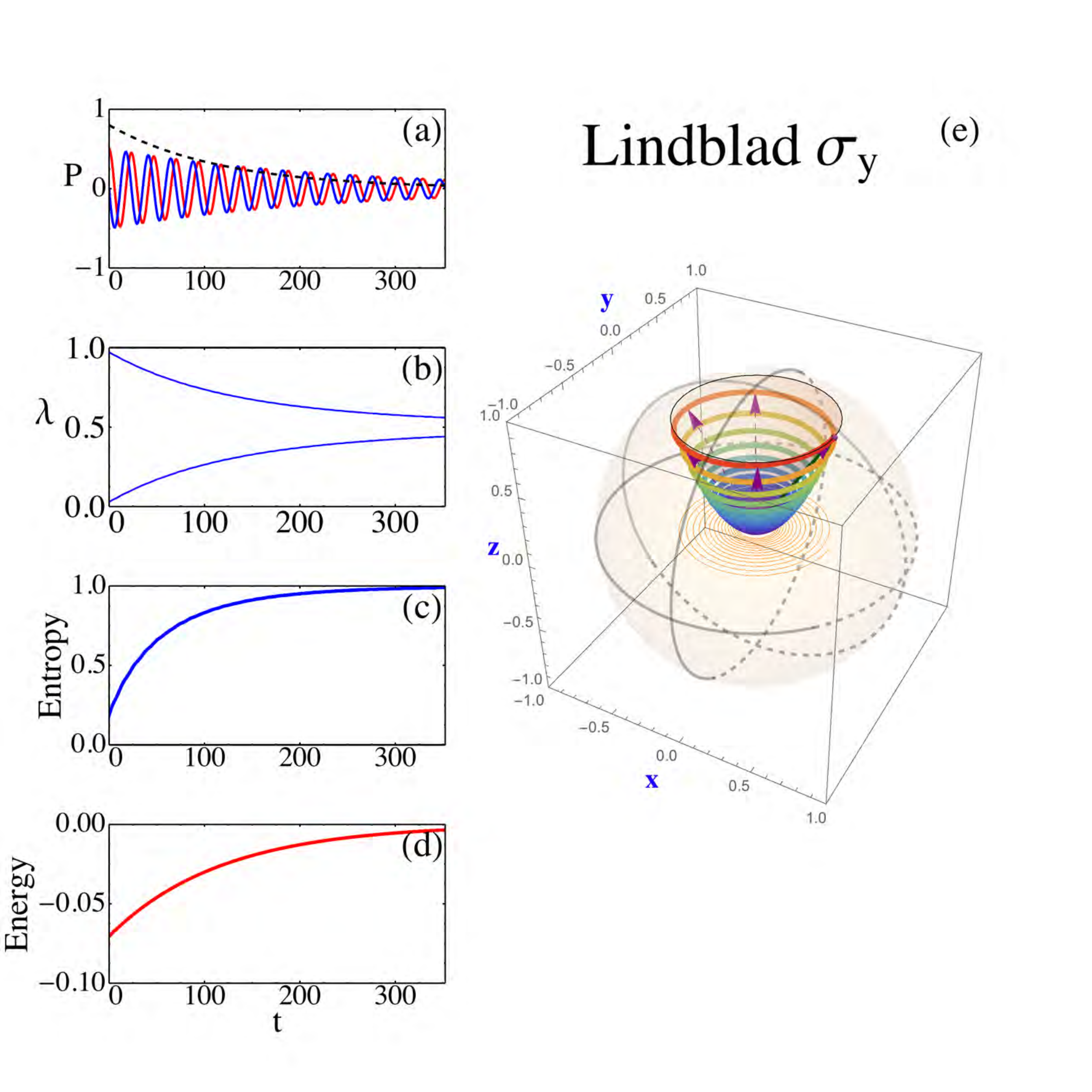}  
        \caption{Steady Lindblad. $\sigma_y$}
\protect\label{LindY}

\end{figure}

\begin{figure}[h]       
    \includegraphics[width=\textwidth]{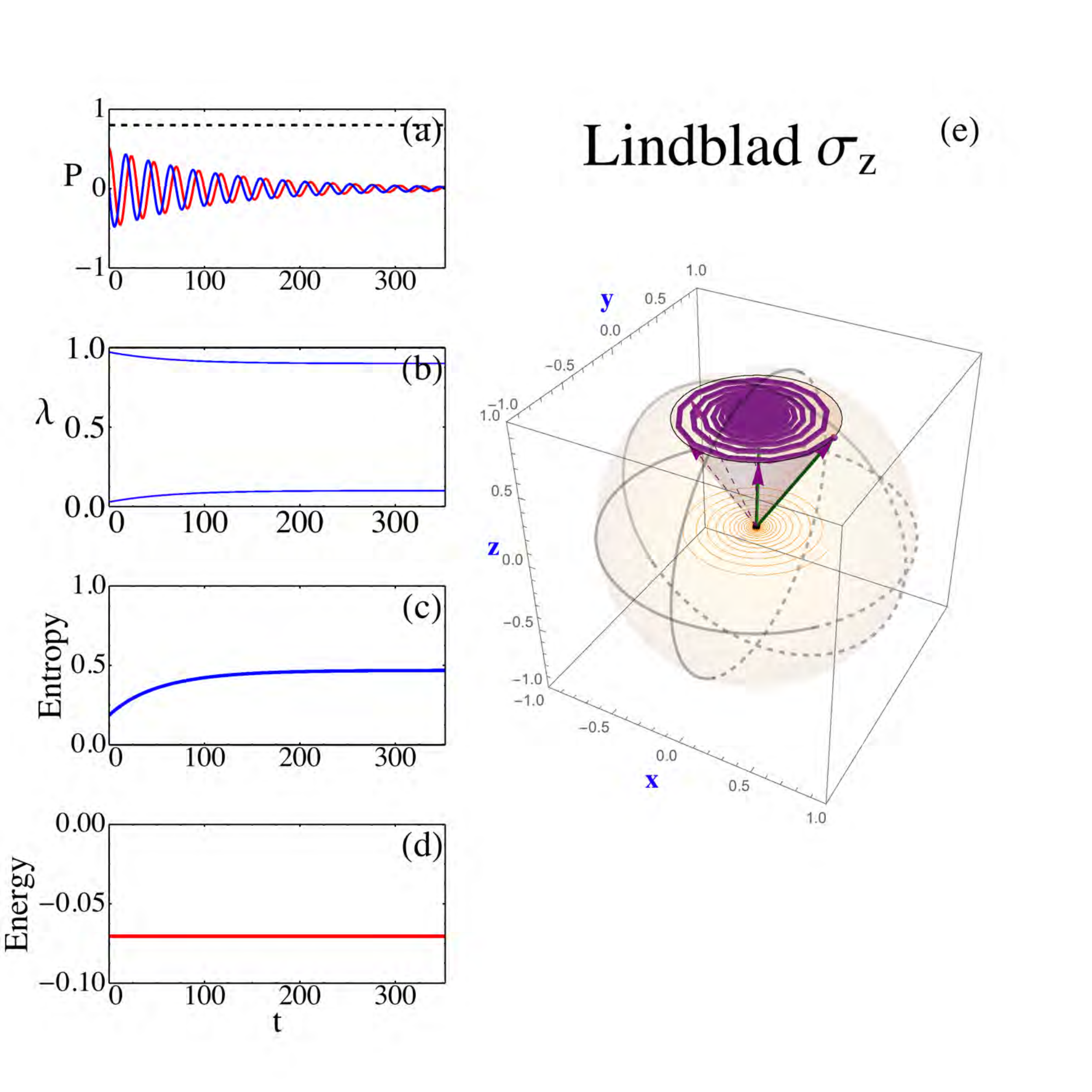}  
        \caption{Steady Lindblad. $\sigma_z$}
\protect\label{LindZ}
\end{figure}

\begin{figure}[h]       
    \includegraphics[width=\textwidth]{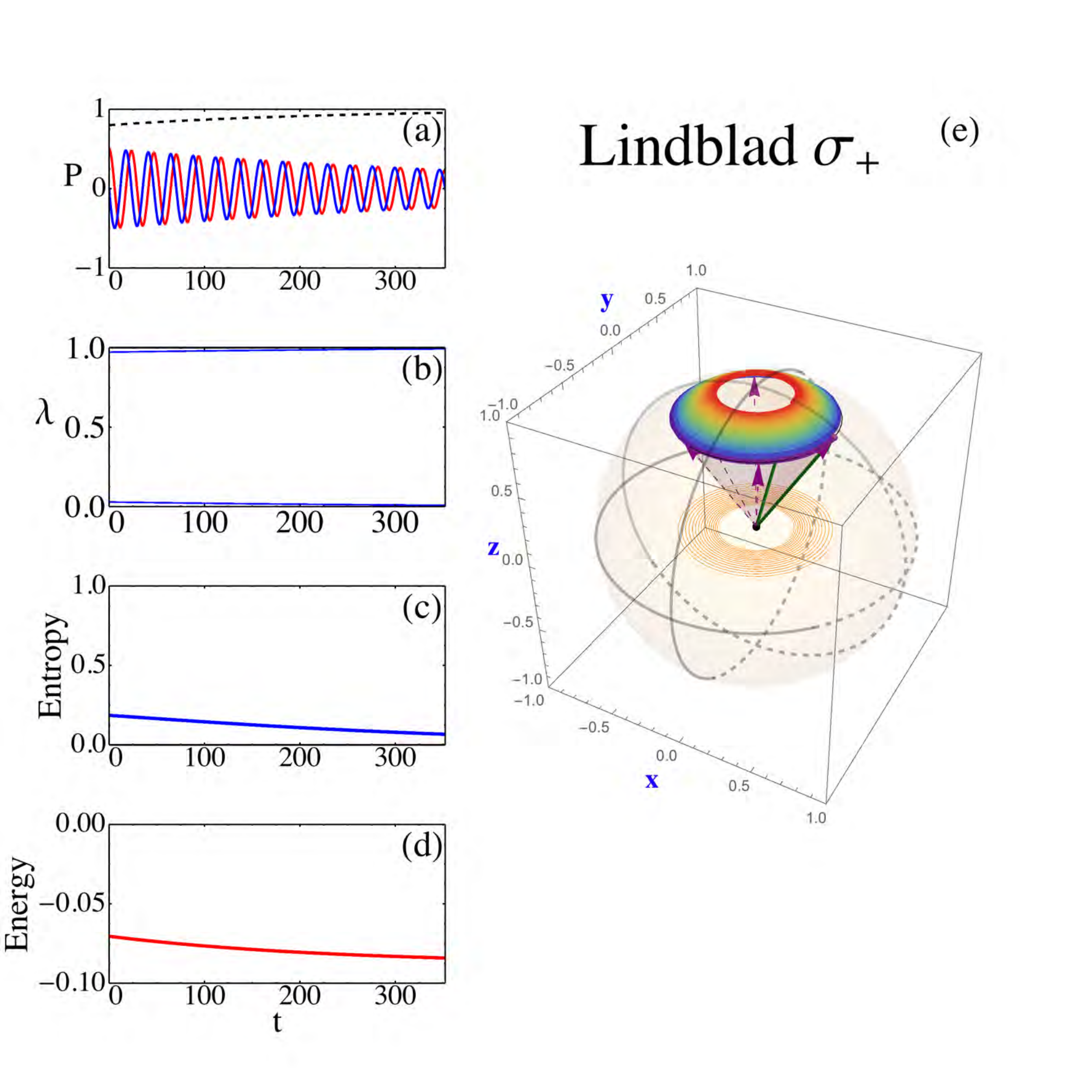}  
       \caption{Steady Lindblad:  $\sigma_{+}$.  This operator drives the system towards a pure $\mid 0 \rangle$ state. }
       \protect\label{LindS+}
\end{figure}

\begin{figure}[h]       
    \includegraphics[width=\textwidth]{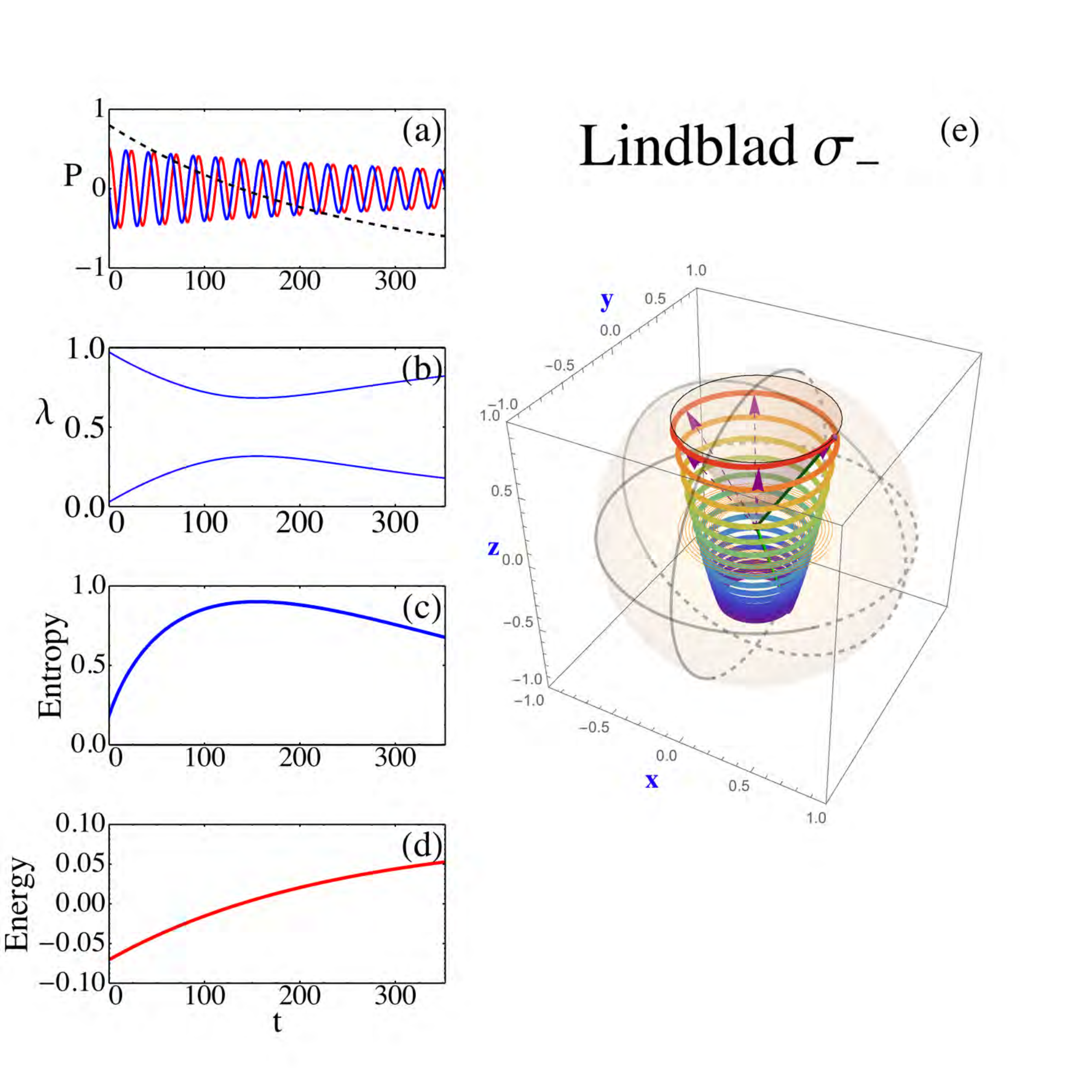}
        \caption{Steady Lindblad: $\sigma_{-}$ This operator drives the system towards a pure $\mid 1 \rangle$ state. Note here entropy first increases and then decreases.}
\protect\label{LindS-}
\end{figure}

  To gain insight into the properties of the Lindblad form, we first consider steady, i.e. time-independent, Lindblad operators.   Results for a master equation
  consisting of the unitary plus the Lindblad term only are
  shown in Figures~\ref{LindX}-\ref{LindS-} for several simple steady Lindblad operators $L(t).$ 
  
  The first case is   $L(t)=\sigma_x,$  for which the polarization vector time dependence is give by:
   $ \dot{P}_i =
 \sum\limits_{j=1,3}  M^{(1)}_{i\,j} P_j,$ with
 \begin{eqnarray}\label{Mmatrix1} 
 M^{(1)}=
 \left(
\begin{array}{lcl}
\ \ 0 & \omega_L &\  0\\
-\omega_L & -2\, \Gamma&\  0\\
\ \  0 &\ 0 &\  -2\, \Gamma
\end{array}\right)   ,
\end{eqnarray}  
   The  polarization vector then varies  as   
\bea
 P_x(t)&=&e^{-\Gamma \, t}{\Big \lbrace} \  P_x(0)  \cos \left( \omega t \right) +  (P_x(0) \frac{\Gamma}{\omega}+P_y(0)  \frac{\omega_L}{\omega})  \sin \left( \omega t \right)  {\Big \rbrace} \\ \nonumber
 P_y(t)&=&e^{-\Gamma   \, t}{\Big \lbrace}  P_y(0)  \cos \left( \omega t \right) -  (P_y(0) \frac{\Gamma}{\omega}+P_x(0) \frac{\omega_L}{\omega})   \sin \left( \omega t \right)  {\Big \rbrace} \\ \nonumber
 P_z(t)&=&e^{-2 \Gamma  \, t}\  P_z(0),
 \eea  with $\omega=\sqrt{\omega_L^2 - \Gamma^2}.$

  The second case is   $L(t)=\sigma_y,$  for which the polarization vector time dependence is give by:
   $ \dot{P}_i =
 \sum\limits_{j=1,3}  M^{(2)}_{i\,j} P_j,$ with
 \begin{eqnarray}\label{Mmatrix2} 
 M^{(2)}=
 \left(
\begin{array}{lcl}
\ \  -2\, \Gamma & \omega_L &\  0\\
-\omega_L &0 &\  0\\
\ \  0 &\ 0 &\  -2\, \Gamma
\end{array}\right)   ,
\end{eqnarray}  
   The  polarization vector then varies  as   
\bea
 P_x(t)&=&e^{-\Gamma \, t}{\Big \lbrace}  P_x(0)  \cos \left( \omega t \right) +  (-P_x(0)\frac{\Gamma}{\omega}+P_y(0) \frac{\omega_L}{\omega})  \sin \left( \omega t \right)  {\Big \rbrace} \\ \nonumber
 P_y(t)&=&e^{-\Gamma   \, t} {\Big \lbrace} P_y(0)  \cos \left( \omega t \right) +  (P_y(0) \frac{\Gamma}{\omega} -P_x(0)  \frac{\omega_L}{\omega})   \sin \left( \omega t \right)  {\Big \rbrace} \\ \nonumber
 P_z(t)&=&e^{-2 \Gamma  \, t}\  P_z(0),
 \eea   with $\omega=\sqrt{\omega_L^2 - \Gamma^2}.$

 Next case is   $L(t)=\sigma_z,$  for which the polarization vector's time dependence is very simple:
   $ \dot{P}_i =
 \sum\limits_{j=1,3}  M^{(3)}_{i\,j} P_j,$ with
 \begin{eqnarray}\label{Mmatrix3} 
 M^{(3)}=
 \left(
\begin{array}{lcl}
\ \  -2\, \Gamma & \omega_L &\  0\\
-\omega_L &-2\, \Gamma &\  0\\
\ \  0 &\ 0 &\  0
\end{array}\right)   ,
\end{eqnarray}  

   The  polarization vector then varies  as   
\bea
 P_x(t)&=&e^{- 2 \Gamma \, t}{\Big \lbrace}  P_x(0)  \cos \left( \omega_L t \right) +  P_y(0)   \sin \left( \omega_L t \right)  {\Big \rbrace} \\ \nonumber
 P_y(t)&=&e^{-2 \Gamma   \, t} {\Big \lbrace} P_y(0)  \cos \left( \omega_L t \right) -  P_x(0)   \sin \left( \omega_L t \right)  {\Big \rbrace} \\ \nonumber
 P_z(t)&=& P_z(0).
 \eea 
 
 All three of the above cases satisfy the condition $[L,L^\dagger] =0,$ and hence have increasing entropy and decreasing purity as displayed in Figures~\ref{LindX}-~\ref{LindZ}.
 
Next case is   $L(t)=\sigma_{+}= \frac{ \sigma_1 + i \sigma_2}{\sqrt{2}}$  for which the polarization vector's time dependence is 
   $ \dot{P}_i = \delta_{i 3} \Gamma+
 \sum\limits_{j=1,3}  M^{(4)}_{i\,j} P_j,$ with
 \begin{eqnarray}\label{Mmatrix4} 
 M^{(4)}=
 \left(
\begin{array}{lcl}
\ \  -\frac{1}{2}\, \Gamma & \omega_L &\  0\\
-\omega_L & -\frac{1}{2}\, \Gamma &\  0\\
\ \  0 &\ 0 &\  -\Gamma
\end{array}\right)   .
\end{eqnarray}  
   The  polarization vector then varies  as   
\bea
 P_x(t)&=&e^{- \frac{1}{2} \Gamma \, t}{\Big \lbrace}  P_x(0)  \cos \left( \omega_L t \right) +  P_y(0)   \sin \left( \omega_L t \right)  {\Big \rbrace} \\ \nonumber
 P_y(t)&=&e^{- \frac{1}{2} \Gamma   \, t} {\Big \lbrace} P_y(0)  \cos \left( \omega_L t \right) -  P_x(0)   \sin \left( \omega_L t \right)  {\Big \rbrace} \\ \nonumber
 P_z(t)&=&1+ e^{-  \Gamma \, t} (-1+ P_z(0)) .
 \eea   Here we see that the z-component changes faster than the other components and increases with time from its initial value to one.  The  $\sigma_{+}$ is a lowering operator and drives the system towards a pure $\mid 0 \rangle$ state and thus decreases entropy.  This decrease is also seen from  $[L,L^\dagger] = +2  \sigma_z$ and thus $\dot{S} \geq -2 \Gamma P_z.$

 Last case is   $L(t)=\sigma_{-}= \frac{ \sigma_1 - i \sigma_2}{\sqrt{2}}$  for which the polarization vector's time dependence is 
   $ \dot{P}_i =  - \delta_{i 3} \Gamma+
 \sum\limits_{j=1,3}  M^{(5)}_{i\,j} P_j,$ with
 \begin{eqnarray}\label{Mmatrix5} 
 M^{(5)}=
 \left(
\begin{array}{lcl}
\ \  -\frac{1}{2}\, \Gamma & \omega_L &\  0\\
-\omega_L & -\frac{1}{2}\, \Gamma &\  0\\
\ \  0 &\ 0 &\  -\Gamma
\end{array}\right)   .
\end{eqnarray}  
   The  polarization vector then varies  as   
\bea
 P_x(t)&=&e^{- \frac{1}{2} \Gamma \, t}{\Big \lbrace}  P_x(0)  \cos \left( \omega_L t \right) +  P_y(0)   \sin \left( \omega_L t \right)  {\Big \rbrace} \\ \nonumber
 P_y(t)&=&e^{- \frac{1}{2} \Gamma   \, t} {\Big \lbrace} P_y(0)  \cos \left( \omega_L t \right) -  P_x(0)   \sin \left( \omega_L t \right)  {\Big \rbrace} \\ \nonumber
 P_z(t)&=&-1+ e^{-  \Gamma \, t}  (1+ P_z(0)).
 \eea
 Here we see that the z-component changes faster than the other components and decreases with time from its initial value to minus one.  The  $\sigma_{-}$ is a raising operator Fig.~\ref{figlevels} and drives the system towards a pure $\mid 1 \rangle$ state. We now have  $[L,L^\dagger] = - 2 \sigma_z$ and thus $\dot{S} \geq +2 \Gamma P_z,$  which is positive initially ($P_z  \rangle 0$)
 and turns negative after $P_z$ flips to negative values.

 These last two cases  displayed in Figures~\ref{LindS+}-~\ref{LindS-}  do not satisfy the condition $[L,L^\dagger] =0,$
 and their entropy and purity evolutions  do not satisfy the  $ \frac{d S}{dt} \geq 0$
 and  $ \frac{d {\cal P}(t)}{d t}\leq 0$ rules.
 
 These various choices for Lindblad operators provide different behavior.  For example, the $L= \sigma_z$ gives attenuation of the x and y
 polarization vectors and leaves the z component fixed. As we see later that case describes a system that has an increasing entropy at a fixed temperature.
 To model a system that evolves with fixed temperature, increasing entropy and no heat flow, we later turn to the Beretta form.
 
\clearpage 
   \subsubsection{Lindblad Projective Measurement}
   
   The Lindblad operator represents a disturbance of the system due to outside effects.  The act of measurement is an important outside effect in that a device is used to act on the system and to record its impact.  As an example of how a Lindblad operator can represent the measurement process, we consider a one qubit system and its associated observable polarization vector.  The Lindblad operator is assumed to be of the form 
    \bea
    L &=&( \vec{\sigma}\cdot \vec{a} ) \  \theta_M(t) \\ \nonumber
    \theta_M(t)&=&\frac{1}{2}\,{\Big \lbrace} {\rm Erf}(\frac{t-t_1}{r})-  {\rm Erf}(\frac{t-t_2}{r} ) {\Big \rbrace}\, ,   \eea
    where $\theta_M(t)$ is a pulse that equals one in the interval $t_1$ to $t_2$ with soft edges with a small $r = \tau_m/100 .$ This Lindblad pulse  acts on the system during a short interval
$\tau_m=t_2-t_1.$  The center of the pulse  is at $t_0 \equiv \frac{t_1+t_2  }{2},$ and the measurement is over at time
$t_m=t_2.$ 
   The  real unit vector  $\vec{a}\equiv a_x \hat{n}_x + a_y \hat{n}_y  + a_z \hat{n}_z \equiv \hat{m}$ is used to
   implement a measurement in the $  \hat{m}$ direction.    Note:  $ \int^{\infty}_{-\infty}\theta_M(t)\, dt = t_2-t_1 .$
   
 In order to minimize any precession effect during the measurement, we need to have a very fast pulse with a strong  Lindblad strength 
$ \Gamma \gg  \omega_L,$~\footnote{The Beretta and Bath terms do not affect the measurement provided  
 $\gamma_2 \ll \Gamma$ and $\gamma_3 \ll \Gamma.$    }. 
 We could simply stop the precession $ \omega_L \rightarrow 0,$ during the short measurement period  $\tau_m.$ That stoppage could be implemented by a  Hamiltonian bias pulse to generate temporary  level degeneracy.

Let us focus on how the polarization evolves during this measurement.  The change in polarization vector is now (see
the  Appendix)  for large $\Gamma$
\begin{eqnarray}\label{measure1}
 \frac{ d }{dt}\vec{P}(t)&=& - \vec{\omega}_L \times \vec{P}(t) + 2\ \Gamma\     \overrightarrow{\delta_{\!P}}\!(t)  \rightarrow + 2\ \Gamma\     \overrightarrow{\delta_{\!P}}\!(t) \\ \nonumber
  \overrightarrow{\delta_{\!P}}\!(t)&=&   \, \theta_M^2(t) \, (- \vec{P}(t) + (\vec{m}\cdot \vec{P}(t))\ \vec{m}).
\end{eqnarray} We have set  the  Lindblad parameters as real $\alpha_i= a_i  $ for $ i= 1 \dots 3,$ and $\hat{a}.\hat{a}=1.$
The condition $[L^\dagger,  L]=0$  holds and thus  the system's entropy increases during the  measurement.
A measurement  along axis  $\hat{m}$  becomes:
\be
\label{measure2}
 \frac{ d }{dt}\vec{P}(t)= - 2\ \Gamma   {\Big \lbrace}    \vec{P}(t) - (\hat{m}\cdot \vec{P}(t))\ \hat{m} {\Big \rbrace}\, \theta_M^2(t) ,
\ee  which displays the measurement property that the polarization in the measurement direction is unchanged in time, whereas the components in the other directions are reduced to zero rapidly.  
Inside the pulse time region the rule for one qubit is simply:
\bea
\label{measure3}
 \frac{ d }{dt}( \hat{m}\cdot\vec{P}(t))&=& 0 \\ \nonumber
 \frac{ d }{dt}( \hat{m}_{ \perp }\cdot\vec{P}(t))&=& - 2 \Gamma   (\hat{m}_{ \perp }\cdot\vec{P}(t))\,   , \eea where $ \hat{m}$ denotes the measurement  direction and  $ \hat{m}_{ \perp }$ directions perpendicular to the
 measurement direction.  In the pulse time region,  clearly $ \hat{m}\cdot\vec{P}(t_m)= \hat{m}\cdot\vec{P}(t_1)\ \  \&\ \ 
 \hat{m}_{ \perp }\cdot\vec{P}(t) =e^{- 2 \Gamma  (t-t_1) } \  \hat{m}_{ \perp }\cdot\vec{P}(t_1),$ where
 $t_1$ is  the measurement starting time.  Here $\Gamma$ is much larger then the Larmor frequency to assure rapid collapse of the density matrix under measurement.   Thus the perpendicular component go to zero rapidly at the end of the measurement $t_m;$  to assure that  limit the value of $\tau_m=t_2-t_1$ is set equal to $\pi/\Gamma.$
 Then $e^{- 2 \Gamma  (t_2-t_1) } =e^{- 2 \pi}\approx 0.0019.$  To obtain a better zero a smaller $\tau_m$ can be used,  but then the numerical evaluation requires increased precision.

The removal of perpendicular polarization components by measurement,  shifts the one-qubit  eigenvalues towards 1/2, with a  corresponding increase in entropy and decrease in purity.  The measurement is completed at time $t_m$  when no further change occurs and the collapsed components have been removed.
Therefore the final polarization is
\be
\vec{P}_F =  (\hat{m}\cdot \vec{P}_I)\, \hat{m}.
\ee

{\bf  Measurement operator}

The above Lindblad dynamical picture of measurement is equivalent to the Copenhagen version of
density matrix collapse.
The one-qubit operator for a projective measurement is in general
\be
M_\alpha = \mid\alpha   \rangle  \langle  \alpha   \mid = \frac{1}{2}  {\Big \lbrace}  \sigma_0 + \vec{\sigma}\cdot \hat{m}_\alpha)  {\Big \rbrace},
\ee  where $\hat{m}_\alpha$ denotes a measurement direction.  For a projective measurement the states
$\mid\alpha   \rangle$ are orthonormal.   Two sequential identical measurements
$M_\alpha M_\alpha =M_\alpha,$ are equivalent to one measurement.  For two distinct measurements
$\beta \neq \alpha$  $M_\beta M_\alpha = \mid\beta \rangle  \langle  \beta   \mid \alpha   \rangle  \langle  \alpha   \mid =0.$
Thus for a basis of distinct measurements  $M_\beta M_\alpha = \delta_{\beta\, \alpha} M_\alpha.$
A complete set of measurements yields $\sum_{ \alpha } M_\alpha =1.$  A simple example of these general remarks is taking $\alpha = \hat{m}$ and $ \beta = -\hat{m}.$ Compounding both of these measurements for polarization in the positive and negative $ \hat{m}$ directions is used to construct a final state density matrix after such measurements.

In line with these remarks,  the usual quantum rule for the final density matrix after projective measurement is
\bea
\rho_F &=&\frac  { \sum_{ \alpha} M_\alpha \rho_I  M_\alpha }{ {\rm Tr}( M_\alpha \rho_I  M_\alpha   )}
\\ \nonumber
 &=& \frac  { \sum_{ \alpha}(  \mid\alpha   \rangle  \langle  \alpha   \mid)\,   \langle  \alpha \mid \rho_I \mid \alpha \rangle}{  \sum_{ \alpha}   \langle  \alpha \mid \rho_I \mid \alpha \rangle},
\eea where the state sum is over the $\pm \hat{m}$ directions.
Note that the probability for a specific measurement is  $ \langle  \alpha \mid \rho_I \mid \alpha \rangle={\rm Tr}(M_\alpha \rho_I)=
\frac{1}{2} (1 + \vec{P}_I\cdot \hat{m}_\alpha)$ and then the probability summed over both directions is 
$\sum_\alpha  \langle  \alpha \mid \rho_I \mid \alpha \rangle=1.$ Evaluation of the numerator, then gives
the final density matrix
\bea
\rho_F &=&  \mid\hat{m}   \rangle  \langle \hat{m}   \mid  \frac{1}{2} (1 + \vec{P}_I\cdot \hat{m})  
 + \mid\hat{-m}   \rangle  \langle \hat{-m}   \mid  \frac{1}{2} (1 - \vec{P}_I\cdot \hat{m}) \\ \nonumber
 &=&  \frac{1}{2} (\sigma_0 + \vec{P}_I\cdot \hat{m}\  ( \mid \hat{m}    \rangle  \langle \hat{m} \mid  - \mid \hat{-m}    \rangle  \langle \hat{-m} \mid )\,  ) \\ \nonumber
  &=& \frac{1}{2} (\sigma_0 +  \vec{P}_I\cdot \hat{m}\  \vec{\sigma}\cdot \hat{m} ) \\ \nonumber
 \vec{P}_F  &=& ( \hat{m} \cdot\vec{P}_I )  \hat{m}.
  \eea  The final result shows that the usual measurement operator collapse is equivalent to a time-dependent Lindblad equation approach to measurement.   If the Lindblad equation can be mapped to a stochastic Schr\"{o}dinger equation,  that would be an additional step towards a dynamic view of state collapse during measurement.~\footnote{ S. Weinberg ~\cite{Weinberg1,Weinberg2} has recently used the Lindblad equation to describe measurement from a more sophisticated view point which he suggests  provides a conceptually improved starting point for formulating quantum mechanics. Also see~\cite{Hatsopoulos,Park} for earlier work based on a formulation of nonequilibirium quantum thermodynamics. }

 A numerical example of a one-qubit Lindblad measurement  is shown in Fig~\ref{measure1}.  In this example the measurement is accomplished by a strong, fast Lindblad operator pulse ($\Gamma= 100\,  \omega_L= 26.752\, GHz $)    in direction $ \hat{m}=(\hat{x}+\hat{z})/\sqrt{2}. $
  The pulse starts at $t_1=2\  T_L = 46.973$\, nsec and ends at time $t_2= t_1 + \tau_m=47.091$\, nsec . 
  The Larmor precession period ( $T_L=23.487$\, nsec )  is much less than the
   pulse width  ($\tau_m= \pi/(2 \Gamma)=
 0.117$\, nsec),  which is set so negligible precession occurs during the measurement pulse. 
 The initial polarization at t=0  equals  $( P_x(t_1), P_y(t_1), P_z(t_1))=(0.2 ,0.4,0.8).$  
    The measurement is completed at time $t_2$  with a Lindblad operator  $\theta_M(t)\ (\sigma_x+ \sigma_z)/\sqrt{2}.$  As expected, the polarization vector after this measurement collapses to  $\vec{P}_z=\vec{P}_x=.5 $ and  $\vec{P}_y=0.$

 In Fig~\ref{measure2}, we take a closer  look at the above measurement example. 
  The left plot shows the initial precession, the rapid measurement starting at $t_1= 2\ T_L==46.97$\, nsec and the post-measurement precession .  The measurement collapses the  z and x components to equal values of  0.5 ;   whereas, in the un-measured perpendicular $\hat{y}$ direction the polarization collapses to zero. Thereafter, normal precession continues starting from the measured $P_x= P_z= .5  ,
 P_y=0 $ values after collapse.   The right plot shows the details during the pulse with the vertical dotted lines indicating the start and end of the Lindblad pulse.
 
 This procedure is readily generalized to multi-qubit, qutrit and hybrid cases, by extension to a wider range of polarization observables.

 \begin{figure}[!tbp]]      
  \includegraphics[width=\textwidth]{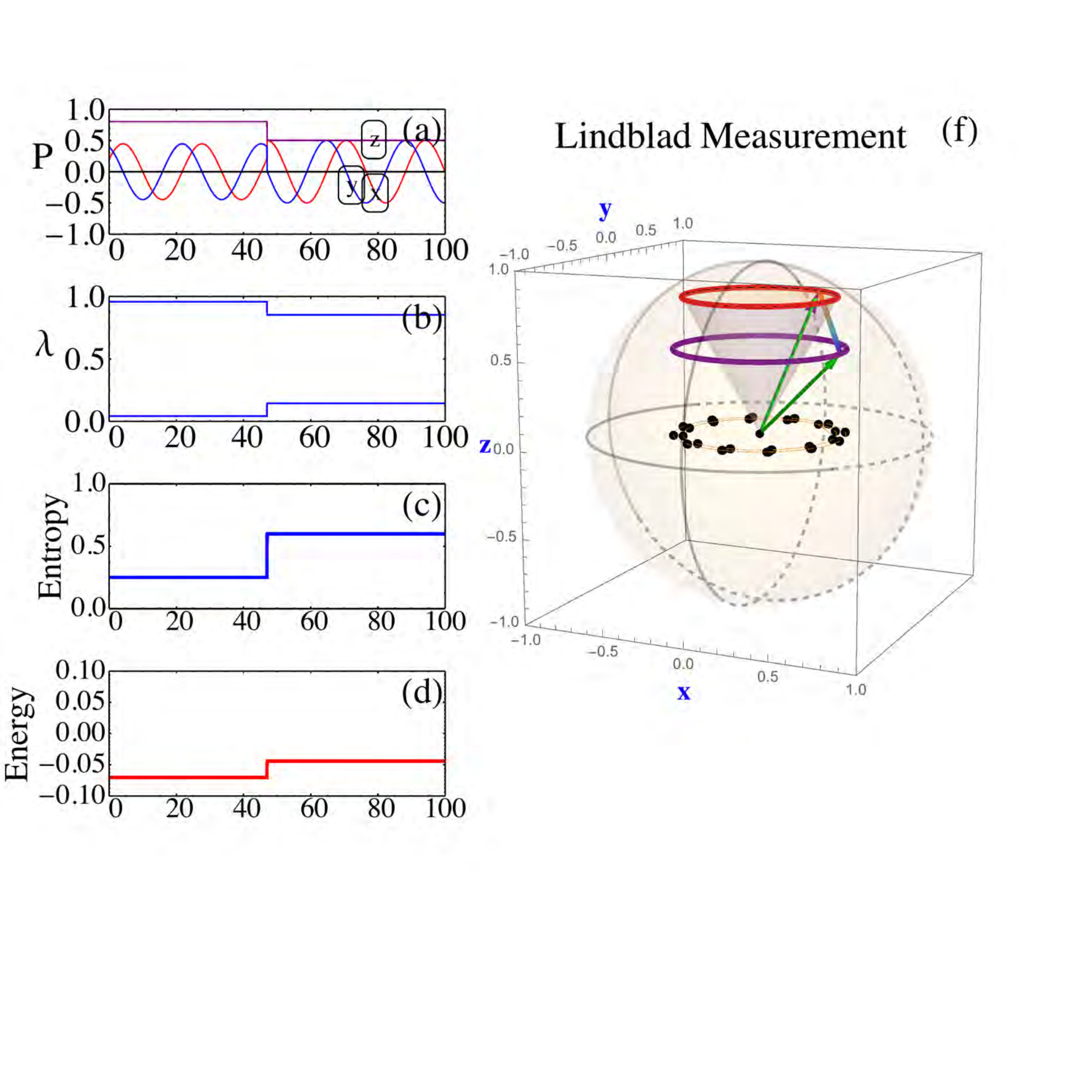} 
    \caption{ Measurement as a strong  Lindblad operator pulse.
    (a) Polarization vector evolution.  Note after measurement polarization vector collapses to $\vec{P}=(\vec{P}_I\cdot \hat{m})\, \hat{m}=(0.50,0.0,0.50),$ with $P_y \rightarrow 0.$   (b) Change in eigenvalues as expected from reduced magnitude of polarization vector;
   (c) Entropy increase due to measurement; (d) Energy change due to heat transfer; (e) Original precession cone followed by collapse to final (solid green) vector and continued precession.   }
\protect\label{measure1}
\end{figure}

\begin{figure}[!tbp]]      
    \includegraphics[width=\textwidth]{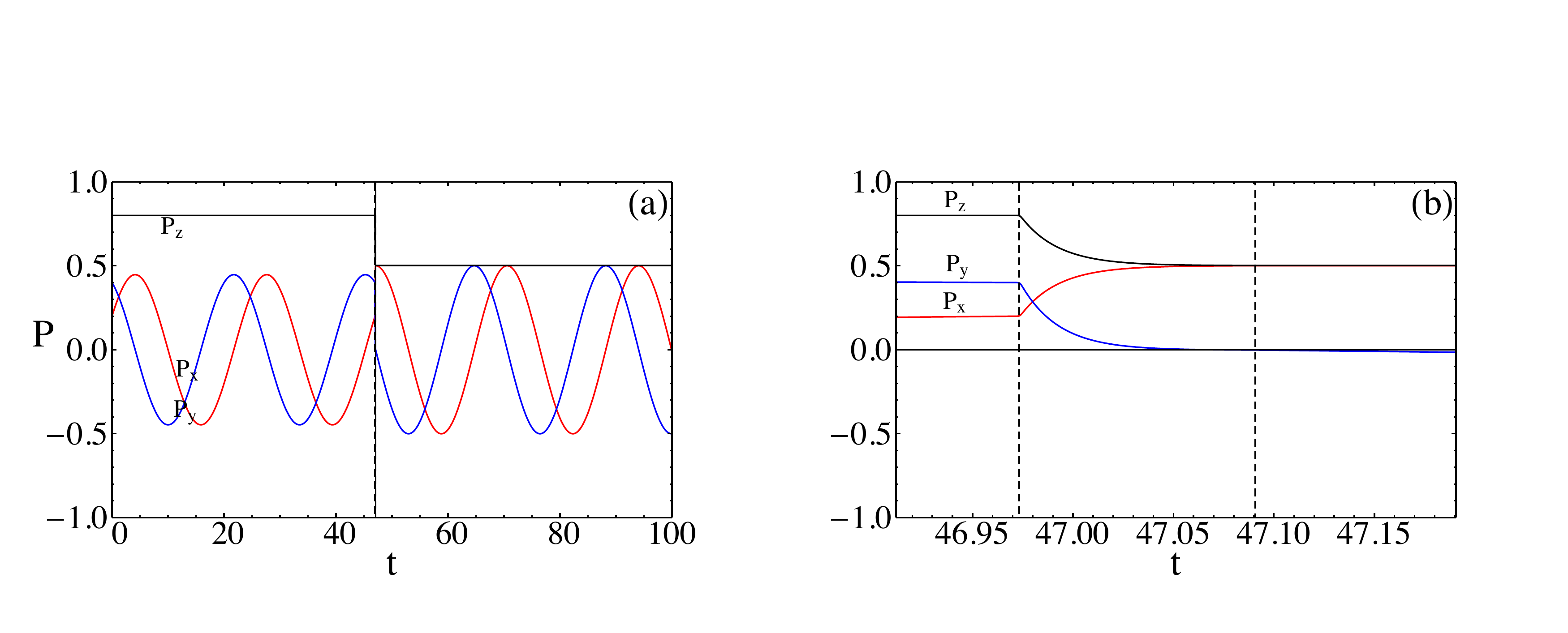}  
    \caption{ The initial precession followed by the rapid measurement and then continued precession:  (a) shows the 
    evolution during the 0 to 100 nsec time; (b) shows the detailed evolution during the $t_1$ to $t_2$ pulse (marked as vertical dotted lines).  In both plots the initial polarization $( P_x(t_1), P_y(t_1), P_z(t_1)) \rightarrow ( P_x(t_2), P_y(t_2), P_z(t_2)) $ with $P_y(t_2)=0$
    and  $P_x(t_2)=P_z(t_2)=(P_x(t_2)+P_z(t_2))/2$ as expected.}    \protect\label{measure2}
\end{figure}

%\clearpage
     \subsubsection{Noise Lindblad Operator }
     We do not use steady Lindblad operators! Instead  the Lindblad form is used only to incorporate noise pulses.  If a noise pulse coincides in time with a gate pulse, we denote those cases as gate friction.   It is possible to design 
     time-dependent Lindblad operators that could drive the system to a specific equilibrium state or  could describe a closed system that evolves with no heat transfer, but increasing entropy.  These would be rather complicated Lindblad operators.  It is much simpler to separate those effects into the Bath and Beretta terms, as we advocate.  Thus in this treatment,  the Lindblad form is only used to include random noise, gate friction and as described earlier, measurements.

\begin{figure}[!tbp]]  
 \includegraphics[width=1.0\textwidth]{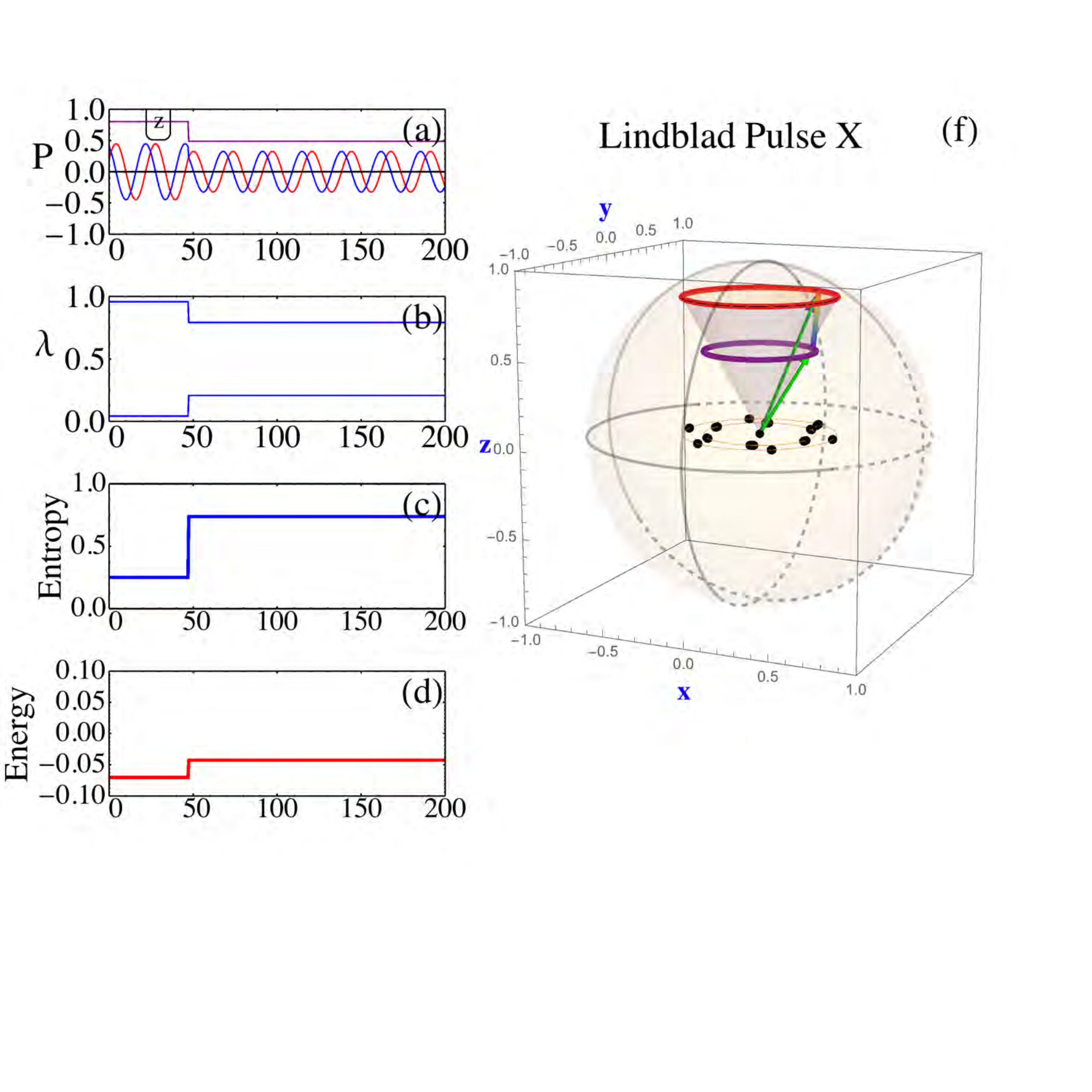} 
   \caption{  Lindblad pulse  $L(t) = \sigma_1\  \theta_L (t),$ acts from $t_1=46.97\,$ to $t_2=47.44$\, nsec\, .
      Strength  $\Gamma=  2\, \omega_L= 0.535\, GHz$ was used to enhance the Lindblad pulse role.
      The initial polarization vector is $ \{0.2,0.4,0.8  \}. $  } 
       \protect\label{pulse1NA}
        \end{figure}

     Some simple examples of Lindblad pulses are now provided.  The first Fig~\ref{pulse1NA} is a  Lindblad pulse  $L(t) = \sigma_1\  \theta_L (t),$ acting at a time $t_0$ with width $\tau_L.$ 
      A  simple soft square shape is used to introduce noise: $$ \theta_L(t)= \frac{1}{2}  \  {\Big  \lbrack} {\rm Erf}(\frac{t-t_1}{\tau})-  {\rm Erf}(\frac{t-t_2}{\tau}) {\Big  \rbrack}  ,$$ where 
      $ \int^{\infty}_{-\infty}\theta_M(t)\, dt = t_2 - t_1 ,$ and
 $ \theta_M(t_0)=1,$  along with $\Gamma$ set the strength of the pulse.   Figure~\ref{pulse1NB} provides a closer look at the action of this Lindblad operator during its pulse. During the pulse, the z and y polarizations are reduced and the x component is unchanged, which is reflected in the increase of entropy and the eigenvalue motion towards equality seen in  Fig~\ref{pulse1NA}.

\begin{figure}[!htbp]]      
    \includegraphics[width=\textwidth]{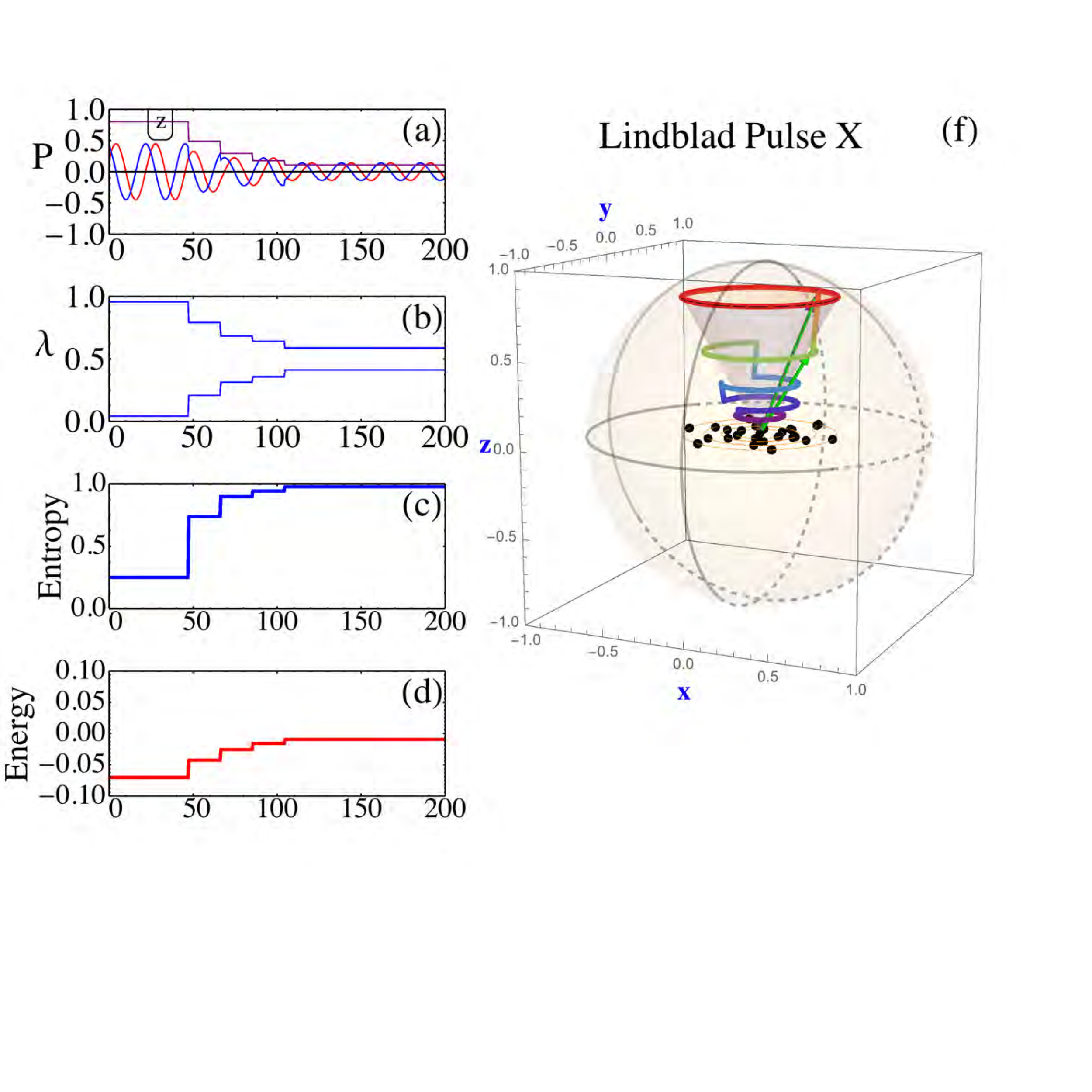} 
    \caption{ Four  equi-spaced Lindblad pulses  $L(t) = \sigma_1\  \theta_L (t),$ acting from $t_1=46.97$ \, nsec to $t_f=104.7$ \, nsec, of width $ 0.47$\, nsec.
  A large $\Gamma=  2\, \omega_L$ was used  to enhance the Lindblad pulse role.  The initial polarization vector is $ \{ 0.2,0.4,0.8  \} .$ }    \protect\label{pulse4NA}
\end{figure}

        The second example Fig~\ref{pulse4NA} consists of four $\sigma_1$ noise pulses  of fixed strength
     $L(t) = \sigma_1\ \sum_{i,1,4}  \theta_L (t -  \Delta_i),$ acting at a times $t_0 +\Delta_i $ with  fixed width $\tau_L.$
          
\begin{figure}[!tbp]]      
    \includegraphics[width=.85\textwidth]{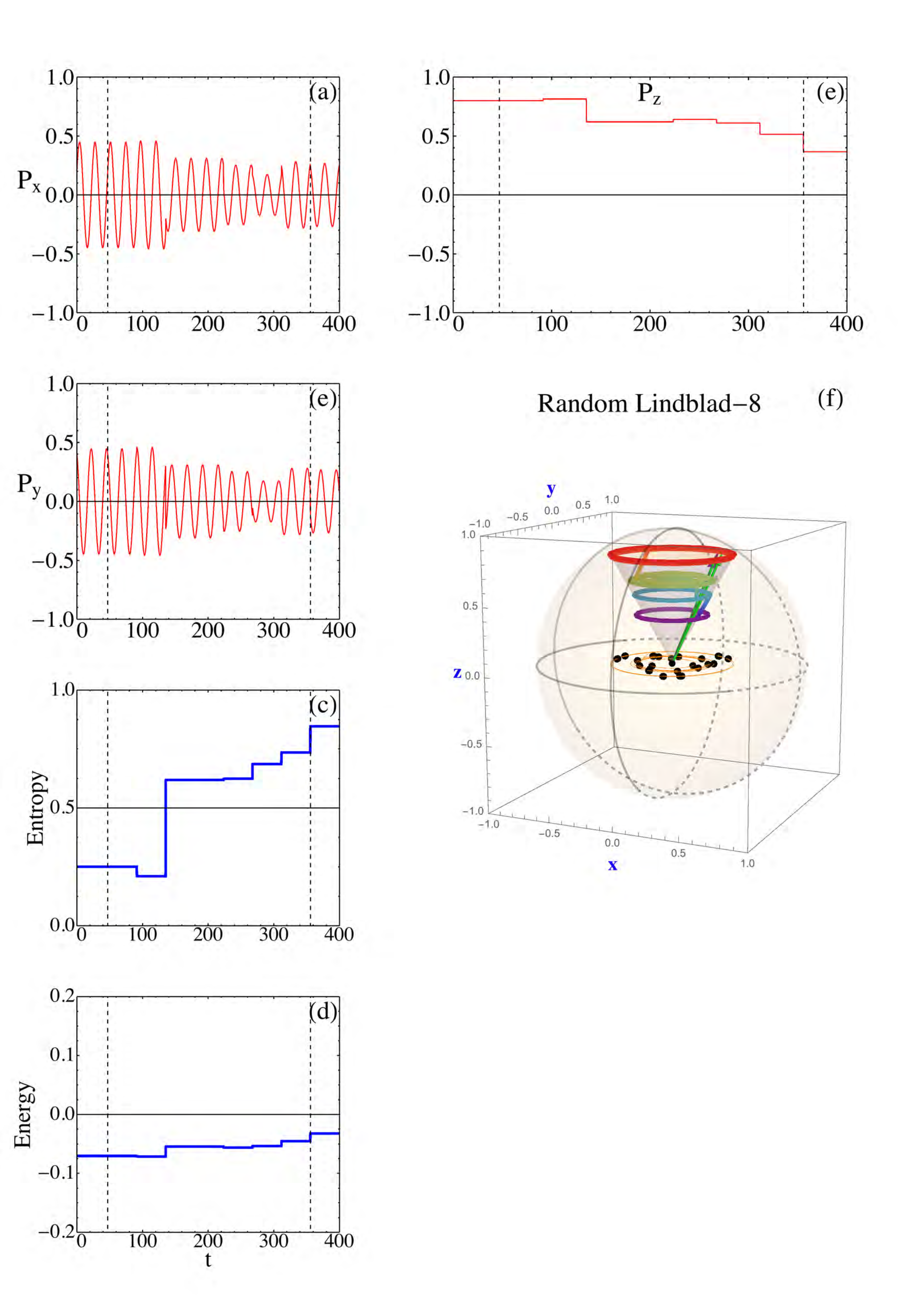}  
    \caption{ Eight random Lindblad pulses of width $ 0.47$\, nsec acting at equally spaced time intervals during $t_1=46.97$\, nsec to $t_f=355.9$\,  nsec. 
 Each pulse is of form
    $L(t) = \vec{\alpha}\cdot \vec{\sigma}  \theta_L (t),$ with complex random  $\vec{\alpha}$.  Note that the entropy decreases in intervals when $[L^\dagger,L]\neq 0.$  Vertical  dotted lines indicate the start of the first and end of the last pulse. A large $\Gamma=  2\,  \omega_L$ was used here to enhance the Lindblad pulse role.  The initial polarization vector is $\{0.2,0.4,0.8  \}.$  }
      \protect\label{pulserand8}
\end{figure}

Finally, we present a case Fig~\ref{pulserand8} with eight random Lindblad pulses each of general form $  \vec{\alpha}\cdot \vec{\sigma}.$
     In that complex case  periods of entropy decrease as well as increase occur.  In general, the Lindblad pulse set 
     is designed to simulate the extant noise impacting the  system.
     
     In the above cases there are no gates, just Lindblad noise.

      These are preliminary tests of the stability of quantum memory under the intrusion of noise.  The stability of quantum memory
      is  illustrated in Fig.~\ref{FpulseNot3} where the dependence of Fidelity on the Lindblad strength is displayed.
      The two density matrices used in the fidelity are $\rho(t)$ at $t_1=2 T_L$ and at a final Larmor grid time after the noise abates, $t_f = {\rm Mod}[\rm{tend}, T_L]\times  T_L,$ i.e. on the Larmor time grid.  This procedure can set limits on allowed noise to maintain a  stable density matrix.
       Later memory stability will also be tested including the Beretta and Bath terms.

\begin{figure}[!tbp]] 
 \protect\label{pulse1NB}     
   \includegraphics[width=\textwidth]{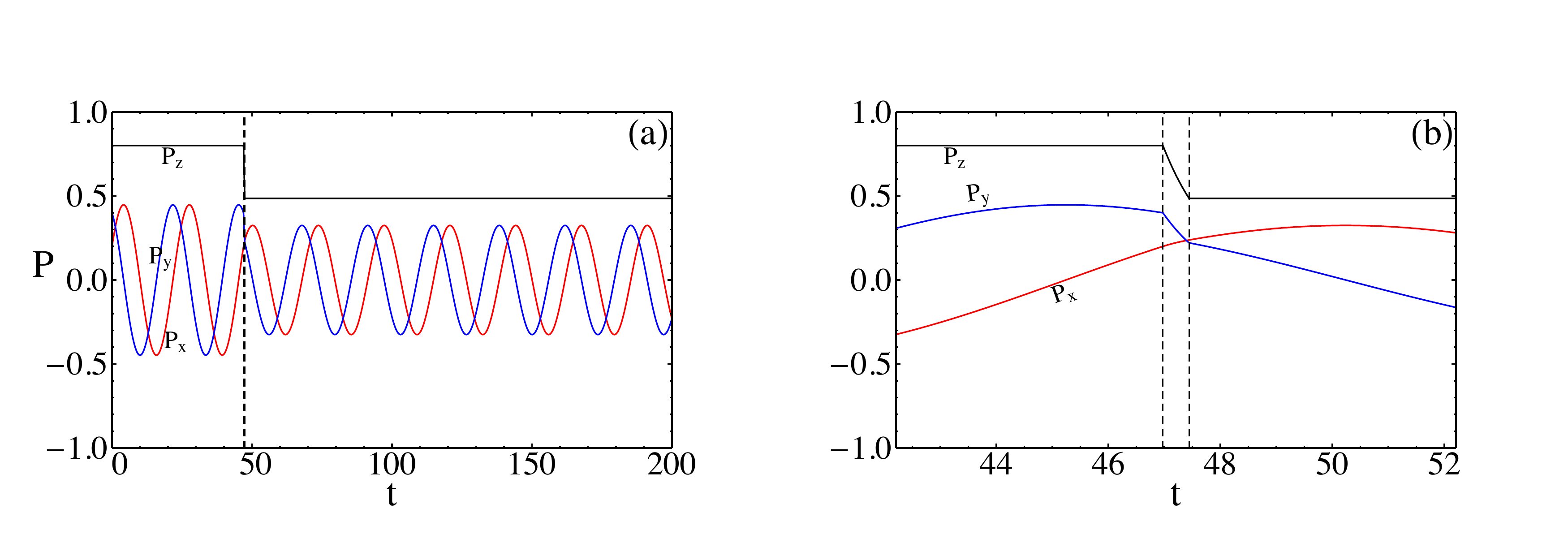}
  \caption{  View of Lindblad  pulse  $L(t) = \sigma_1\  \theta_L (t)\,$ (a) over the 0 to 200 nsec time span (b) during the pulse period . Vertical dashed lines indicate pulse start $t_1=46.97$\, nsec and finish at $t_2=47.44$\, nsec. Change of polarization due to this Lindblad pulse is given by
  $\delta \vec{P} = - 2\, \Gamma\, \{ 0,P_y,P_z\},$ see the Appendix.
    Again $\Gamma=  2\, \omega_L$ enhances the Lindblad pulse role. The initial polarization vector is $\{0.2,0.4,0.8  \}.$ } 
    \protect\label{pulse1NB}  
\end{figure}

     In all cases, we use a positive strength $\Gamma\geq 0$ for the Lindblad noise model.  However, it has been noted in the literature~\cite{feedback} that a negative $\Gamma\leq 0$ can simulate non-Markov  effects.  This would represent an environment
     that is affected by the system and after a delay feeds back some of the information to the system, thereby 
     ameliorating the detrimental effects of noise.  This kind of feed-back is well known in the optical model of nuclear reactions as formulated by projection operator methods.  Design of such an environment could be the key to
     achieving stable memory and operations in a QC.  This will be explored in a future study.

     \subsubsection{Gates plus Lindblad Operators}
     
\begin{figure}[!tbp]]      
    \includegraphics[width=.85\textwidth]{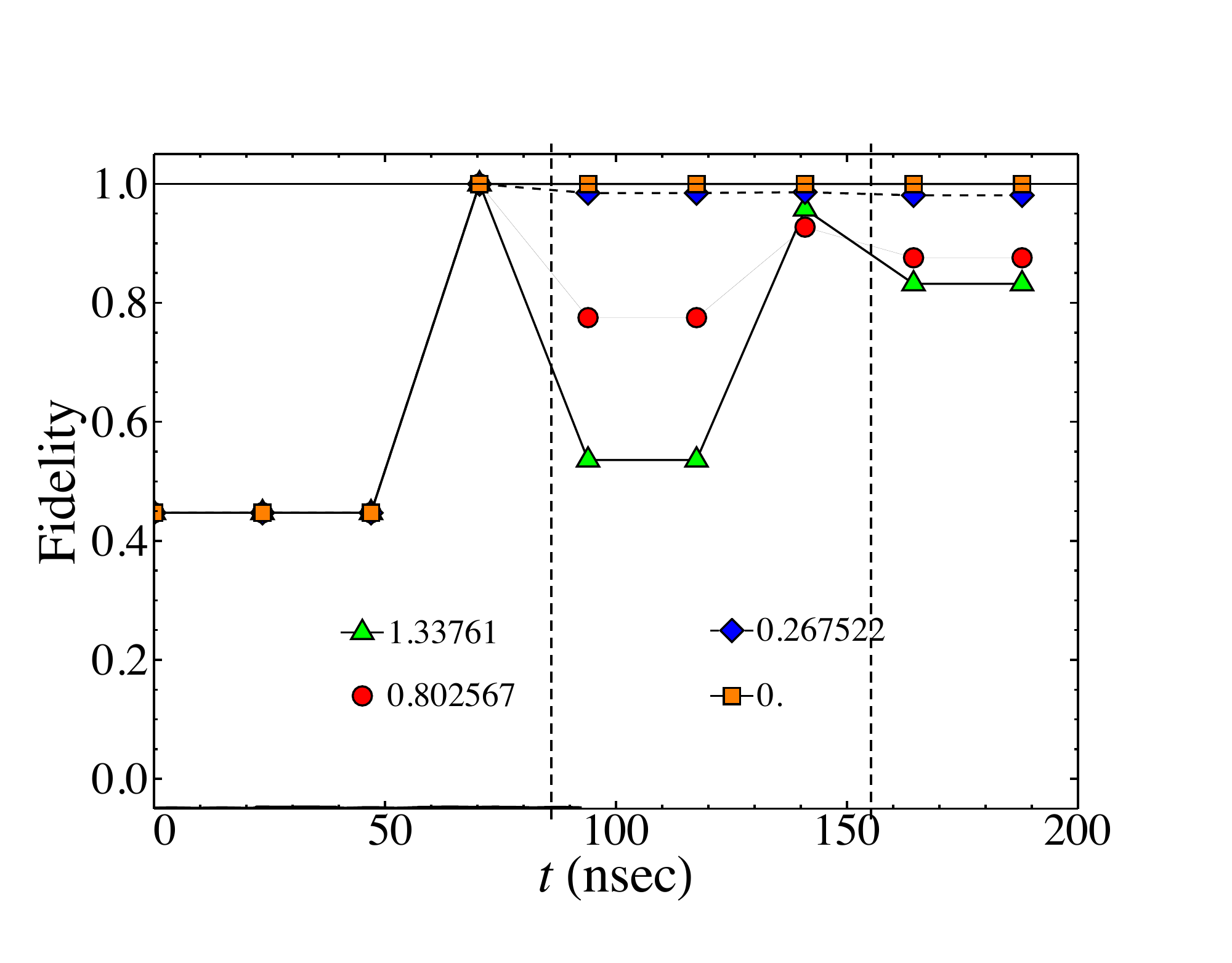} 
     \includegraphics[width=.85\textwidth]{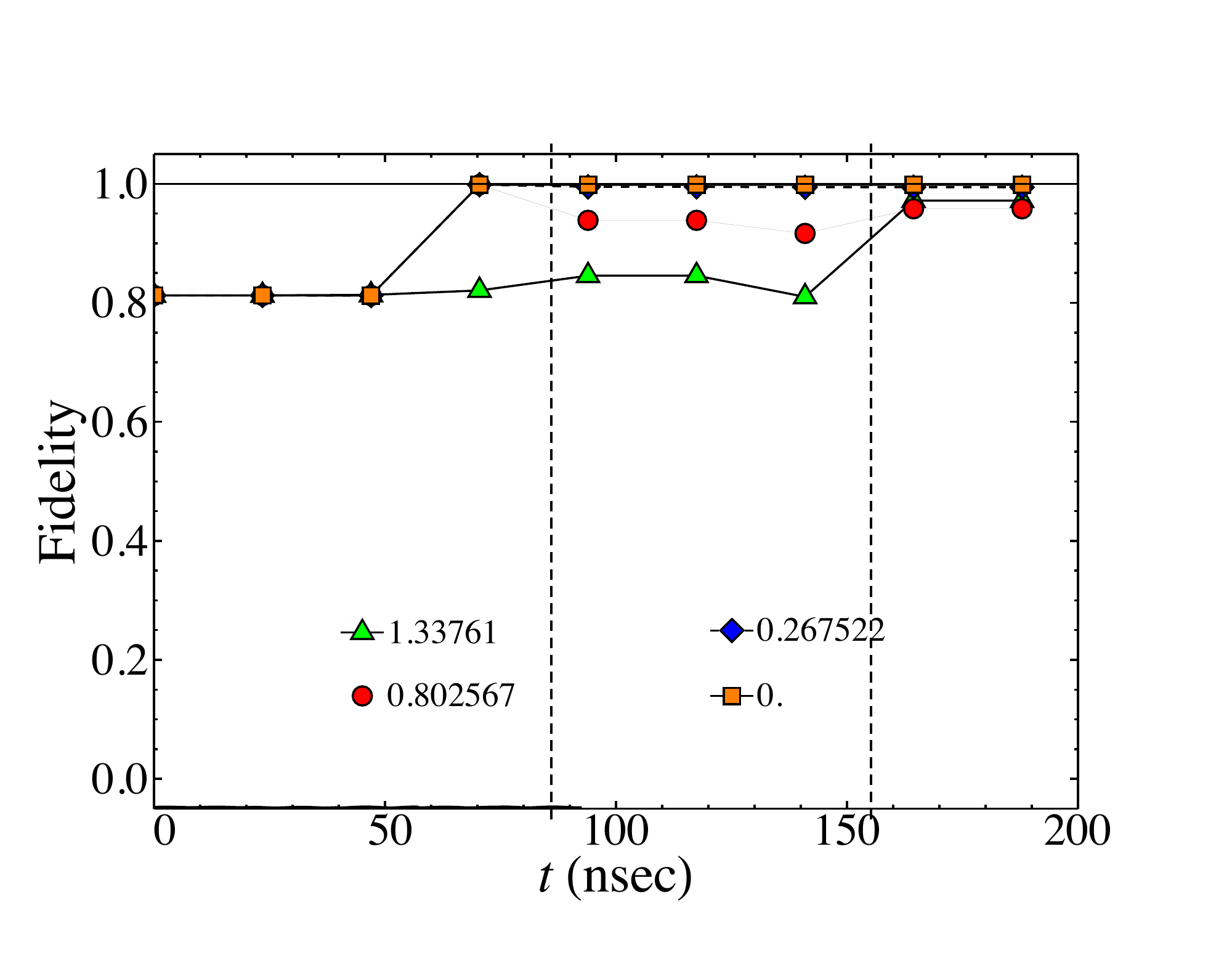} 
    \caption{Top( Not),Bottom(Hadamard) gates  followed by three  random Lindblad pulses at equally spaced  time intervals.  
   Gate pulse is from $t_1=46.97$\, nsec to $t_2=47.443$\,  nsec, of width $ 0.47$\, nsec.
     Lindblad pulses are 
    $L(t) = \vec{\alpha}\cdot \vec{\sigma}  \theta_L (t),$ with complex random  $\vec{\alpha}$.   Vertical  dotted lines mark  start/end of the Lindblad noise. Different $\Gamma's$ show effect of noise on the gate fidelity. Symbols are on the Larmor period grid. }
      \protect\label{FpulseNot3}
\end{figure}

     In addition to evolution of a non-degenerate system subjected to Lindblad noise, we can include quantum gates.  The cases of a NOT gate and then of a Hadamard gate along with three subsequent random noise pulses are shown in Fig.~\ref{FpulseNot3}. 
      As a figure of merit, the fidelity, evaluated on the Larmor period  $T_L$ grid is displayed  in Fig.~\ref{FpulseNot3}  for various  Lindblad strengths $\Gamma,$  which illustrates how 
     noise can affect efficacy and how to ascertain the allowed noise level to accomplish a simple process.
      
 \subsection{Equilibrium-Closed System}
 \label{sec4d}
\begin{figure}[h]       
    \includegraphics[width=\textwidth]{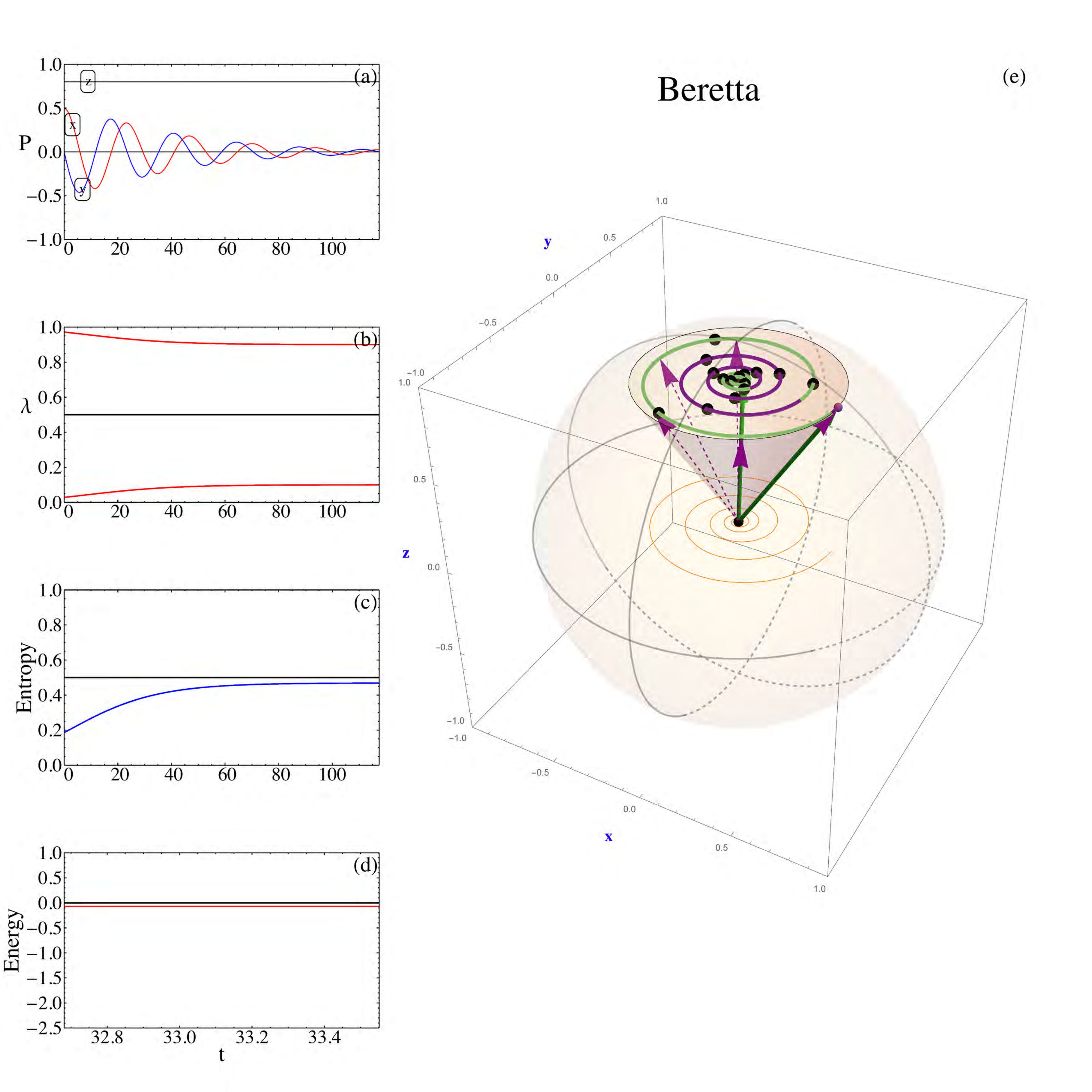}  
    \caption{Density matrix evolution with unitary plus Beretta closed-system term.  The polarization in the z-direction stays fixed, while the perpendicular components decay.  No energy flows into or out of the system and the temperature is fixed by the initial value.  The entropy increases steadily.  }
\protect\label{beretta1}
\end{figure}

 Now let us consider the case of unitary evolution, absent noise but subject to the Beretta term
 ${\cal L}_2  $.
 For a simple initial state as stipulated in Table 1, we see in Figure~\ref{beretta1} that for a closed system as described by ${\cal L}_2$, the entropy increases uniformly and no heat is transferred.  This is accomplished by a steady reduction
 of  the polarization vector components in directions perpendicular the axis associated with the level splitting, in this case the $\hat{z}$ axis.  Since both $P_x\  \&\  P_y$ decrease, while $P_z$ remains fixed, it follows that the entropy increases and the purity decreases since they both depend on the length of the polarization vector, which gets reduced.
 The temperature of the system is dictated by the unchanging  $P_z$ component, so the closed system has a fixed temperature
 determined by the initial condition.  That is consistent with  the no heat transfer property of the Beretta term.  
 
\begin{figure}[h]       
    \includegraphics[width=\textwidth]{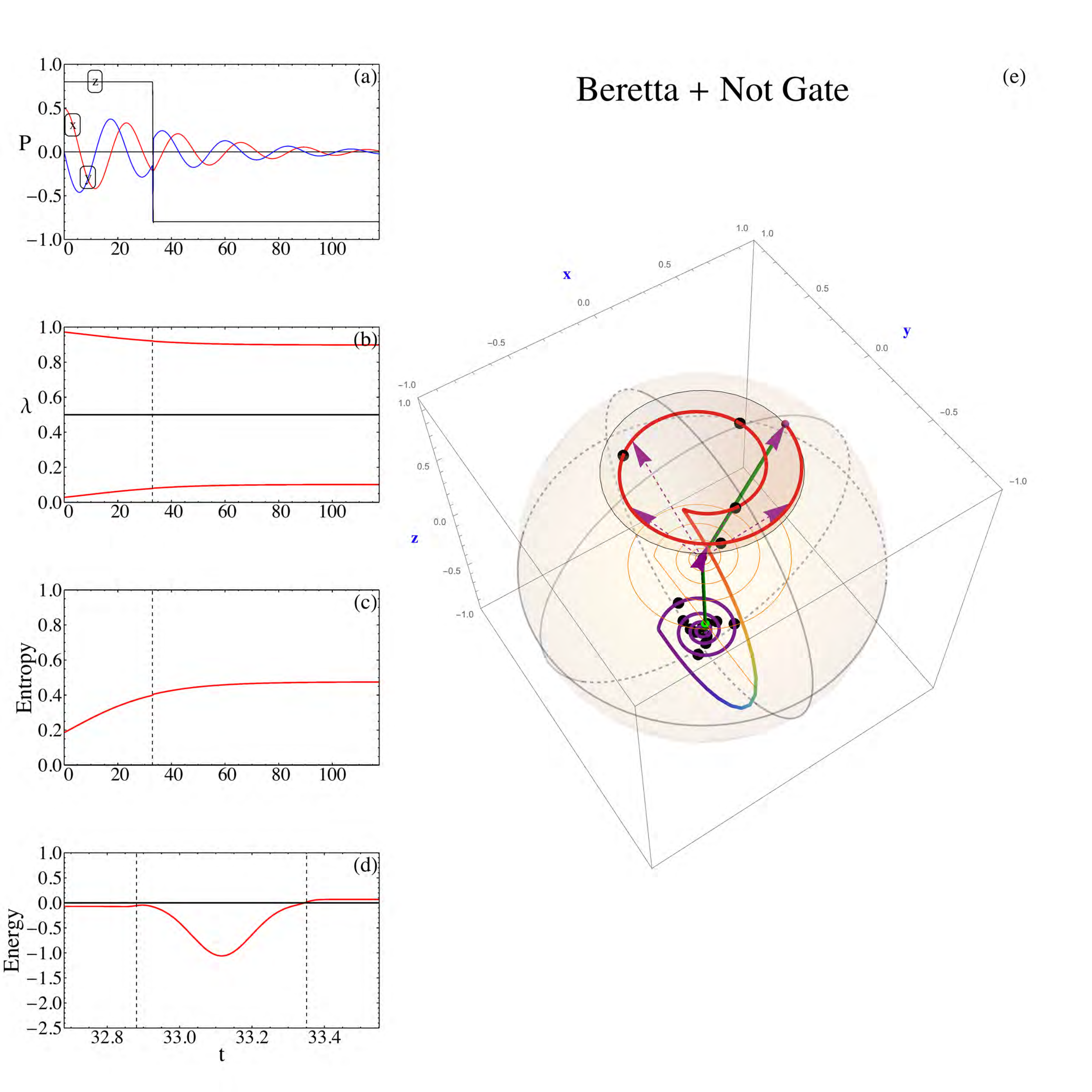}  
    \caption{ Density matrix evolution with unitary evolution including a Not gate, plus the Beretta closed-system term.   The polarization in the z-direction flips, while the perpendicular components decay. Energy of the system changes because of the Not gate. The entropy increases steadily.  Temperature experiences an inversion.  Vertical dashed lines indicate the start and end times of the Not gate.}
\protect\label{beretta2}
\end{figure}

  Figure~\ref{beretta2}  shows the same situation when a Not gate is included.
  
  These statements generalize to multi-qubit and qutrit system, which involve more spin observables.

\clearpage

 \subsection{Equilibrium-Thermal Bath }
 \label{sec4e}
 
\begin{figure}[h]       
    \includegraphics[width=\textwidth]{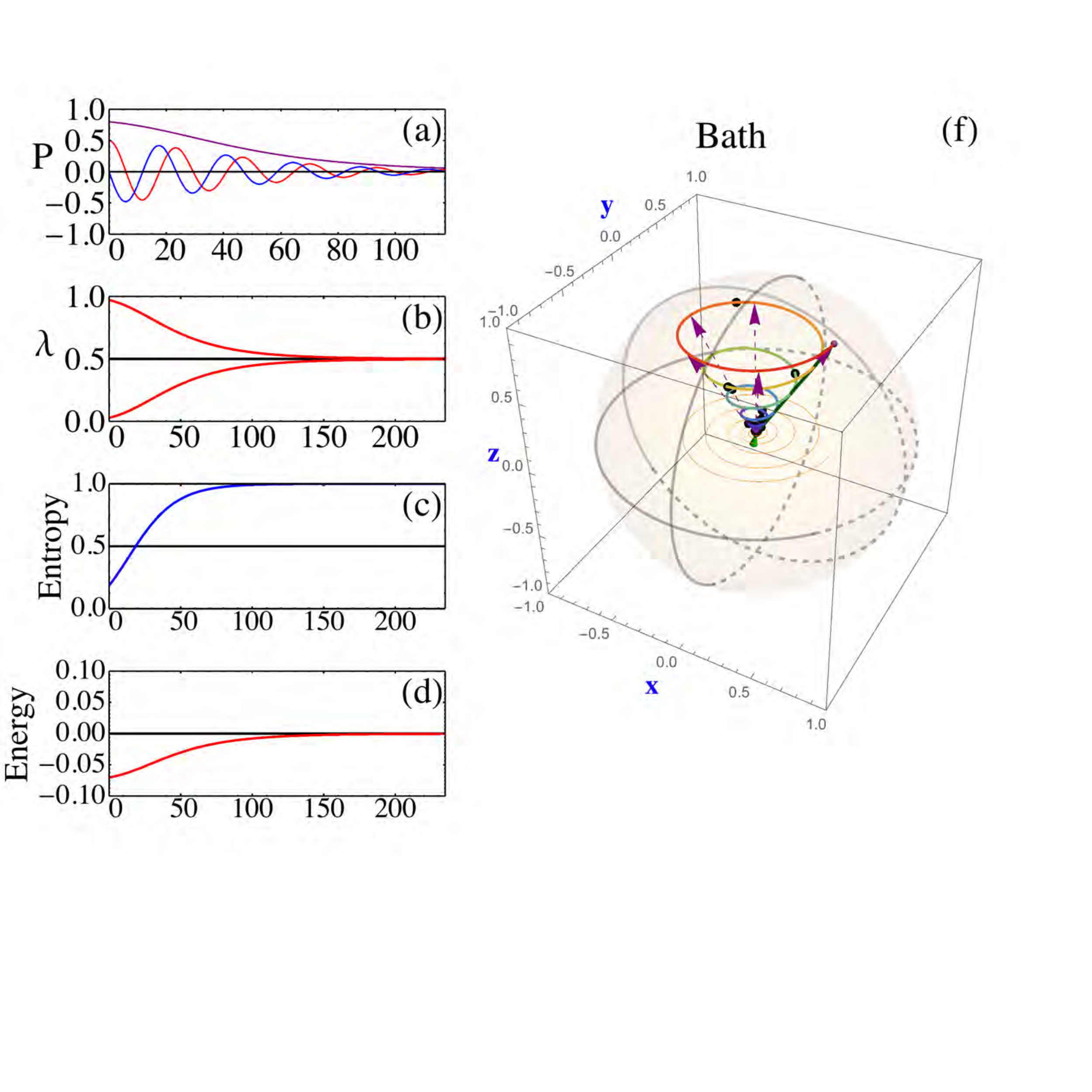}  
    \caption{ Evolution with unitary plus Korsch bath  terms. Bath temperature is  $T =  0.273\, {\rm Kelvin}$ and initial state temperature is  $T_0 = 0.00093\, {\rm  Kelvin}.$  Equilibrium eigenvalues are both close to 1/2, corresponding to a maximum entropy of almost 1.  The final polarization vector is $\vec{P}=\{0, 0 , 0.0037\},$  which is determined by the bath $T =  0.273\,{\rm Kelvin}.$   }
\protect\label{bath1}
\end{figure}

 In Figure~\ref{bath1}, the case of unitary evolution absent gates, but with a  Korsch (K) type bath ${\cal L}_3 $ is shown.  The temperature of the bath is stipulated
 by setting $T =  0.273\,  {\rm Kelvin}$ and thus $\beta=\frac{1}{k_B T}= 0.0425\,  (\mu ev)^{-1}$  .  Since $T$  is higher than the temperature obtained from the initial density matrix $T_0= 0.00093\, {\rm  Kelvin},$   heat flows into the system, polarization moves closer to zero, entropy goes closer to 1.
 (When the bath temperature is lower than the initial temperature $T_0$, heat flows out and the final
 z-polarization is larger than the initial value.)
  The final equilibrium density matrix
 is of Gibbs form, as expected. 
 
\begin{figure}[h]       
  \includegraphics[width=\textwidth]{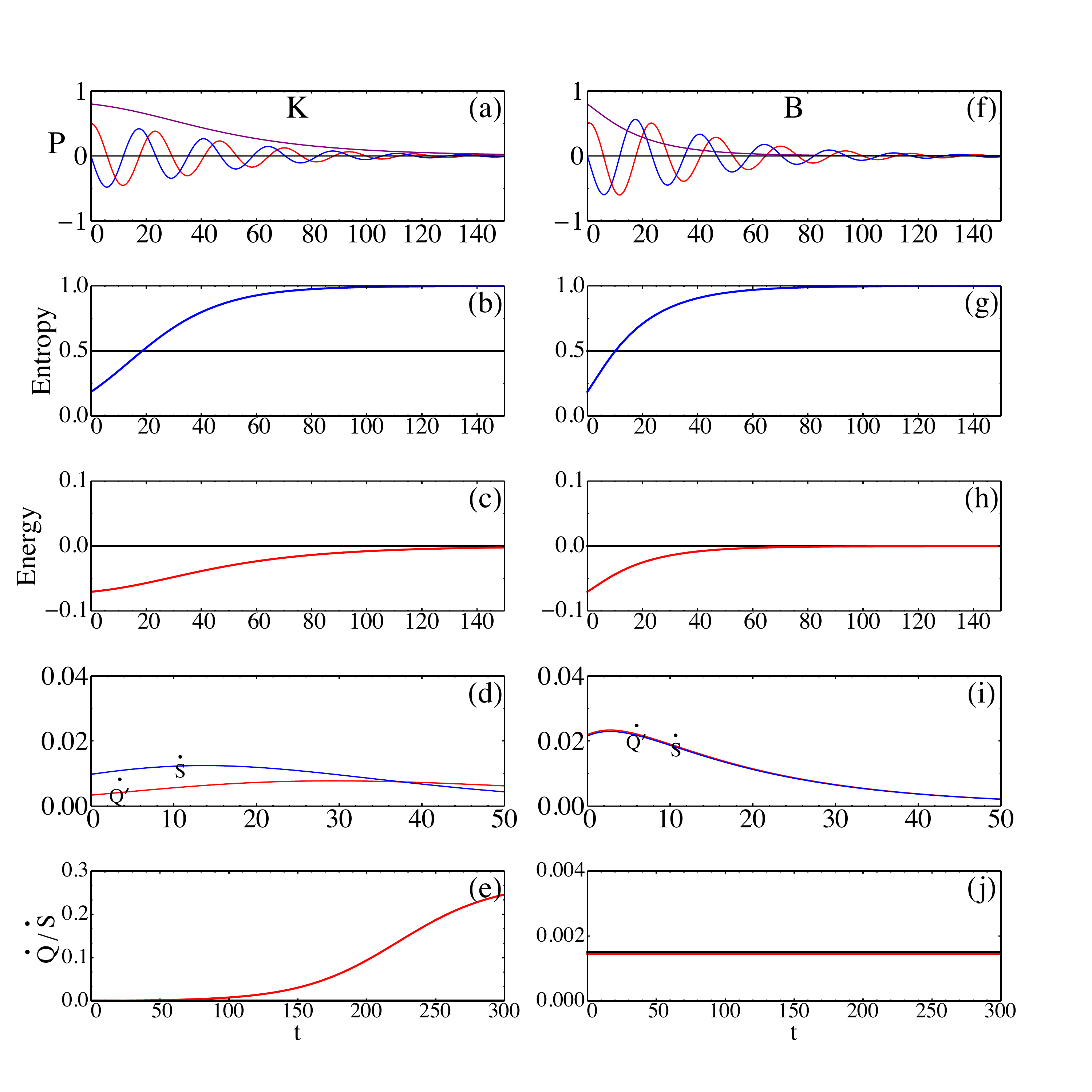}  
  \caption{ Comparision of the Korsch (K) and Beretta (B) baths.  Here there are no gate, Lindblad or closed system terms.  The Beretta clearly displays a steady value for $ \dot{Q}/ \dot{S}$ of $k_B T=   .$   The Korsch bath has a constant value $\beta_3=  ,$  but the   $ \dot{Q}/ \dot{S}$ is very different.  Energy unit is  $\mu$eV  and time in nsec .  Note: Beretta bath equilibriates first.  }
\protect\label{bathnew1}
\end{figure}

In Figure~\ref{bathnew1}, a special comparison of  Korsch(K) and  Beretta(B)  cases are compared for the same situation,  $T_Q$  was selected
~\footnote{ The value of $T_Q$ was set by the overall ratio $(Q(t_f)- Q(0) ) / (S(t_f)- S(0) ),$ where the final values are determined by $\beta_3.$  } to give similar results for the polarization, entropy, and energy evolutions. 
 Recall that for a (B) bath the ratio   $ \dot{Q}/ \dot{S}$ is held fixed by $T_Q $, while
the parameter $\boldsymbol{\beta_3(t)}$ is state and hence time-dependent; while, for a (K) bath  $\boldsymbol{\beta_3(t)}$ is kept fixed,  but the 
$ \dot{Q}/ \dot{S}$  ratio varies with time as seen in Fig.~\ref{bathnew1}. The $ \dot{Q}',  \  \dot{S}$ and $ \dot{Q}/ \dot{S}$ display interesting differences.
A rescaling  $\dot{Q}'=  8.2 \dot{Q}$ is used for a closer look at the detailed evolution, where for (B) the energy and entropy rates  are instantaneously  correlated.  This seems a correct property for an ideal bath, in contrast to  a time lag between an entropy rate peak followed by a later energy transfer peak as seen for the Korsch case.  The (K) shows a smooth and substantially increasing
  $ \dot{Q}/ \dot{S},$  but an expected steady value for the
(B) bath.  The net change in energy and in energy are the same for both cases by design.
Note that equilibrium is reached somewhat earlier for the (B) bath.

%\clearpage
\subsection{Equilibrium- Thermal Bath plus Lindblad }
 \label{sec4f}
\begin{figure}[h]       
  \includegraphics[width=\textwidth]{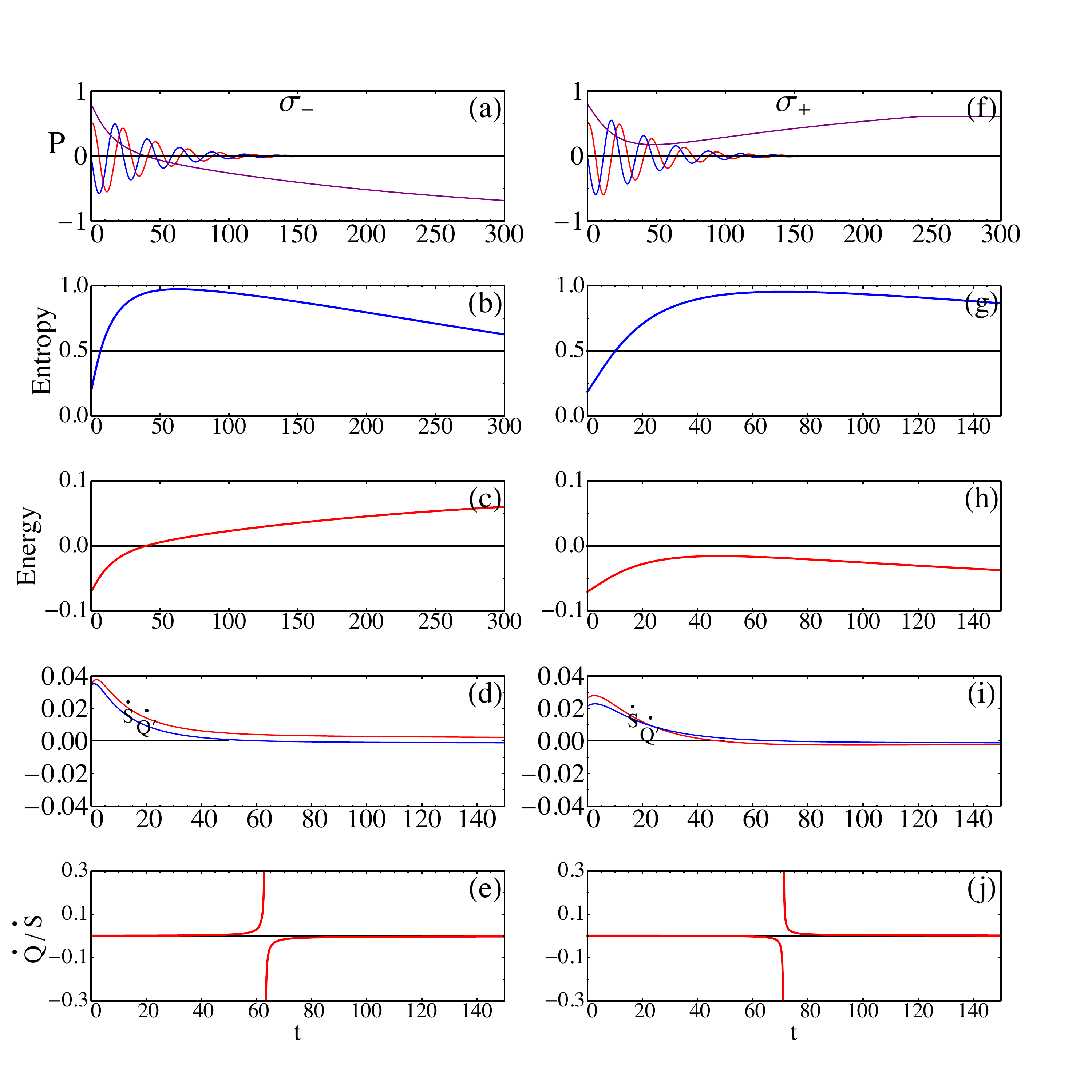}  
  \caption{ Sample result for a Beretta (B) bath with steady $\sigma_{-}$ and  $\sigma_{+}$  Lindblad operators.  Here  gate and closed system terms
  are omitted.  The Beretta  displays a steady value for $ \dot{Q}/ \dot{S}$ except in $\bar \dot{S}_1 +\dot{S}_3  \bar\rightarrow 0 $ regions induced by 
  $\sigma_{\pm }$ Lindblad operators.  }
\protect\label{bathnew2}
\end{figure}

 To gain additional insight concerning the Beretta (B) bath density matrix evolution, consider a simple case with no gates,  and steady Lindblad operators 
of either  $\sigma_{-}$ or  $\sigma_{+}$ form; these are compared  Figure~\ref{bathnew2}.
 The $ \dot{Q}/ \dot{S}$ ratio now consists of the fraction~\footnote{ The subscript labels are 1: Lindblad,  2:  closed system,  3:  Bath.} 
 \be
\frac{\dot{Q}} {\dot{S}}=  \frac{\dot{Q}_1 + \dot{Q}_2+\dot{Q}_3}  { \dot{S}_1 + \dot{S}_2+\dot{S}_3  }=
  \frac{\dot{Q}_1 +\dot{Q}_3}  { \dot{S}_1 + \dot{S}_2+\dot{S}_3  } \rightarrow  \frac{\dot{Q}_1 +\dot{Q}_3}  { \dot{S}_1 +\dot{S}_3  }\, , 
\ee where the closed system term is now omitted (when on, it has:  $\dot{Q}_2=0,$ and    $\dot{S}_2 > 0$).

For a $\sigma_{-}$ Lindblad term Figure~\ref{LindS-}, $\dot{Q}_1$ is positive and decreasing and $\dot{S}_1$  starts positive, decrease to zero
at $t \approx 40$  nsec, then goes to a small negative value. 
For a $\sigma_{+}$ Lindblad term Figure~\ref{LindS+},  both $\dot{Q}_3$  and $\dot{S}_3$  are both negative decreasing  Meanwhile, the (B) bath contributions has positive  $\dot{Q}_3 , \dot{S}_3.$

 For  $\sigma_{+},$ starts positive and becomes negative as the upper level
 gets more occupied and the entropy heads for zero.
   Recall that a $\sigma_{-}$ Lindblad first increase the entropy and then drives it towards zero,  while also flipping the z-polarization.  In contrast the $\sigma_{+}$ Lindblad depopulates the upper level and drives the entropy down from its original value, with a corresponding increase in the z-polarization.    
   
   As a result of these properties, the denominator changes sign when the Lindblad entropy slope become strong enough to
 exceed the  positive  bath entropy rate of increase.
 
 As a result, the $\frac{\dot{Q}} {\dot{S}}$ changes sign and deviates from its steady value where such cancelations occur.
 The up-down-up spike for $\sigma_{-},$  near 62 nsec and the  down-up-down spike for $\sigma_{+},$  near 70 nsec follow
  from the differing  $\dot{Q}_1,\dot{Q}_3$ and   $\dot{S}_1,\dot{S}_3$ signs and trajectories near the entropy nodal regions.  It would be interesting to see if such blips appear with realistic noise.

%\clearpage 

\section{One Qubit System and the Full Model Master Equation}
\label{sec5}
The full model master equation includes unitary evolution with  gate pulses,  the Lindblad ${\cal L}_1 $  with noise pulses,
 the Beretta ${\cal L}_2 $  to describe a closed system, and  a bath ${\cal L}_3 $ term to include contact with a bath of fixed temperature.   This provides a flexible model that can be used to gain insight into QC dynamics and gauge the requisite condition for a successful QC process.  We give a simple example here,  with additional cases and tools to be posted. 
 
\subsection{Full master equation Not gate}
\label{sec5a}
\begin{figure}[!tbp]       
    \includegraphics[width=\textwidth]{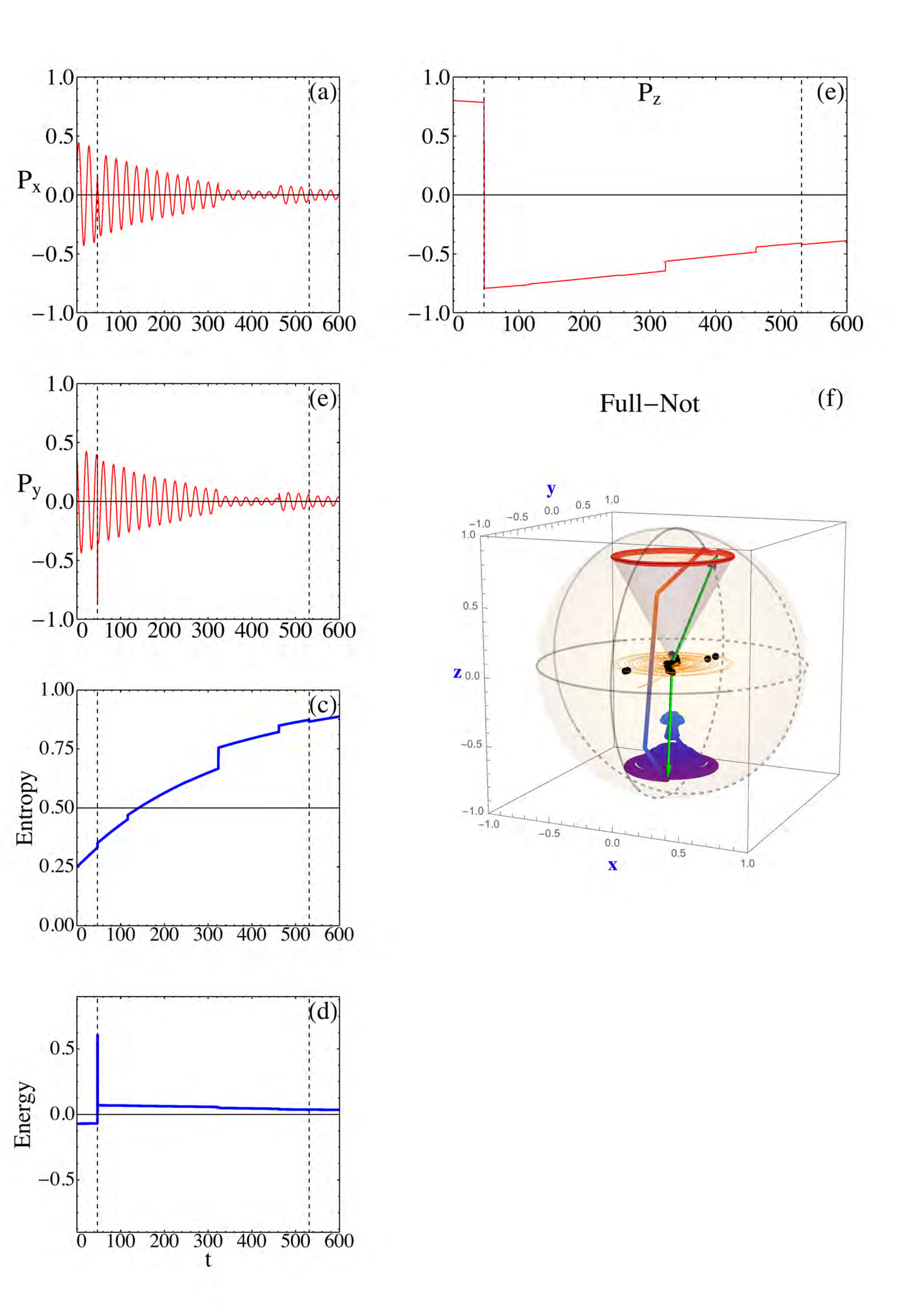}  
    \caption{ A Full master equation case with a Not gate, Lindblad noise and Beretta and Bath terms.  }
\protect\label{full1}
\end{figure}

In Figure~\ref{full1}, a full model master equation case is displayed with a single Not gate.
The initial polarization $\vec{P} =  \{0.2  ,0.4 , 0.8\}$ precesses about the z-axis with a Larmor angular frequency of
$\omega_L= 0.2675\ GHz.$   Eight random equi-spaced Lindblad pulses act during the t=46.9  to  531 nsec interlude;
the overall Lindblad strength is set 
as $\Gamma =  0.4\, \omega_L .$ 
 At 46.9 nsec a Not gate acts.
The Beretta (closed system) strength is set as $\gamma_2=    .01\,  \omega_L ,$  The bath term strength 
$\gamma_3=    .005\, \omega_L ,$ and the bath temperature is $  27.3\, {\rm Kelvin}.$  The evolution of the polarization
shows a  $P_z$  gate flip  followed by attenuation and noise alterations,  as expected.  The Lindblad noise shows up as jagged entropy evolution, where the random nature of the Lindblad pulses allows for entropy decreases as well as increases.  The energy plot shows the Not and bias pulse work
and the energy flip to increased energy occupation.

\begin{figure}[!tbp]       
    \includegraphics[width=\textwidth]{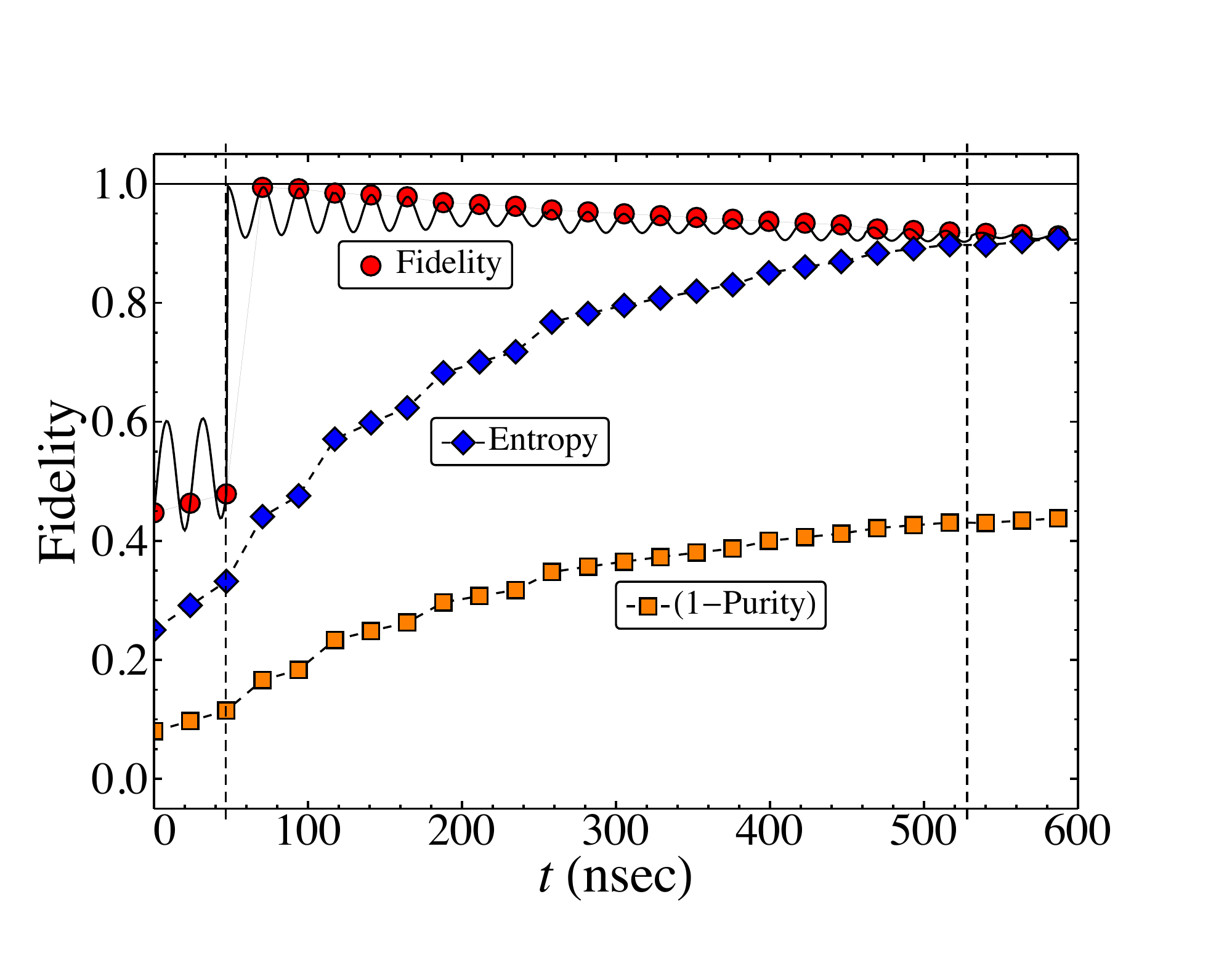}  
    \caption{ Fidelity with a Not gate, noise and Beretta and Bath terms.
 Between the vertical dashed lines eight Lindblad noise pulses occur.
Fidelity , entropy and purity are shown on the Larmor period grid. The full master equation density matrix is compared to the expected $\rho_e=\sigma_1\cdot \rho(0)\cdot\sigma_1,$
    by examining the Fidelity($\rho(t),\rho_e$). The fidelity jumps to one after the gate, then drifts down due to noise, Beretta and Bath terms. }
\protect\label{full2}
\end{figure}

In Fig.~\ref{full2}, the fidelity, entropy and purity evolutions are presented with values on the Larmor grid (integer multiples of $T_L$) indicated by red dots (fidelity), blue diamonds (entropy) and  orange squares (1-purity).  There is a clear reduction in fidelity due to noise and gradual fall off from the Beretta and bath effects. Both entropy and purity  reveal the 
affect of Lindblad noise.

\begin{figure}[!tbp]       
      \includegraphics[width=\textwidth]{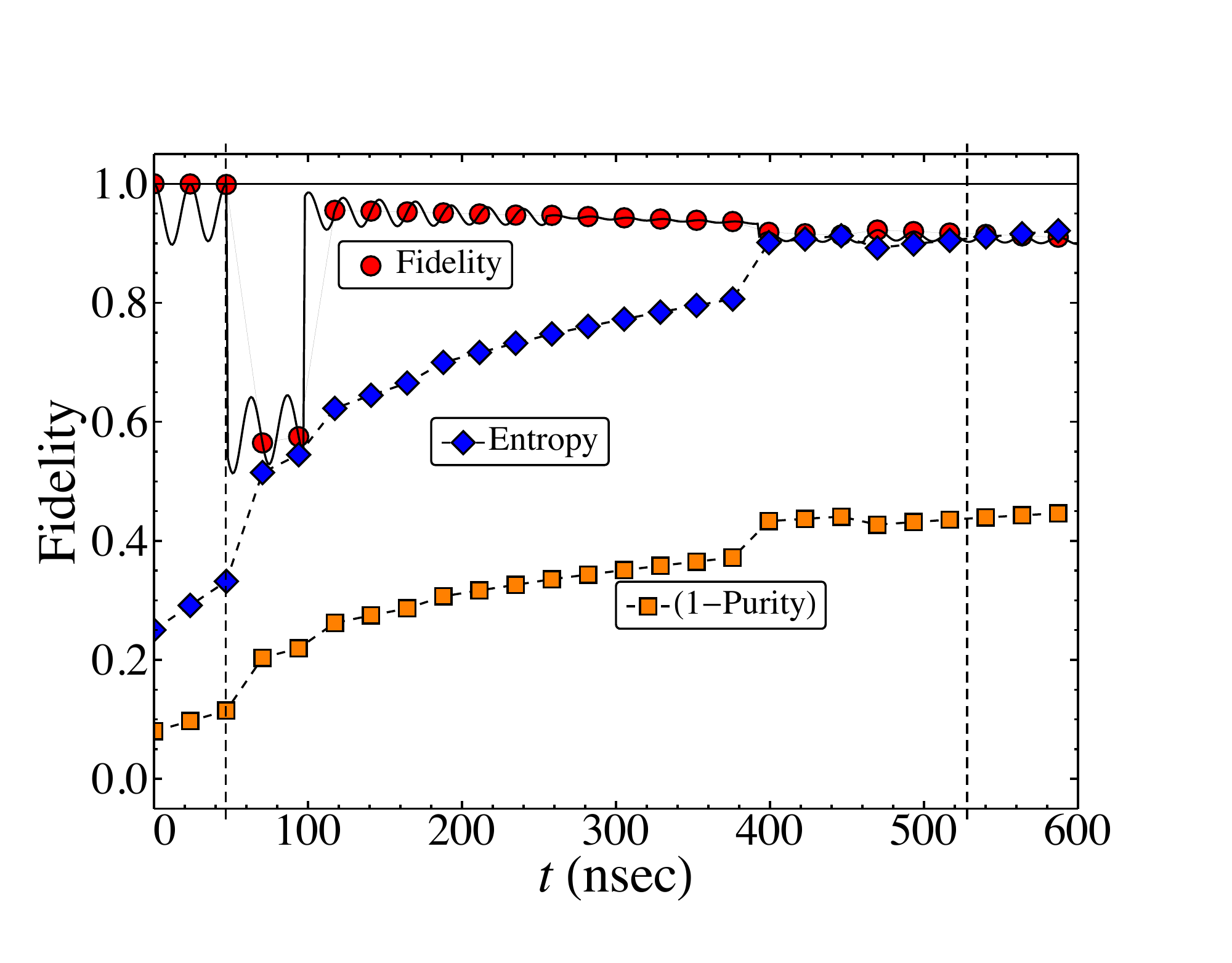}
    \caption{ Fidelity for two sequential not gates, Beretta and bath terms as in Fig.~\ref{full2}.  Noise occurs between the vertical dashed lines, but differs from prior plot. After the second not gate, the fidelity does not equal 1, due to the noise, Beretta and bath effects. }
\protect\label{full3}
\end{figure}

In  Fig.~\ref{full3}, the case of two sequential Not gates is shown.  Studies of other gate sequences and alternate noise scenarios reveal similar properties.  Cases of no gates, but noise, Beretta and bath terms can be used to identify quantum memory losses.


\clearpage 
\section{Conclusions and Future Steps}
\label{sec6}
\subsection{Conclusions}

The main result from this study is the design of a dynamical  density matrix model that incorporates the essential features of a quantum computer.
Although much of the input is well-known,  it is shown here how to implement unitary gate pulses, plus an associated bias pulse, to replicate the usual quantum gates.  The bias pulse is introduce to obviate the accumulation of detrimental phase accumulations due to qubit non-degeneracy.
To replicate the QC static gate network for a sequence of gates, it is shown that the various gates also need to be applied on the Larmor time grid.  The model also includes dissipative, decoherence and thermodynamic effects.  The Lindblad addition to unitary dynamics has the essential feature of maintaining the unit trace, Hermiticity and positive definite nature of the evolving density matrix.  A general  Lindblad form, after examining various static Lindblad operators, is used to incorporate random noise and gate friction effects; these are input as non-static pulses.  In addition, strong rapid  Lindblad operators are implemented as measurements, and the associated restriction to be valid measurements examined.  Although many other effects can also  be cast into Lindblad form,  it is much simpler to design separate forms for closed systems and for system-Bath interactions.  The closed system form
is one developed by Beretta based on a general study of non-equilibrium thermodynamics.  Two types of system-Bath interactions are examined,
one that has been studied before, and another one also originated by Beretta that is illustrated herein to have physical advantages.  Another result of this study is provided by several examples of how to apply the full model including random noise, gate friction, closed system entropy increase, and system-Bath interactions to a set of unitary gates.  Fidelity is used to gauge the stability of such a QC setup.  This illustrates how the model can be used as a tool  to examine and design valid experiments to achieve stable quantum computation.

\subsection{Future Steps}
Clearly, there is much more to explore.  Extension to two or more qubit systems are planned along the lines delineated in this work, as well as generalization to qutrit and hybrid qubit/qutrit systems.  Application to small Carnot and Otto cycle
 qubit engines  would aid in clarification of non-equilibrium dynamics and of non-ideal QC~\cite{engines1,engines2}.   Such studies will be greatly influenced by the preferred choice of a Beretta type bath model. 
 
 Lindblad pulses as measurement operators warrants further exploration.  One possibility is to generate a stochastic  Schr\"odinger equation that approximates this density matrix model, as a means of exploring quantum wave function collapse.
Of course, the main application should be to QC algorithms using realistic parameter settings based on extant experiments
to explore the requisite conditions for good fidelity results.  Error correction methods can also be evaluated for their efficacy.

It might also be possible to invent pulses or chirps to simulate  non-Markovian dynamics as a potential
mechanism for enhancing QC memory and algorithm efficiency.

It is hoped that the methods explored here will help in these directions.

\section*{Acknowledgments}
    We gratefully acknowledge the participation of  Dr. Yangjun Ma  and Dr. Victor Volkov at an early stage of this work.  One of the reviewers enhanced this paper considerably by excellent suggestions and in-depth insights, for which the author is deeply grateful.
    The figures for this article have been created using the  SciDraw-LevelScheme scientific figure preparation system [M. A. Caprio, Comput. Phys. Commun. 171, 107 (2005) and http://scidraw.nd.edu]. The author also thanks Prof. Edward Gerjuoy for stressing the need to examine non-degenerate qubits, and Dr. Robert A. Eisenstein for encouragement and excellent advice during the completion of this paper.
    %This project was supported earlier by the U.S. National Science Foundation.  
\clearpage
\appendix*

\section{One-Qubit Lindblad in Simplified Form}

To gain insight into the effect of the Lindblad operators on the evolution of the density matrix as stipulated by the motion of the polarization vector, let us generate the equation for the time dependence of $\vec{ P}(t).$   The general form of the Lindblad term is

\begin{eqnarray}
	\mathcal{L}_1&=&  \Gamma\    \overrightarrow{\delta_{\!P}}\!(t) \cdot \vec{\sigma}  \\ \nonumber
\end{eqnarray}where $   \overrightarrow{\delta_{\!P}}\!(t)$ is a real vector in (x,y,z) spin space.  That form arises from the fact that $ \mathcal{L}_1 $ is a traceless and Hermitian scalar.  Our task is to express  $   \overrightarrow{\delta_{\!P}}\!(t)$ in terms of the Lindblad parameters $\alpha_i.$

 The first step in extracting that result is to take the ensemble average
	\begin{eqnarray}\label{Pdot1} 
	\frac{ d }{dt} {\rm Tr }{\Big \lbrack} \vec{\sigma} \rho(t){\Big \rbrack} &=& 
{\rm Tr }{\Big \lbrack}  \vec{\sigma}  \frac{ d }{dt} \rho(t){\Big \rbrack}=\frac{ d }{dt}\vec{P}(t)  \\ \nonumber
 \frac{ d }{dt}\vec{P}(t) &=& - \frac{i}{\hbar}  {\rm Tr }{\Big \lbrack}  \vec{\sigma}   {\Big  \lbrack}  H(t), \rho(t){\Big  \rbrack}       {\Big \rbrack} 
         + {\rm Tr }{\Big \lbrack}  \vec{\sigma}   \mathcal{L}_1 \        {\Big \rbrack} .    
\end{eqnarray}  For a Hamiltonian $ H \equiv - \frac{\hbar}{2} \  \omega_L \   \vec{\sigma}\cdot \hat{h}$, the first term in Eq.~\ref{Pdot1} becomes
$$  - \frac{i}{\hbar} {\rm Tr}{\Big \lbrack}  \vec{\sigma} \,   {\Big  \lbrack}  H(t), \rho(t){\Big  \rbrack}       {\Big \rbrack} =
- \omega_L \   \hat{h}\times \vec{P}(t),$$  which yields a Larmor precession of the polarization vector about the direction $\hat{h}.$  We then have
\be \label{Lrole} \frac{ d }{dt}\vec{P}(t)= - \omega_L \   \hat{h}\times \vec{P}(t) + 2\ \Gamma\     \overrightarrow{\delta_{\!P}}\!(t).\ee
For the second  term in Eq.~\ref{Pdot1} , after inserting the expansions of  both the density matrix and the Lindblad and taking the traces, one arrives at the following result:
 \begin{eqnarray}\label{Omega}
  \overrightarrow{\delta_{\!P}}\!(t)&=& -a^2 \, [  \vec{P}(t) -(\hat{a}\cdot \vec{P}(t))\ \hat{a}  ] 
  -b^2 \, [  \vec{P}(t) - (\hat{b}\cdot \vec{P}(t))\ \hat{b}  ]  \\ \nonumber
	&+&	\vec{P}(t)\times(a_0\  \vec{b}-b_0\  \vec{a})
	 + 2 (\vec{a}\times \vec{b}).
\end{eqnarray} We have expressed  the eight complex Lindblad parameters $\alpha_0, \vec{\alpha}$ in terms of  eight real quantities by the definition  
$\alpha_j= a_j + i\, b_j $ for $ j= 0 \dots 3.$ Recall that $L = \alpha_0\, \sigma_0 + \vec{\alpha}\cdot\vec{\sigma}.$

The three components of the vector $  \overrightarrow{\delta_{\!P}}\!(t)$  give the change in the associated component of the polarization vector induced by the Lindblad term and allows one to identify how the Lindblad coefficients ( and the corresponding  Pauli operators) affect the motion of the polarization vector within the Bloch sphere.  For example for $b_0\rightarrow0,$ and
$\vec{b}\rightarrow 0$ The result simplifies to
 \begin{equation}\label{Omega2}
  \overrightarrow{\delta_{\!P}}\!(t)\rightarrow -a^2 \, [  \vec{P}(t) -(\hat{a}\cdot \vec{P}(t))\ \hat{a}  ],  \end{equation} which yields a reduction in the polarization components in the direction perpendicular to $\hat{a}\, .$  For  $\hat{a}$ in the x-y plane, this would take a polarization vector precessing about say the z-axis~\footnote{In this case,  we have assumed that the Hamiltonian is of the simple form
$ H = - \frac{\hbar}{2} \  \omega_L \   \vec{\sigma}_3$ } and move it into the Bloch sphere in a spiral motion towards
 a limit of  zero polarization or maximum entropy.  This is clearly a dissipative or friction situation. 
 
 Equation~\ref{Lrole} also gives the time derivative
  of $\vec{P}(t)\cdot\vec{P}(t),$ as
  \begin{eqnarray}\label{Pmag}
 \frac{ d }{dt}\vec{P}(t)\cdot\vec{P}(t)&=&  2\, \vec{P}(t)\cdot\frac{ d }{dt}\vec{P}(t)  \\ \nonumber
 &=&4\ \Gamma\ \vec{P}(t)\cdot \overrightarrow{\delta_{\!P}}\!(t)  \\ \nonumber
 &=&  4\ \Gamma\ \{ -a^2 \, {\bf P}^2 [ 1 -(\hat{a}\cdot \hat{n})^2 \ \ ] 
 -b^2 \, {\bf P}^2 [ 1 -(\hat{b}\cdot \hat{n})^2 \ \ ]   \\ \nonumber
	&+&		  2\,a\, b\,   (\hat{a}\times \hat{b})\cdot\vec{P}
	\ \}.
 \end{eqnarray}  Since  $ (\hat{a}\cdot \hat{n})^2 $ and
  $ (\hat{b}\cdot \hat{n})^2 $ are both less than 1, the first two terms reduce ${\bf P}.$  The third term contributes only when 
 $ (\hat{a}\times \hat{b})\neq 0,$ and it can be positive or negative.
 Since $ [ L , L^\dagger ] = 4 \,a\, b\, (\hat{a}\times \hat{b})\cdot\vec{\sigma},$ we see that the polarization vector decreases 
 when  $ [ L , L^\dagger  ] = 0,$  but can increase when
 $ [ L , L^\dagger  ] \neq 0.$  since the purity, entropy and eigenvalues depend only on {\bf P}, this shows that purity decreases, entropy increases and eigenvalues move towards 1/2, when  $ [ L , L^\dagger ] = 0,$ and 
  purity can increases, entropy can decrease and eigenvalues can move towards 1 and 0, when  $ [ L , L^\dagger ] \neq 0.$  For example, when
 $ \vec{a}= n_x\, / \sqrt{2}, $ and $ \vec{b}= \pm\, n_y\, / \sqrt{2}, $ 
 the third term becomes  $  \pm \, 2 \,a\, b\, P_z ,$ which increases ${\bf P}$ when $\pm P_z\,  \rangle\,0$ and decreases ${\bf P}$ when $\pm P_z\,  \langle\,0.$  This explains the numerical results for the role of Lindblad operators.

 The role of each Lindblad  operator, as stipulated by the $a_j, b_j$ parameters, can thus be understood based on Eqn~\ref{Lrole}.  Equation~\ref{Omega2} plays an important role when Lindblad is a  measurement operator.
  
 The eight  real parameters $a_j, b_j, j=0 \dots 3$ are used as time-dependent pulses
 to simulate external dissipative and decoherence effects.


\clearpage

\begin{thebibliography}{9}

\bibitem{Nielsen} Michael A. Nielsen and Isaac I. Chuang,
``Quantum Computation and Quantum Information'', Cambridge
University Press (2000).

\bibitem{Preskill}  J. Preskill's remarkable lectures available at:
 http://www.theory.caltech.edu/people/preskill/ph229/  .

 \bibitem{Joos} E. Joos, H. D. Zeh, C. Kiefer, D. Giulini, J. Kupsch, and I.-O. Stamatescu, "Decoherence
and the Appearance of a Classical World in Quantum Theory," Springer,
Berlin, 2nd edition, 2003.
  \bibitem{Zurek}  ``Quantum Theory and Measurement,"
 John Archibald Wheeler \& Wojciech Hubert Zurek (1983).
\bibitem{Hornberger}  ``Introduction to decoherence theory," K. Hornberger, Lect. Notes Phys. 768, 221-276 (2009).

\bibitem{Kraus} K. Kraus, ``States, Effects and Operations: Fundamental Notions of Quantum Theory", Springer Verlag 1983.
\bibitem{Carmichael}Howard Carmichael, ``Statistical Methods in Quantum Optics 1: Master Equations
and Fokker-Planck Equations," (Springer, Berlin, 1999).
 \bibitem{Alicki} Alicki, Lidar, and  Zanardi, Physical Review A 73, 052311 2006.
\bibitem{Adler} Adler, S. L., Physics Letters A,  265,  58 (2000).
and ÒCorrigendum to: ÕDerivation of the Lindblad generator structure by use of the It™ stochastic calculusÕ Ó, Phys. Lett. A 267, 212 (2000).
\bibitem{Peres} Asher Peres,
 Physical Review A 61 022116  (2000).

\bibitem{Lindblad1} Kossakowski, A.  ``On quantum statistical mechanics of non-Hamiltonian systems". Rep. Math. Phys. 3 (4): 247(1972). 
  \bibitem{Lindblad2}
   Lindblad, G. ``On the generators of quantum dynamical semigroups," Commun. Math. Phys. 48 (2): 119  (1976). 
   \bibitem{Lindblad3}Gorini, V., Kossakowski, A., Sudarshan, E.C.G. , ``Completely positive semigroups of N-level systems," J. Math. Phys. 17 (5): 821(1976). 
\bibitem{Lindblad4}Lindblad, G. ,``Non-Equilibrium Entropy and Irreversibility," Dordrecht:  Reidel, (1983).
 \bibitem{Beretta1} G.P. Beretta, "Nonlinear Quantum Evolution Equations to Model Irreversible Adiabatic Relaxation With Maximal Entropy Production and Other Nonunitary Processes," Reports on Mathematical Physics, Vol. 64, pp. 139-168 (2009).

 \bibitem{Beretta2}  G.P. Beretta, "Well-behaved nonlinear evolution equation for steepest-entropy-ascent dissipative quantum dynamics," International Journal of Quantum Information, Vol. 5, 249-255 (2007).

 \bibitem{Beretta3} G.P. Beretta, "Maximum entropy production rate in quantum thermodynamics," Journal of Physics: Conference Series, Vol. 237, 012004, pp. 1-32 (2010). See also G.P. Beretta, "Time-energy and time-entropy uncertainty relations in dissipative quantum dynamics," ArXiv-quant-ph-0511091, 2005.

\bibitem{Korsch}H.J.Korsch and H.Steffen, J.Phys.A 20, 3787 (1987) and 
  M.Hensel and H.J.Korsh, J.Phys.A 25, 2043 (1992) .
 


\bibitem{Neumann}Von Neumann, John (1932). Mathematische Grundlagen der Quantenmechanik. Berlin: Springer. ISBN 3-540-59207-5.; Von Neumann, John (1955). Mathematical Foundations of Quantum Mechanics. Princeton University Press.
\bibitem{Landau} Landau, L. (1927). "Das D\"{a}mpfungsproblem in der Wellenmechanik". Zeitschrift f\"{u}r Physik 45 (5Ð6): 430Ð464. 
\bibitem{Fano}
U. Fano,``Description of States in Quantum Mechanics by Density Matrix and Operator Techniques,"
Rev. Mod. Phys. 29, 74 (1957).
 \bibitem{Weinberg1}Steven Weinberg,
``Quantum mechanics without state vectors,"
Phys. Rev. A 90, 042102  (2014).
\bibitem{Weinberg2}
Steven Weinberg,``What happens in a measurement?"
Phys. Rev. A 93, 032124 (2016).

\bibitem{Hatsopoulos} G.N. Hatsopoulos and E.P. Gyftopoulos, "A unified quantum theory of mechanics and thermodynamics," Found. Phys., Vol. 6, 15-31, 127-141, 439-455, 561-570 (1976).

\bibitem{Park} J.L. Park, "Nature of quantum states," Am. J. Phys., Vol. 36, 211 (1968). See also R.F. Simmons, Jr. and J.L. Park, "On completely positive maps in generalized quantum dynamics," Foundations of Physics, Vol. 11, 47 (1981); and  R.F. Simmons, Jr. and J.L. Park, "The essential nonlinearity of N-level quantum thermodynamics," Foundations of Physics, Vol. 11, 297 (1981).

\bibitem{qdensity} Bruno Juli\'a-D\'{\i}az, Joseph M. Burdis and Frank Tabakin,  
``QDENSITY - A Mathematica Quantum Computer simulation,"
Comp. Phys. Comm., 174 (2006) 914-934. Also see:
Comp. Phys. Comm.,180, (2009) 474,   Comp. Phys. Comm., 182, (2011)1693,  and 
 Comp. Phys. Comm., 201 , (2016) 171.

\bibitem{feedback}``Effect of non-Markovianity on the dynamics of coherence, concurrence and Fisher information,"
Samyadeb Bhattacharya, Subhashish Banerjee, Arun Kumar Pati,
arXiv:1601.04742.
\bibitem{Tinversion1}Norman F. Ramsey
Phys. Rev. 103, 20  1 July 1956


\bibitem{Tinversion2}
S. Braun, J. P. Ronzheimer, M. Schreiber, S. S. Hodgman, T. Rom, I. Bloch, U. Schneider,
``Negative Absolute Temperature for Motional Degrees of Freedom,"
Science  339, Issue 6115,  04 Jan 2013.
Thermodynamics and Statistical Mechanics at Negative Absolute Temperatures.

\bibitem{engines1}
Alicki, R. The quantum open system as a model of the heat engine. J. Phys. A: Math. Gen. 12, L103 (1979);
Felipe Barra,    Scientific Reports 5, Article number: 14873 (2015).
         
\bibitem{engines2}
D Gelbwaser-Klimovsky, W Niedenzu, G Kurizki,
         Advances In Atomic, Molecular, and Optical Physics, 64, 329 (2015).


 \end{thebibliography}
\end{document}